\documentclass[%
  manyauthors,
  nocleardouble,
  COMPASS,
  american,
  fleqn,  %
]{cernphprep}

\usepackage[nottoc]{tocbibind}  %
\usepackage{cite}               %
\usepackage{caption}            %
\usepackage{multirow}           %
\usepackage[%
]{packages}
\makeatletter
\@ifpackageloaded{tikz}%
{}%
{\usepackage{tikz}}%
\makeatother

\hypersetup{%
  pdfborder={0 0 0.1}  %
}

\makeatletter
\g@addto@macro\bfseries{\boldmath}
\makeatother

\addto\captionsamerican{}  %
\crefname{section}{Sec.}{Secs.}                             %
\Crefname{section}{Section}{Sections}
\crefname{footnote}{footnote}{footnote}                     %
\Crefname{footnote}{Footnote}{Footnotes}
\renewcommand{\thesection}{\Roman{section}}%
\renewcommand{\thesubsection}{\Alph{subsection}}%
\renewcommand{\thesubsubsection}{\arabic{subsubsection}}%
\makeatletter
\renewcommand{\p@section}{}%
\renewcommand{\p@subsection}{\thesection\,}%
\renewcommand{\p@subsubsection}{\thesection\,\thesubsection\,}%
\makeatother
\usepackage[titles,subfigure]{tocloft}

\makeatletter
\renewcommand{\appendix}{%
  \par
  \setcounter{section}\z@%
  \setcounter{subsection}\z@%
  \setcounter{subsubsection}\z@%
  \renewcommand{\thesection}{\Alph{section}}%
  \renewcommand{\theHsection}{Dummy.\Alph{section}}%
  \renewcommand{\thesubsection}{\arabic{subsection}}%
  \renewcommand{\thesubsubsection}{\alph{subsubsection}}%
  \renewcommand{\p@section}{}%
  \renewcommand{\p@subsection}{\thesection\,}%
  \renewcommand{\p@subsubsection}{\thesection\,\thesubsection\,}%
  \renewcommand{\theequation}{\thesection\arabic{equation}}%
  \addtocontents{toc}{\protect\appendix}%
  \crefalias{section}{appendix}%
}%
\makeatother

\graphicspath{%
  {./},%
  {../tikz/},%
  {../images/},%
  {../plots/},%
  {../plots/mass_dependent_fit_14waves/},%
  {../plots/3pi_analyses},%
}

\usepackage{macros}

\newlength{\totalPlotWidth}
\newlength{\twoPlotWidth}
\newlength{\twoPlotSpacing}
\newlength{\threePlotWidth}
\newlength{\threePlotSpacing}
\newlength{\threePlotSmallWidth}
\newlength{\threePlotSmallSpacing}
\newlength{\fourPlotWidth}
\newlength{\fourPlotSpacing}
\begin{document}

\begin{titlepage}
  \PHnumber{2021--162}
  \PHdate{\today}
\title{Exotic meson \PpiOne[1600] with $\JPC = 1^{-+}$ and its decay into $\Prho \pi$}
   \makeatletter
  \ShortTitle{\@title}
  \makeatother
  \Collaboration{The COMPASS Collaboration}
  \ShortAuthor{}

\begin{abstract}
  We study the spin-exotic $\JPC = 1^{-+}$ amplitude in
  single-diffractive dissociation of \SI{190}{\GeVc} pions into
  \threePi using a hydrogen target and confirm the $\PpiOne[1600] \to
  \Prho\pi$ amplitude, which interferes with a nonresonant $1^{-+}$
  amplitude.  We demonstrate that conflicting conclusions from
  previous studies on these amplitudes can be attributed to different
  analysis models and different treatment of the dependence of the
  amplitudes on the squared four-momentum transfer and we thus
  reconcile these experimental findings.  We study the nonresonant
  contributions to the \threePi final state using pseudodata
  generated on the basis of a Deck model.  Subjecting pseudodata and
  real data to the same partial-wave analysis, we find good agreement
  concerning the spectral shape and its dependence on the squared
  four-momentum transfer for the $\JPC = 1^{-+}$ amplitude and also
  for amplitudes with other \JPC quantum numbers.  We investigate for
  the first time the amplitude of the \twoPi subsystem with $\JPC =
  1^{--}$ in the $3\pi$ amplitude with $\JPC = 1^{-+}$ employing the
  novel freed-isobar analysis scheme.  We reveal this \twoPi amplitude
  to be dominated by the \Prho for both the \PpiOne[1600] and the
  nonresonant contribution.  These findings
  largely confirm the underlying assumptions for the isobar model used
  in all previous partial-wave analyses addressing the $\JPC = 1^{-+}$
  amplitude.
\end{abstract}

  \vspace*{20pt}
  \begin{flushleft}
    PACS numbers:
    11.80.Et,    %
    13.25.Jx,    %
    13.85.Hd,    %
    14.40.Be \\  %
    Keywords:
experimental results, magnetic spectrometer;
hadron spectroscopy, meson, light;
CERN Lab;
CERN SPS;
COMPASS Experiment;
beam, pi-, 190 GeV/c;
target, hydrogen;
pi- p, inelastic scattering, exclusive reaction;
pi-, hadroproduction, meson resonance;
pi-, diffraction, dissociation;
pi-, multiple production, (pi+ 2pi-);
pi- p {-}{-}\textgreater\ p pi+ 2pi-;
data analysis method;
partial-wave analysis;
isobar model;
hadronic decay, amplitude analysis;
spin, density matrix;
meson resonance, exotic;
pi1(1600);
vector meson, isovector;
rho(770)
   \end{flushleft}
  \Submitted{(to be submitted to Physical Review D)}
\end{titlepage}

{\pagestyle{empty}
  %%%%%%%%%%%%%%%%%%%%%%%%%%%%%%%%%%%%%%%%%%%%%%%%%%%%%%%%%%%%%%%
%
% 2016_auththorlist.tex (default list, updated 26.10.2016)
%
%%%%%%%%%%%%%%%%%%%%%%%%%%%%%%%%%%%%%%%%%%%%%%%%%%%%%%%%%%%%%%%
\section*{The COMPASS Collaboration}
\label{app:collab}
\renewcommand\labelenumi{\textsuperscript{\theenumi}~}
\renewcommand\theenumi{\arabic{enumi}}
\begin{flushleft}
G.D.~Alexeev\Irefn{dubna}, %1
M.G.~Alexeev\Irefnn{turin_u}{turin_i},
A.~Amoroso\Irefnn{turin_u}{turin_i},
V.~Andrieux\Irefnn{cern}{illinois},
V.~Anosov\Irefn{dubna}, %3
K.~Augsten\Irefnn{dubna}{praguectu}, %2 phd
W.~Augustyniak\Irefn{warsaw},
C.D.R.~Azevedo\Irefn{aveiro},
B.~Bade{\l}ek\Irefn{warsawu},
F.~Balestra\Irefnn{turin_u}{turin_i},
M.~Ball\Irefn{bonniskp},
J.~Barth\Irefn{bonniskp},
R.~Beck\Irefn{bonniskp},
Y.~Bedfer\Irefn{saclay},
J.~Berenguer~Antequera\Irefnn{turin_u}{turin_i},
J.~Bernhard\Irefnn{mainz}{cern},
M.~Bodlak\Irefn{praguecu},
F.~Bradamante\Irefn{triest_i},
A.~Bressan\Irefnn{triest_u}{triest_i},
V.~E.~Burtsev\Irefn{tomsk},
W.-C.~Chang\Irefn{taipei},
C.~Chatterjee\Irefnn{triest_u}{triest_i},
M.~Chiosso\Irefnn{turin_u}{turin_i},
A.~G.~Chumakov\Irefn{tomsk},
S.-U.~Chung\Irefn{munichtu}\Aref{B}\Aref{B1},
A.~Cicuttin\Irefn{triest_i}\Aref{C},
P.~M.~M.~Correia\Irefn{aveiro},
M.L.~Crespo\Irefn{triest_i}\Aref{C}
% ...
\end{flushleft}
%%%%%%%%%%%%%%%%%%%%%%%%%%%%%%%%%%%%%%%%%%%%%%%%%%%%%%%%%%%%%%%%%%%%%%%%%%%%%%%%%%%%%%%%%%%%%%%%%%%%%%%%%%%%%%%%%%%%%%%
%
% institutes
%
%%%%%%%%%%%%%%%%%%%%%%%%%%%%%%%%%%%%%%%%%%%%%%%%%%%%%%%%%%%%%%%%%%%%%%%%%%%%%%%%%%%%%%%%%%%%%%%%%%%%%%%%%%%%%%%%%%%%%%%
%\item \Idef{bielefeld}{Universit\"at Bielefeld, Fakult\"at f\"ur Physik, 33501 Bielefeld, Germany\Arefs{l}}
%\item \Idef{munichlmu}{Ludwig-Maximilians-Universit\"at M\"unchen, Department f\"ur Physik, 80799 Munich, Germany\Arefs{l}\Arefs{r}}
\begin{Authlist}
\item \Idef{aveiro}{University of Aveiro, Department of Physics, 3810-193 Aveiro, Portugal}
\item \Idef{bochum}{Universit\"at Bochum, Institut f\"ur Experimentalphysik, 44780 Bochum, Germany\Arefs{Q}$^,$\Arefs{R}}
\item \Idef{bonniskp}{Universit\"at Bonn, Helmholtz-Institut f\"ur  Strahlen- und Kernphysik, 53115 Bonn, Germany\Arefs{Q}}
\item \Idef{bonnpi}{Universit\"at Bonn, Physikalisches Institut, 53115 Bonn, Germany\Arefs{Q}}
\item \Idef{brno}{Institute of Scientific Instruments of the CAS, 61264 Brno, Czech Republic\Arefs{S}}
\item \Idef{calcutta}{Matrivani Institute of Experimental Research \& Education, Calcutta-700 030, India\Arefs{T}}
\item \Idef{dubna}{Joint Institute for Nuclear Research, 141980 Dubna, Moscow region, Russia\Arefs{T1}}
\item \Idef{freiburg}{Universit\"at Freiburg, Physikalisches Institut, 79104 Freiburg, Germany\Arefs{Q}$^,$\Arefs{R}}
\item \Idef{cern}{CERN, 1211 Geneva 23, Switzerland}
\item \Idef{liberec}{Technical University in Liberec, 46117 Liberec, Czech Republic\Arefs{S}}
\item \Idef{lisbon}{LIP, 1649-003 Lisbon, Portugal\Arefs{U}}
\item \Idef{mainz}{Universit\"at Mainz, Institut f\"ur Kernphysik, 55099 Mainz, Germany\Arefs{Q}}
\item \Idef{miyazaki}{University of Miyazaki, Miyazaki 889-2192, Japan\Arefs{V}}
\item \Idef{moscowlpi}{Lebedev Physical Institute, 119991 Moscow, Russia}
\item \Idef{munichtu}{Technische Universit\"at M\"unchen, Physik-Department, 85748 Garching, Germany\Arefs{Q}}
\item \Idef{nagoya}{Nagoya University, 464 Nagoya, Japan\Arefs{V}}
\item \Idef{praguecu}{Charles University, Faculty of Mathematics and Physics, 12116 Prague, Czech Republic\Arefs{S}}
\item \Idef{praguectu}{Czech Technical University in Prague, 16636 Prague, Czech Republic\Arefs{S}}
\item \Idef{protvino}{State Scientific Center Institute for High Energy Physics of National Research Center ``Kurchatov Institute,'' 142281 Protvino, Russia}
\item \Idef{saclay}{IRFU, CEA, Universit\'e Paris-Saclay, 91191 Gif-sur-Yvette, France\Arefs{R}}
\item \Idef{taipei}{Academia Sinica, Institute of Physics, Taipei 11529, Taiwan\Arefs{W}}
\item \Idef{telaviv}{Tel Aviv University, School of Physics and Astronomy, 69978 Tel Aviv, Israel\Arefs{X}}
\item \Idef{triest_u}{University of Trieste, Department of Physics, 34127 Trieste, Italy}
\item \Idef{triest_i}{Trieste Section of INFN, 34127 Trieste, Italy}
\item \Idef{turin_u}{University of Turin, Department of Physics, 10125 Turin, Italy}
\item \Idef{turin_i}{Torino Section of INFN, 10125 Turin, Italy}
\item \Idef{tomsk}{Tomsk Polytechnic University, 634050 Tomsk, Russia\Arefs{Y}}
\item \Idef{illinois}{University of Illinois at Urbana-Champaign, Department of Physics, Urbana, Illinois 61801-3080, USA\Arefs{Z}}
\item \Idef{warsaw}{National Centre for Nuclear Research, 02-093 Warsaw, Poland\Arefs{a} }
\item \Idef{warsawu}{University of Warsaw, Faculty of Physics, 02-093 Warsaw, Poland\Arefs{a} }
\item \Idef{warsawtu}{Warsaw University of Technology, Institute of Radioelectronics, 00-665 Warsaw, Poland\Arefs{a} }
\item \Idef{yamagata}{Yamagata University, Yamagata 992-8510, Japan\Arefs{V} }
%\item \Idef{retired}{Retired}
\end{Authlist}
%%%%%%%%%%%%%%%%%%%%%%%%%%%%%%%%%%%%%%%%%%%%%%%%%%%%%%%%%%%%%%%%%%%%%%%%%%%%%%%%%%%%%%%%%%%%%%%%%%%%%%%%%%%%%%%%%%%%%%%
%
% Notes
%
%%%%%%%%%%%%%%%%%%%%%%%%%%%%%%%%%%%%%%%%%%%%%%%%%%%%%%%%%%%%%%%%%%%%%%%%%%%%%%%%%%%%%%%%%%%%%%%%%%%%%%%%%%%%%%%%%%%%%%%
%\vspace*{-\baselineskip}
\renewcommand\theenumi{\alph{enumi}}
\begin{Authlist}
\item [{\makebox[2mm][l]{\textsuperscript{\#}}}] Corresponding authors
%\item [{\makebox[2mm][l]{\textsuperscript{*}}}] Deceased
%\item \Adef{A}{Also at Instituto Superior T\'ecnico, Universidade de Lisboa, Lisbon, Portugal}
\item \Adef{B}{Also at Department of Physics, Pusan National University, Busan 609-735, Republic of Korea}
\item \Adef{B1}{Also at Physics Department, Brookhaven National Laboratory, Upton, NY 11973, USA}
\item \Adef{C}{Also at Abdus Salam ICTP, 34151 Trieste, Italy}
\item \Adef{D}{Supported by the DFG cluster of excellence `Origin and Structure of the Universe' (www.universe-cluster.de) (Germany)}
\item \Adef{E}{Deceased}
%\fntext[E]{Supported by CERN-RFBR Grant 12-02-91500}
\item \Adef{F}{Supported by the DFG Research Training Group Programmes 1102 and 2044 (Germany)}
\item \Adef{G}{Also at Chubu University, Kasugai, Aichi 487-8501, Japan}
\item \Adef{H}{Also at Department of Physics, National Central University, 300 Jhongda Road, Jhongli 32001, Taiwan}
\item \Adef{I}{Also at KEK, 1-1 Oho, Tsukuba, Ibaraki 305-0801, Japan}
\item \Adef{J}{Present address: Universit\"at Bonn, Physikalisches Institut, 53115 Bonn, Germany}
\item \Adef{K}{Also at Moscow Institute of Physics and Technology, Moscow Region, 141700, Russia}
\item \Adef{L}{Also at Yerevan Physics Institute, Alikhanian Brothers Street, Yerevan, Armenia, 0036}
\item \Adef{M}{Also at Department of Physics, National Kaohsiung Normal University, Kaohsiung County 824, Taiwan}
\item \Adef{M1}{Supported by ANR, France with P2IO LabEx (ANR-10-LBX-0038) in the framework ``Investissements d'Avenir'' (ANR-11-IDEX-003-01)}
\item \Adef{N}{Also at Institut f\"ur Theoretische Physik, Universit\"at T\"ubingen, 72076 T\"ubingen, Germany}
%\item \Adef{N1}{Retired}
%\item \Adef{N2}{Present address: Brookhaven National Laboratory, Brookhaven, USA}
\item \Adef{O}{Also at University of Eastern Piedmont, 15100 Alessandria, Italy}
\item \Adef{O1}{Present address: Universit\"at Hamburg, 20146 Hamburg, Germany}
\item \Adef{O2}{Present address: Institue of Modern Physics, Chinese Academy of Sciences, Lanzhou 730000, China}
\item \Adef{P}{Present address: RWTH Aachen University, III.\ Physikalisches Institut, 52056 Aachen, Germany}
\item \Adef{Q}{Supported by BMBF - Bundesministerium f\"ur Bildung und Forschung (Germany)}
\item \Adef{R}{Supported by FP7, HadronPhysics3, Grant 283286 (European Union)}
\item \Adef{S}{Supported by MEYS, Grant LM20150581 (Czech Republic)}
\item \Adef{T}{Supported by B.~Sen fund (India)}
\item \Adef{T1}{Supported by CERN-RFBR Grant 12-02-91500}
\item \Adef{U}{Supported by FCT, Grants CERN/FIS-PAR/0007/2017 and  CERN/FIS-PAR/0022/2019 (Portugal)}
\item \Adef{V}{Supported by MEXT and JSPS, Grants 18002006, 20540299, 18540281 and 26247032, the Daiko and Yamada Foundations (Japan)}
\item \Adef{W}{Supported by the Ministry of Science and Technology (Taiwan)}
\item \Adef{X}{Supported by the Israel Academy of Sciences and Humanities (Israel)}
\item \Adef{Y}{Supported by the Tomsk Polytechnic University within the assignment of the Ministry of Science and Higher Education (Russia)}
\item \Adef{Z}{Supported by the National Science Foundation, Grant no. PHY-1506416 (USA)}
\item \Adef{a}{Supported by NCN, Grant 2017/26/M/ST2/00498 (Poland)}
\item \Adef{b}{Retired from Ludwig-Maximilians-Universit\"at, 80539 M\"unchen, Germany}

\end{Authlist}

   \clearpage
}

\setcounter{page}{1}
 \tableofcontents

\setlist{noitemsep}
\setlist{nolistsep}
\setlist[enumerate]{label=\textit{\roman*})}

\clearpage
\section{Introduction}%
\label{sec:introduction}

The presently known meson spectrum is to a large extent attributed to
quark-antiquark (\qqbarPrime) states.  These states, \ie the ground
states and their excitations, are described by the constituent-quark
model and are classified using $\text{SU(3)}_\text{flavor} \times
\text{SU(2)}_\text{spin}$ symmetry.  However, QCD in principle allows
for a richer spectrum of excitations including multiquark
configurations as well as gluonic excitations, called
\textquote{exotic} mesons hereafter.  Such states are expected to be
different from \qqbarPrime states in terms of either their quantum
numbers and/or their couplings to initial or final states, thus
leaving their own fingerprints.  In case of quantum numbers allowed
for constituent-quark model states, they may, however, mix in
configuration space.

Many model calculations for light-quark exotic states consisting
of~$u$, $d$, or $s$~quarks exist, predicting a variety of features
(see, \eg\
\refsCite{Tanimoto:1982eh,Isgur:1985vy,Iddir:1988jc,Close:1994hc,Page:1998gz}),
but no clear signatures exist.  More recently, first calculations of
the excitation spectrum of light mesons have been performed by the
authors of
\refsCite{Dudek:2009qf,Dudek:2010wm,Dudek:2011bn,Dudek:2013yja} using
lattice QCD\@.  They find exotic states with large contributions from
excited gluonic field configurations, \ie hybrid mesons, the lightest
of which having so-called spin-exotic quantum numbers $\JPC = 1^{-+}$
that are forbidden for \qqbar states.\footnote{Here, $J$~is the spin,
$P$~the parity, and $C$~the charge conjugation quantum number of the
state.}  However, the predictive power of these calculations is
currently limited by the fact that all states are considered to be
quasistable.  Recently, the authors of \refCite{Woss:2020ayi} have
performed the first lattice QCD calculation of the hadronic decays of
the lightest $1^{-+}$ resonance.  This calculation was performed using
up, down, and strange-quark masses that approximately match the
physical strange-quark mass.  At this SU(3)$_\text{flavor}$ symmetric
point, which corresponds to a pion mass of about \SI{700}{\MeVcc}, the
scattering amplitudes of eight meson-meson systems were studied in a
coupled-channel approach.  Extrapolating the extracted resonance pole
and its couplings to the physical light-quark masses suggests a broad
\PpiOne* resonance that decays predominantly into $\PbOne\pi$ and that
has much smaller partial widths into $\PfOne\pi$, $\Prho\pi$,
$\eta'\pi$, and $\eta\pi$.  The present state-of-the-art method to
calculate multi-body decays and scattering processes on the lattice
requires using large pion masses to limit the analysis to coupled
two-body channels only (see, \eg\ \refCite{Briceno:2017max} and
references therein).  However, the extension of these calculations to
three-body final states is under active development (see, \eg\
\refCite{Hansen:2019nir}).

The field of exotic hadrons has changed dramatically with the
observations of the~$X$, $Y$, $Z$~states involving heavy quarks.  In
particular, the observation of charged quarkoniumlike states,
$Z_c^\pm$~\cite{Ablikim:2013mio,Liu:2013dau} and
$Z_b^\pm$~\cite{Belle:2011aa}, has been considered as clear evidence
for the existence of exotic hadrons.  They are characterized by an
exotic combination of presumed flavor content and isospin quantum
numbers.  In addition, the~$P_c$ states are considered as the first
observation of pentaquark states~\cite{Aaij:2015tga,Aaij:2019vzc}.
The nature and internal structure of these states are discussed widely
in the literature (see, \eg\ \refsCite{Lebed:2016hpi,Olsen:2017bmm}).

In the sector of light-quark mesons, several candidates for
non-\qqbarPrime states with conventional \qqbarPrime quantum numbers
are discussed in the literature, \eg\ \PfZero[1500], \Ppi[1300],
\Ppi[1800], \PaOne[1420], and \PfOne[1420], although none of them was
conclusively identified as such.  While production and decay patterns
constitute a mandatory but often strongly model-dependent signature,
spin-exotic \JPC quantum numbers are generally considered the cleanest
path to prove the existence of mesonlike objects beyond \qqbarPrime.
Three such states with $\JPC = 1^{-+}$, the \PpiOne[1400],
\PpiOne[1600], and the \PpiOne[2015], have been discussed frequently
as first evidence for exotic mesons and their observation was reported
by various experiments~\cite{Tanabashi:2018zz}.  Their masses agree
qualitatively with lattice QCD calculations~\cite{Dudek:2013yja}.
However, the existence of these states is disputed and the
experimental situation requires clarification and further studies.
The \PpiOne[1400] has been observed by several
experiments~\cite{Alde:1988bv,Aoyagi:1993kn,Thompson:1997bs,Chung:1999we,Dorofeev:2001xu,Adams:2006sa}
in the $\eta \pi$ final state produced in $\pi^-$ diffraction at beam
momenta ranging from \SIrange{6.3}{100}{\GeVc}.  It has also been
observed in the $\eta \pi$ final state produced in \pbarp and \pbarn
annihilations studied by the Crystal Barrel
experiment~\cite{Abele:1998gn,Abele:1999tf,Albrecht:2019ssa} and in
the $\Prho \pi$ channel by the Obelix
experiment~\cite{Salvini:2004gz}.  The \PpiOne[2015] has so far been
observed only by the BNL~E852 experiment in the $\PfOne
\pi$~\cite{Kuhn:2004en} and $\PbOne \pi$~\cite{Lu:2004yn} decay modes.

\subsection{Status of the \PpiOne[1600]}%
\label{sec:status_piOne1600}

The \PpiOne[1600] is the most extensively studied spin-exotic meson.
Indications were found in $\eta'
\pi$~\cite{Khokhlov:2000tk,Ivanov:2001rv,Amelin:2005ry,Adams:2011sq},
in $\PfOne \pi$~\cite{Zaitsev:2000rc,Kuhn:2004en,Amelin:2005ry}, and
in $\PbOne
\pi$~\cite{Zaitsev:2000rc,Khokhlov:2000tk,Lu:2004yn,Amelin:2005ry,Baker:2003jh}.
Recently, the COMPASS collaboration has published further studies on
the $\eta' \pi$ and $\eta \pi$ final states in diffractive production
relevant to the search for
\PpiOne[1600]~\cite{Adolph:2014rpp}.  A reanalysis of these
data performed by the JPAC collaboration revealed a clear resonance pole in the
spin-exotic wave~\cite{Rodas:2018owy}.  This analysis could even
reconcile the previous observations of the two spin-exotic
states \PpiOne[1400] and \PpiOne[1600] to be the result of only a
single pole with parameters that are consistent with the
\PpiOne[1600].

In this paper, we focus on the \threePi final state including the
$\Prho \pi$ intermediate state.  An observation of the decay
$\PpiOne[1600] \to \Prho \pi$ was first reported by
BNL~E852~\cite{Adams:1998ff,Chung:2002pu} followed by
VES~\cite{Zaitsev:2000rc}.  Later, the COMPASS experiment confirmed
some of the previous findings~\cite{Alekseev:2009aa}.  For a review,
we refer to \refCite{Meyer:2015eta}.  All the above
experiments studied diffractive pion dissociation, but at different
beam energies, using different target materials, and in
various ranges of the four-momentum transfer squared.  Previous
investigations of the $3\pi$ final state yielded contradicting
conclusions on what concerns the proof of existence of the
\PpiOne[1600] or the determination of its properties.  While
BNL~E852~\cite{Adams:1998ff,Chung:2002pu} and COMPASS have stressed
the observation of the \PpiOne[1600], VES~\cite{Zaitsev:2000rc}, and
an analysis of BNL~E852 data by
Dzierba~\etal~\cite{Dzierba:2005jg} have been inconclusive on its
existence or even refuted it.

Recently, COMPASS published an extensive study of isovector mesons
using a large dataset on the \threePi final
state~\cite{Akhunzyanov:2018lqa}.  In this analysis, we observed a
strong modulation of the shape of the intensity distribution of the
spin-exotic $\Prho \pi$ $P$-wave carrying $\JPC = 1^{-+}$ with the
squared four-momentum~\tpr transferred from the beam to the target
particle [see
\cref{fig:piOne1600_resModelFit_int_t1,fig:piOne1600_resModelFit_int_t11}
and definition in \cref{eq:tPrime}].  This modulation is described by
the resonance model as a \tpr-dependent interference between a
\PpiOne[1600] resonance (blue curves) and a nonresonant wave component
(green curves).  The nonresonant component was found to dominate at
low~\tpr whereas the \PpiOne[1600] signal emerged at high~\tpr.  The
resonance characteristics of the \PpiOne[1600] signal was clearly
demonstrated through its phase variation \wrt 13~other waves.  As an
example, we show in \cref{fig:piOne1600_resModelFit_phase_t11} the
phase motion \wrt the $\Prho \pi$ $S$-wave carrying $\JPC = 1^{++}$.
In addition, the dashed red curves in \cref{fig:piOne1600_resModelFit}
show the result of a fit, where the \PpiOne[1600] resonance was
omitted from the fit model.  Although at low~\tpr, where the
nonresonant component is dominant, this model is in fair agreement
with data, it fails to describe the data at high~\tpr.  This
demonstrates that a \PpiOne[1600] resonance is needed to describe the
COMPASS data.

\begin{wideFigureOrNot}[tbp]
  \centering%
  \subfloat[][]{%
    \includegraphics[width=\threePlotWidth]{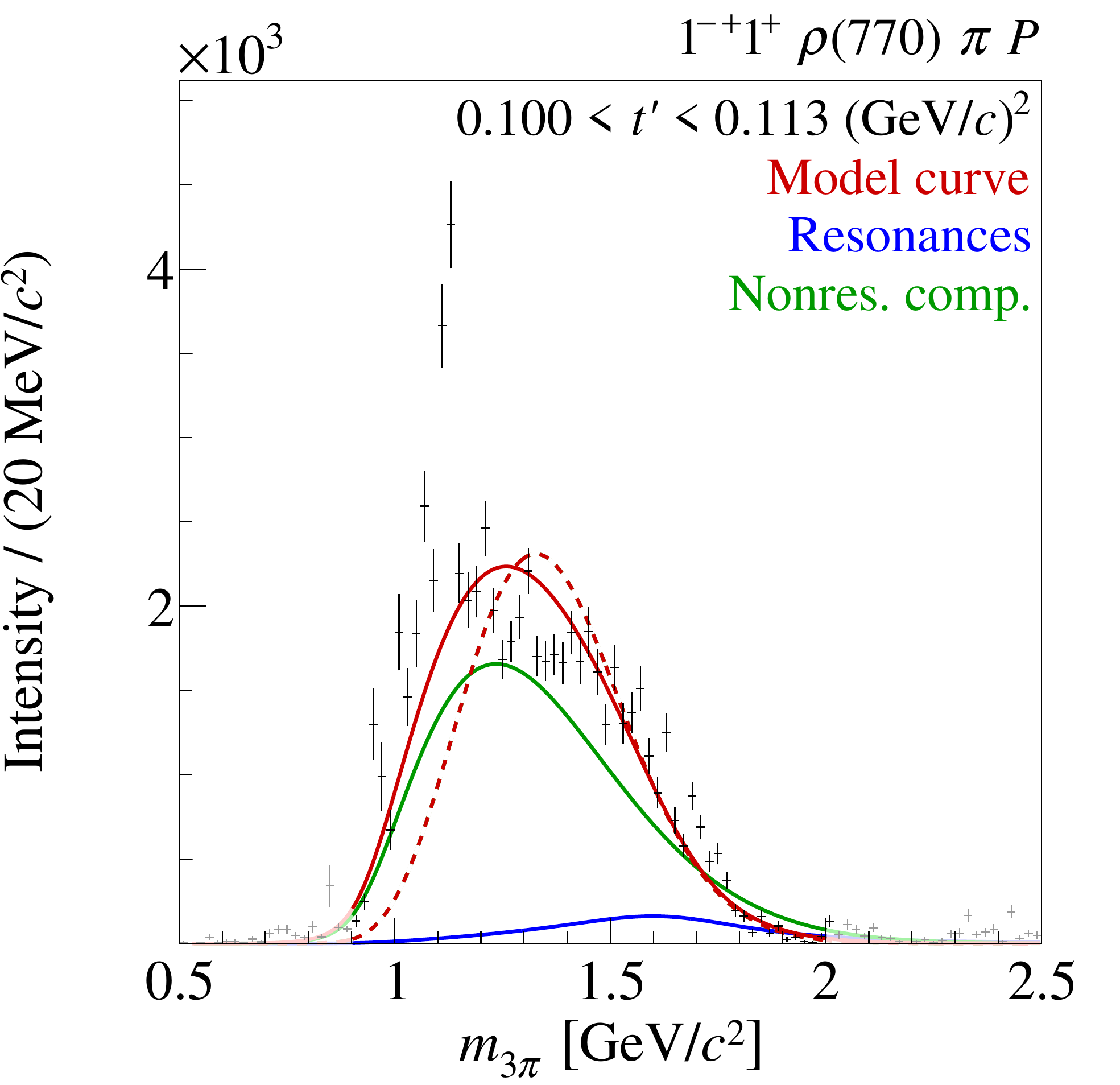}%
    \label{fig:piOne1600_resModelFit_int_t1}%
  }%
  \hspace*{\threePlotSpacing}%
  \subfloat[][]{%
    \includegraphics[width=\threePlotWidth]{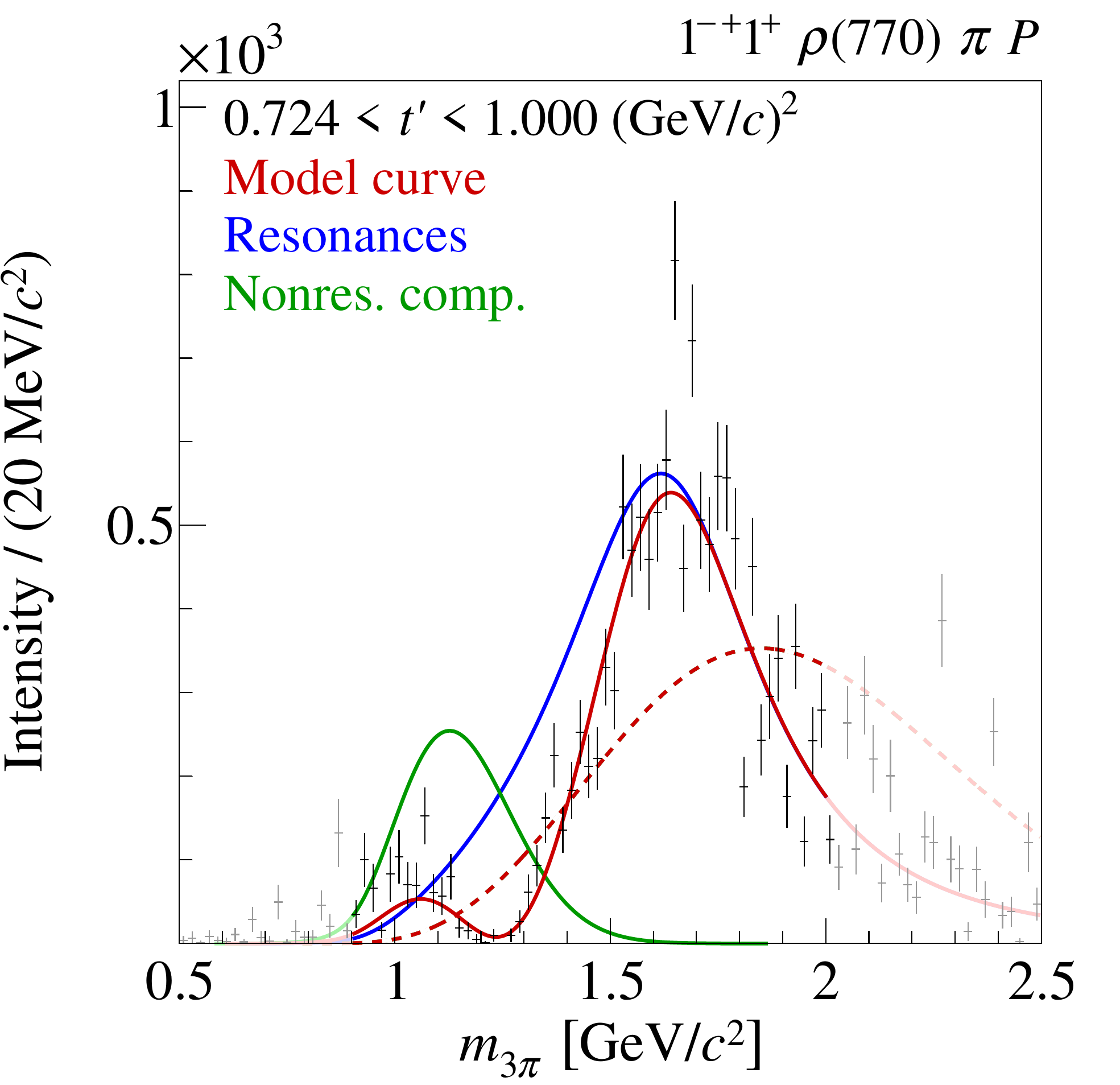}%
    \label{fig:piOne1600_resModelFit_int_t11}%
  }%
  \hspace*{\threePlotSpacing}%
  \subfloat[][]{%
    \includegraphics[width=\threePlotWidth]{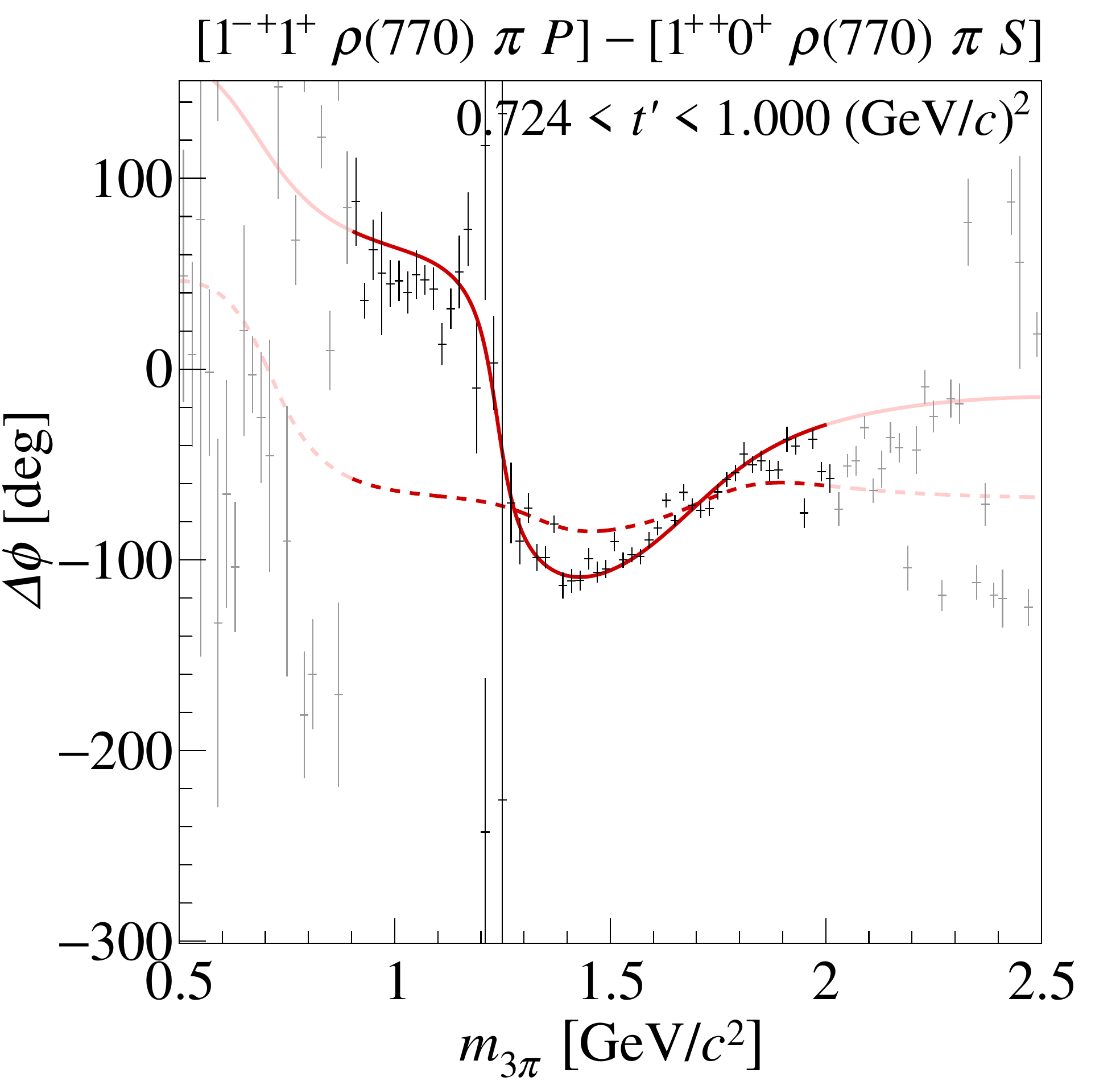}%
    \label{fig:piOne1600_resModelFit_phase_t11}%
  }%
  \caption{Excerpt from the results of a previous study of resonance
    production in $\pi^-\; p \to \threePi\; p$ at \SI{190}{\GeVc}
    pion-beam momentum by the COMPASS
    collaboration~\cite{Adolph:2015tqa,Akhunzyanov:2018lqa}.  The
    partial-wave intensities of the spin-exotic $\Prho \pi$ $P$-wave
    carrying $\JPC = 1^{-+}$ are shown
    in~\subfloatLabel{fig:piOne1600_resModelFit_int_t1}
    and~\subfloatLabel{fig:piOne1600_resModelFit_int_t11} for the
    lowest and highest \tpr~bins, respectively, covered by the
    experiment.
    Panel~\subfloatLabel{fig:piOne1600_resModelFit_phase_t11} shows
    the phase relative to the $\Prho \pi$ $S$-wave carrying $\JPC =
    1^{++}$.  The red solid curve represents the full resonance model
    (see Table~II in \refCite{Akhunzyanov:2018lqa}), which is the
    coherent sum of wave components that are represented by the other
    solid curves: \PpiOne[1600] resonance (blue curves) and
    nonresonant component (green curves).  The extrapolation of the
    model and the wave components beyond the fit range are shown in
    lighter colors.  The narrow enhancement at \SI{1.1}{\GeVcc}
    in~\subfloatLabel{fig:piOne1600_resModelFit_int_t1} is likely an
    artifact induced by imperfections in the analysis method (see
    \cref{sec:comparison}).  The dashed red curves represent a fit,
    where the \PpiOne[1600] resonance was omitted from the fit model.
    This curve hence corresponds to a purely nonresonant
    $\Prho \pi$ $P$-wave.}%
  \label{fig:piOne1600_resModelFit}
\end{wideFigureOrNot}

The discrepancy of results and interpretations on the \PpiOne[1600]
signal from the analyses discussed above requires detailed studies as
the origin could be either inconsistent datasets or analysis
artifacts.  This paper aims at understanding three different aspects
of the spin-exotic $\Prho \pi$ $P$-wave carrying $\JPC = 1^{-+}$ based
on the large COMPASS data sample: \one~can the different and partially
inconsistent observations from previous analyses be reconciled through
studies of the model dependence of the analyses?  \two~Are structures
observed in this wave an artifact of the partial-wave analysis model?
Since the resonant nature of the \PpiOne[1600] has been already
studied extensively in~\refCite{Akhunzyanov:2018lqa}, we will not
readdress the determination the \PpiOne[1600] resonance parameters
here.  \three~Can we model nonresonant production through the
so-called Deck effect~\cite{Deck:1964hm}?

This paper is organized as follows: after a short description of the
COMPASS experiment in \cref{sec:setup_and_event_selection}, we will
briefly review the analysis of our data in \cref{sec:pwa}.
\Crefrange{sec:other_models}{sec:deck_model} each will address one of
the three questions that we posed above.  In \cref{sec:other_models},
we will reconcile our analyses and previous ones performed by the
BNL~E852~\cite{Adams:1998ff,Chung:2002pu,Dzierba:2005jg} and
VES~\cite{Zaitsev:2000rc} experiments and trace the discrepancies to
the different analysis schemes used.  Next, in \cref{sec:freed_isobar}
we will extract the amplitude of the \twoPi subsystem present in the
$\JPC = 1^{-+}$ wave using the new scheme of freed-isobar
analysis~\cite{Krinner:2017dba} proving the decay of $\PpiOne[1600]
\to \Prho \pi$.  Finally, we compare in \cref{sec:deck_model} the
observed intensity distributions (diagonal elements of the
spin-density matrix) of selected partial waves to model calculations
on nonresonant $3\pi$ production.  Each of the three sections will
provide evidence that further confirms the \PpiOne[1600] resonance and
its decay into $\Prho \pi$.  Since the three result sections are
linked only weakly, we will summarize and conclude them individually
and refrain from an additional summary and conclusion at the end of
the paper.
\section{Analyzed data sample}%
\label{sec:setup_and_event_selection}

\begin{figure}[tbp]
  \centering
  \includegraphics[width=\linewidthOr{0.6\textwidth}]{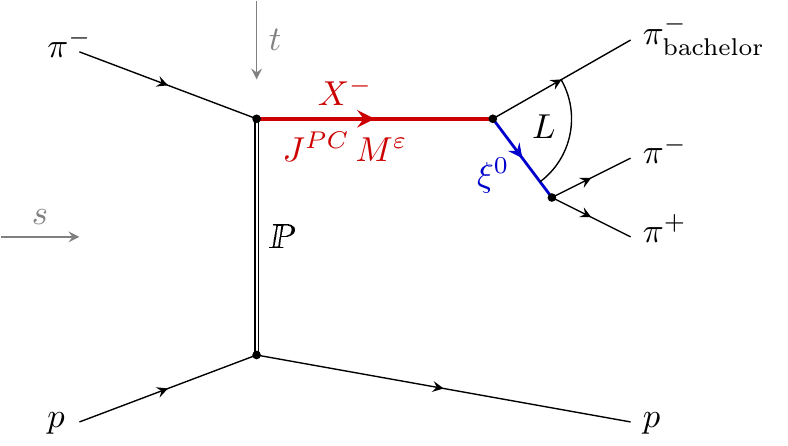}
  \caption{Single-diffractive dissociation of a beam pion on a target
    proton into the \threePi final state via exchange of a
    Pomeron~\Ppom.  In this scattering process, an intermediate $3\pi$
    state~$X^-$ with well-defined quantum numbers is produced.  The
    decay of~$X^-$ is described using the isobar model, which assumes
    that the decay proceeds via intermediate \twoPi states~$\xi^0$,
    the so-called isobars, which also have well-defined quantum
    numbers.  See \cref{sec:pwa} for details.}%
  \label{fig:3pi_reaction_isobar}
\end{figure}

The present study is based on a dataset of \num{46E6} exclusive
events of diffractively produced mesons decaying into three charged
pions.  The data were obtained by the COMPASS experiment and were
already presented in detail in~\refCite{Adolph:2015tqa}.  They contain
exclusive events from the inelastic reaction
\begin{equation}
  \reaction,
  \label{eq:reaction}
\end{equation}
which is induced by a high-energy $\pi^-$~beam impinging on a hydrogen
target. The dominant reaction mechanism is single-diffractive
scattering, where the target particle scatters elastically and the
beam pion is excited via the exchange of a Pomeron with the target
nucleon to a short-lived intermediate state~$X^-$ that then decays
into \threePi as shown in \cref{fig:3pi_reaction_isobar}.  The
experimental setup and the criteria that were applied to select
exclusive events of reaction~\eqref{eq:reaction} are described in
detail in \refsCite{Adolph:2015tqa,Haas:2014bzm}.  Here, we give only
a brief summary.

The COMPASS experiment~\cite{Abbon:2007pq,Abbon:2014aex} is located at
the M2~beam line of the CERN Super Proton Synchrotron. A beam of
negatively charged secondary pions with \SI{190}{\GeVc} momentum was
incident on a \SI{40}{\cm} long liquid-hydrogen target.  The data
selection required a recoil-proton signal and an exclusive measurement
was ensured through a variety of criteria~\cite{Adolph:2015tqa}.

Reaction~\eqref{eq:reaction} depends on two Mandelstam variables: the
squared $\pi^- p$ center-of-momentum energy~$s$, which is fixed to
about $(\SI{19}{\GeV})^2$ by the beam momentum, and the squared
four-momentum~$t$ transferred from the beam to the target particle.
It is convenient to define the \emph{reduced four-momentum transfer
squared}
\begin{equation}
  \label{eq:tPrime}
  \tpr \equiv \tabs - \tmin \geq 0,
\end{equation}
where
\begin{equation}
  \label{eq:tMin}
  \tmin \approx \rBrk{\frac{\mThreePi^2 - m_\pi^2}{2 \Abs{\vec{p}_\text{beam}}}}^2
\end{equation}
is the minimum absolute value of the four-momentum transfer needed to
excite the beam pion to a $3\pi$ state with invariant
mass~\mThreePi.\footnote{For the kinematic range considered here,
\tmin is well below \SI{e-3}{\GeVcsq} and hence $\tpr \approx -t$.}
The beam momentum~$\vec{p}_\text{beam}$ is defined in the laboratory
frame.  For the present analysis, \tpr~was chosen to be in the range
from \SIrange{0.1}{1.0}{\GeVcsq}, where the lower bound is dictated by
the acceptance of the recoil-proton detector and the upper bound by
the exponential decrease of the number of
events with~\tpr.

Since reaction~\eqref{eq:reaction} is dominated by Pomeron exchange,
which conserves isospin~$I$ and $G$~parity of the beam pion, only
intermediate states~$X^-$ with $\IG = 1^-$ can be produced.  This
limits the analysis to meson states that belong to the~\piJ and
\aJ~families with spin~$J$.  This analysis focuses on $3\pi$ resonances with masses
up to about \SI{2}{\GeVcc}.  We hence selected the \mThreePi~range
from \SIrange{0.5}{2.5}{\GeVcc}.
\section{Partial-wave analysis method}%
\label{sec:pwa}

We extract the \PpiOne[1600] contribution with $\JPC = 1^{-+}$ from
the COMPASS data through a partial-wave analysis (PWA) using a model
comprising 88~partial waves (see \cref{tab:wave_sets} in
\cref{sec:wave_sets}).  The PWA model has already been described in
detail in \refsCite{Adolph:2015tqa,Akhunzyanov:2018lqa_suppl}, so we
will provide here only a brief description.

We subdivide the data into 100~equidistant \SI{20}{\MeVcc} wide bins
of the invariant mass \mThreePi of the $3\pi$ system and into
11~nonequidistant bins of the reduced four-momentum transfer squared
\tpr (see Table~IV in \refCite{Adolph:2015tqa}) resulting in
1100~kinematic $(\mThreePi, \tpr)$ cells.  We fit each of these cells
independently with a PWA model for the intensity distribution,
\begin{multlineOrEq}
  \label{eq:intensity_bin}
  \intensity(\tau; \mThreePi, \tpr)
  = \sum_{\refl = \pm 1} \sum_{r = 1}^{N_r^\refl} \Abs[3]{\sum_a^{N_\text{waves}^\refl}
  \prodAmp_a^{r \refl}(\mThreePi, \tpr)\,
  \decayAmp_a^\refl(\tau; \mThreePi)}^2 \newLineOrNot
  + \intensity_\text{flat}(\mThreePi, \tpr),
\end{multlineOrEq}
using an unbinned extended maximum likelihood approach.  Here,
$\tau$~represents the five three-body phase-space variables in a given
$(\mThreePi, \tpr)$ cell (see Sec.~III in \refCite{Adolph:2015tqa} for
a concrete choice for~$\tau$).  The indices \refl and~$r$ are
explained below.  The transition amplitude~$\prodAmp_a^{r \refl}$
encodes the (unknown) strength and phase of partial wave~$a$, while
the decay amplitude~$\decayAmp_a^\refl(\tau)$ encodes the (known)
dependence on~$\tau$.  Within a given $(\mThreePi, \tpr)$ cell, we
neglect the dependence on \mThreePi and \tpr; \ie $\prodAmp_a^{r
\refl}$ is constant and $\decayAmp_a^\refl$ depends only on~$\tau$.
The term $\intensity_\text{flat}$ is the intensity of the so-called
flat wave, which represents three uncorrelated final-state pions that
are distributed isotropically in the three-body phase space.  The flat
wave contributes only \SI{3.1}{\percent} to the total intensity.

The partial waves that enter \cref{eq:intensity_bin} are uniquely
defined by the quantum numbers of the intermediate state~$X^-$ and its
decay mode (see \cref{fig:3pi_reaction_isobar}).  The $X^-$~quantum
numbers are isospin~$I$, $G$~parity, spin~$J$, parity~$P$, $C$~parity,
and the projection~$M$ of~$J$ along the beam axis.\footnote{Although
the $C$~parity is not defined for charged systems, it is customary to
quote the \JPC quantum numbers of the corresponding neutral partner
state in the isospin triplet.  For nonstrange light mesons, the
$C$~parity is related to the $G$~parity via $G = C\, e^{i \pi I_y}$,
where $I_y$ is the $y$~component of the isospin.}  We express the
amplitudes in \cref{eq:intensity_bin} in the reflectivity
basis~\cite{Chung:1974fq}.  As a consequence, $M \geq 0$ and an
additional quantum number of~$X^-$, the reflectivity $\refl = \pm 1$,
is introduced.  The formulation in the reflectivity basis allows us to
take into account parity conservation in the scattering process by
summing incoherently over~\refl.  In addition, at high~$s$ and
neglecting corrections of order $1 / s$, \refl corresponds to the
naturality of the exchange particle in the scattering
process~\cite{Gottfried:1964nx,Cohen-Tannoudji:1968eoa,Mathieu:2019fts}.
Since at high~$s$ the scattering process is dominated by Pomeron
exchange, which has $\refl = +1$, partial-wave amplitudes with $\refl
= -1$ are suppressed.  Hence the two reflectivity sectors are in
general described using wave sets with different
numbers~$N_\text{waves}^\refl$ of waves.

For the COMPASS data, we find that a PWA model with
$N_\text{waves}^{\refl = +1} = 80$, $N_\text{waves}^{\refl = -1} = 7$,
and the flat wave describes the data well~\cite{Adolph:2015tqa}.  This
88-wave set is listed in \cref{tab:wave_sets} in \cref{sec:wave_sets}.
As we will show in \cref{sec:other_models}, the wave set has a strong
influence on the shape and intensity of the spin-exotic $\JPC =
1^{-+}$ wave with $\Mrefl = 1^+$, which contains a potential
\PpiOne[1600] signal.

The incoherent sum over the index~$r$ in \cref{eq:intensity_bin} is
used to model the incoherence in the scattering process.  Incoherences
may, for example, arise due to spin flip and spin nonflip of the
target proton.  Also performing the PWA over wide \tpr~ranges may lead
to effective incoherence because the transition amplitudes of the
various waves have different \tpr~dependences (see discussion below).
The number~$N_r^\refl$ of incoherent terms corresponds to the rank of
the spin-density submatrix with reflectivity~\refl.  Since the two
values of~\refl correspond to different production mechanisms, the
rank may be different for different~\refl.  For the COMPASS data, we
find that a PWA model with $N_r^{\refl = +1} = 1$ and $N_r^{\refl =
-1} = 2$ describes the data well~\cite{Adolph:2015tqa}.  This means
that all positive-reflectivity waves are assumed to be fully coherent.
The sum of the negative-reflectivity amplitudes contributes only
\SI{2.2}{\percent} to the total intensity confirming the dominance of
positive-reflectivity waves.

We construct the decay amplitudes in \cref{eq:intensity_bin} using the
helicity formalism and the isobar model (see Sec.~III in
\refCite{Adolph:2015tqa} for details), \ie we assume that the decay
$X^- \to \threePi$ proceeds via two subsequent two-particle decays,
$X^- \to \xi^0\, \pi^-$ and $\xi^0 \to \twoPi$, with intermediate
two-pion states~$\xi^0$, which are called isobars (see
\cref{fig:3pi_reaction_isobar}).  The decay amplitude of a partial
wave contains a propagator term $\Delta_a(\mTwoPi)$ that describes
this isobar resonance and that we refer to as dynamic isobar
amplitude.  In the case of the \Prho resonance, which dominates the
$\JPCMrefl = 1^{-+}1^+$ wave, we use a relativistic Breit-Wigner
amplitude with mass-dependent width as given by Eqs.~(31) and~(40) in
\refCite{Adolph:2015tqa} as the dynamic isobar amplitude.

In the following, we adopt the partial-wave notation
\wave{J}{PC}{M}{\refl}{\xi^0}{L}, where $\xi^0 \pi L$ defines the
decay mode of~$X^-$ and $L$~is the orbital angular momentum between
the isobar and the bachelor~$\pi^-$ (see
\cref{fig:3pi_reaction_isobar}).  This means that the wave index in
\cref{eq:intensity_bin} is given by
\begin{equation}
  \label{eq:wave_index}
  a = \wave{J}{PC}{M}{\refl}{\xi^0}{L}.
\end{equation}

The \tpr~dependence of the transition amplitudes $\prodAmp_a^{r
\refl}(\mThreePi, \tpr)$ in \cref{eq:intensity_bin} is in general
unknown and may be different for different waves~$a$.  In diffractive
reactions, the \tpr~spectra of the transition amplitudes exhibit an
approximately exponential decrease with~\tpr in the range $\tpr
\lesssim \SI{1}{\GeVcsq}$.  This behavior can be explained in the
framework of Regge theory~\cite{Perl:1974}.  For partial waves with $M
\neq 0$, the \tpr~spectra are modified by an additional factor
$(\tpr)^{\abs{M}}$, which is given by the forward limit of the Wigner
$D$~functions~\cite{Perl:1974}.  This factor suppresses the intensity
of the waves toward small~\tpr, \ie the transition amplitude is
approximately proportional to $(\tpr)^{\abs{M} / 2}$.  Diffractive
production of $\JPC = 1^{-+}$ waves requires $M = 1$.  This follows
from parity conservation and the dominance of natural-parity exchange
in hadronic high-energy scattering
reactions~\cite{Chung:1974fq}.\footnote{$M = 0$ would be allowed only
in unnatural-parity exchange with $\refl = -1$.}  As a consequence,
$1^{-+}$ partial-wave amplitudes with positive reflectivity are
suppressed at low~\tpr.

In the analyses of small datasets (some of which will be discussed in
\cref{sec:other_models} below), where a binning in~\tpr is not
possible, the \tpr~dependence of the transition amplitudes is often
modeled by replacing the transition amplitudes via
\begin{equation}
  \label{eq:trans_amp_t_model}
  \prodAmp_a^{r \refl}(\mThreePi, \tpr)
  \to \prodAmp_a^{r \refl}(\mThreePi)\, f_a^\refl(\tpr),
\end{equation}
where the $f_a^\refl(\tpr)$ are empirical real-valued functions.  The
parameters of these functions are usually determined from data by
performing the PWA in wide \mThreePi~ranges and narrow \tpr~bins.
This approach assumes that the shapes of the \tpr~spectra of the
partial waves are largely independent of~\mThreePi and also does not
take into account possible \tpr~dependences of the relative phases
between the partial waves.  However, we have shown in
\refCite{Adolph:2015tqa}, by performing the PWA in 11~narrow \tpr~bins
and extracting the \tpr~dependences in a model-independent way, that
for some waves the above assumptions do not hold.
\section{Previous results on $\PpiOne[1600] \to \Prho\pi$}%
\label{sec:other_models}

In the past two decades, several experiments studied the
\wave{1}{-+}{1}{+}{\Prho}{P} wave in the $3\pi$ final state.  The key
parameters of the analyzed data samples and the employed PWA models
are listed in \cref{tab:exotic_diff_models}.  A list of the wave sets
can be found in \cref{tab:wave_sets} in \cref{sec:wave_sets}.  In
\cref{fig:exotic_different_fits}, we show the intensity distributions
of the \wave{1}{-+}{1}{+}{\Prho}{P} wave as obtained in the previous
analyses (blue data points).  Based on these distributions,
the previous experiments arrived at seemingly contradictory
conclusions concerning the existence of a \PpiOne[1600] signal in the
$\Prho\pi$ channel.  We will briefly summarize these findings in the
following.

\begin{wideTableOrNot}[tbp]
  \begin{minipage}{\linewidth}
    \setcounter{mpfootnote}{\value{footnote}}%
    \refstepcounter{mpfootnote}%
    \renewcommand{\thempfootnote}{\ifMultiColumnLayout{\arabic{mpfootnote}}{[\alph{mpfootnote}]}}%
    \renewcommand{\arraystretch}{1.2}%
    \caption{Key parameters of the datasets and the PWA models used in
      analyses of diffractively produced $3\pi$ events studying a
      possible spin-exotic $\JPC = 1^{-+}$ resonance in the $\Prho\pi$
      channel.  The table also indicates whether the model takes into
      account different \tpr~dependences of the partial-wave
      amplitudes either by binning in~\tpr or by modeling
      according to \cref{eq:trans_amp_t_model}.  The wave sets
      are listed in \cref{tab:wave_sets} in \cref{sec:wave_sets}.}%
    \label{tab:exotic_diff_models}
    \begin{\ifMultiColumnLayout{small}{scriptsize}}
      \begin{tabular}{p{0.19\linewidth}p{0.36\linewidth}p{0.43\linewidth}}
        \toprule
        \textbf{Experiment} &
        \textbf{Dataset} &
        \textbf{PWA model} \\

        \midrule
        BNL~E852~\cite{Adams:1998ff,Chung:2002pu} &
        \SI{18.3}{\GeVc} $\pi^-$~beam on proton target \newline
        \num{250e3}~\threePi events &
        21~waves, rank 1, \SIvalRange{0.05}{\tpr}{1.0}{\GeVcsq} \newline
        same \tpr~dependence for all partial-wave amplitudes \\

        \midrule
        VES~\cite{Zaitsev:2000rc} &
        \SI{36.6}{\GeVc} $\pi^-$~beam on beryllium target \newline
        \num{3.0e6}~\threePi events &
        44~waves, \enquote{maximum} rank, \SIvalRange{0.03}{\tpr}{1.0}{\GeVcsq} \newline
        $f_a^\refl(\tpr) = (\tpr)^{\abs{M} / 2}\, \sqrt{e^{-b_1\, \tpr} + A\, e^{-b_2\, \tpr}}$ \\

        \midrule
        BNL~E852~\cite{Dzierba:2005jg} &
        \SI{18.3}{\GeVc} $\pi^-$~beam on proton target \newline
        \num{2.6e6}~\threePi events \newline
        \num{3.0e6}~\makebox[\widthof{\threePi}][l]{$\pi^- \pi^0 \pi^0$} events &
        36~waves, rank 1, \SIvalRange{0.08}{\tpr}{0.53}{\GeVcsq} \newline
        12~\tpr bins \\

        \midrule
        COMPASS~\cite{Alekseev:2009aa} &
        \SI{190}{\GeVc} $\pi^-$~beam on lead target \newline
        \num{420e3}~\threePi events &
        42~waves, rank 2, \SIvalRange{0.1}{\tpr}{1.0}{\GeVcsq} \newline
        $f_a^\refl(\tpr)$ for each partial wave~$a$ \\

        \midrule
        COMPASS~\cite{Adolph:2015tqa,Akhunzyanov:2018lqa} &
        \SI{190}{\GeVc} $\pi^-$~beam on proton target \newline
        \num{46e6}~\threePi events & 88~waves, rank 1,\protect\footnotemark[\value{mpfootnote}] \SIvalRange{0.1}{\tpr}{1.0}{\GeVcsq} \newline
        11~\tpr bins \\

        \bottomrule
      \end{tabular}
    \end{\ifMultiColumnLayout{small}{scriptsize}}
    \footnotetext[\value{mpfootnote}]{A rank-1 spin-density matrix was
      used for the 80~waves with positive reflectivity.  For the seven
      waves with negative reflectivity (see \cref{tab:wave_sets} in
      \cref{sec:wave_sets}), which together contribute only
      \SI{2.2}{\percent} to the total intensity, we used a rank-2
      spin-density matrix.}%
    \setcounter{footnote}{\value{mpfootnote}}%
  \end{minipage}
\end{wideTableOrNot}

\begin{wideFigureOrNot}[tbp]
  \centering%
  \subfloat[][]{%
    \includegraphics[width=\twoPlotWidth]{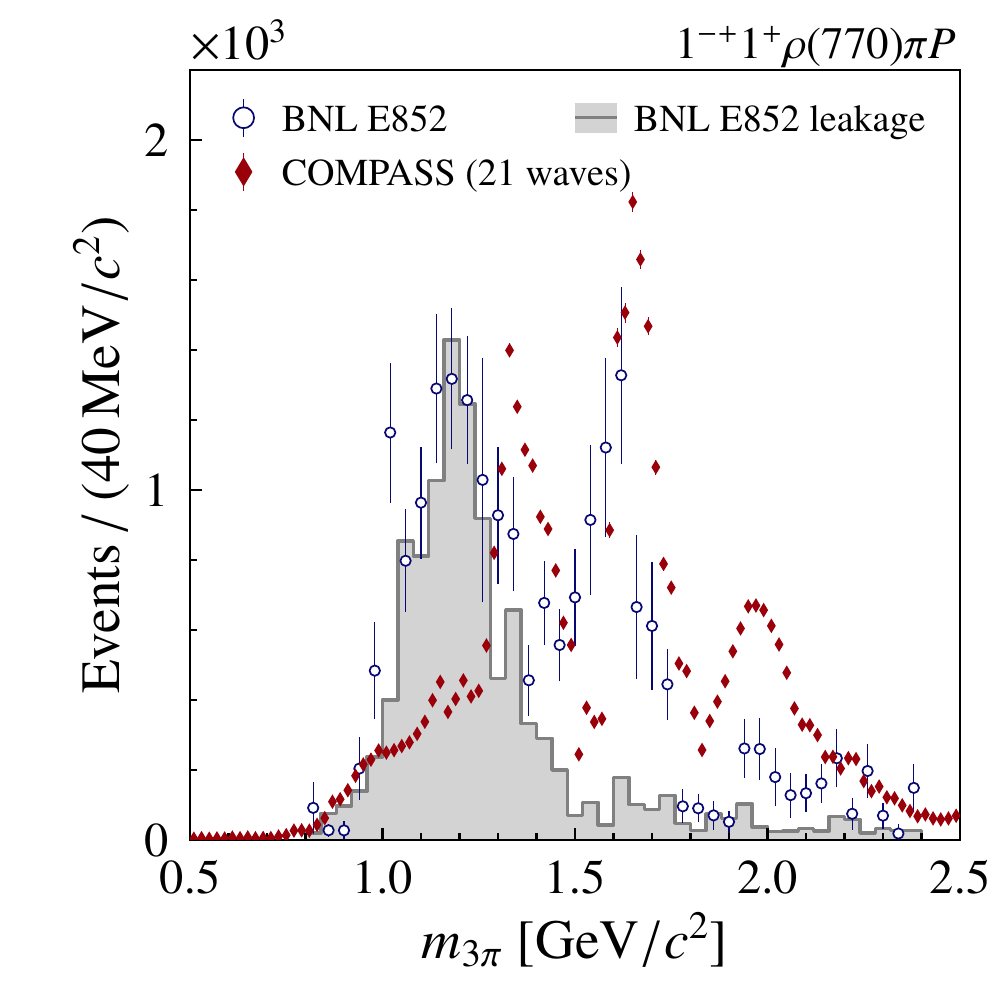}%
    \label{fig:exotic_comparison_bnlA}%
  }%
  \hspace*{\twoPlotSpacing}%
  \subfloat[][]{%
    \includegraphics[width=\twoPlotWidth]{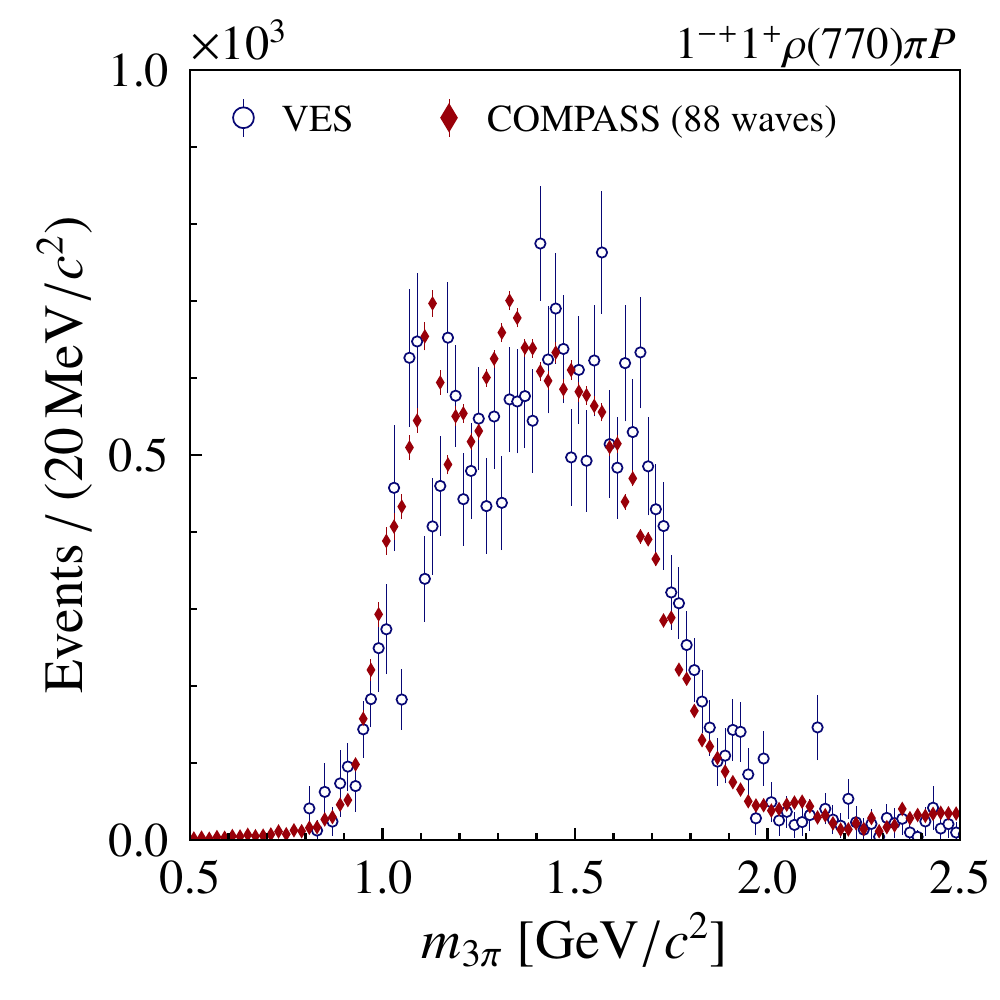}%
    \label{fig:exotic_comparison_VES}%
  }%
  \\%
  \subfloat[][]{%
    \includegraphics[width=\twoPlotWidth]{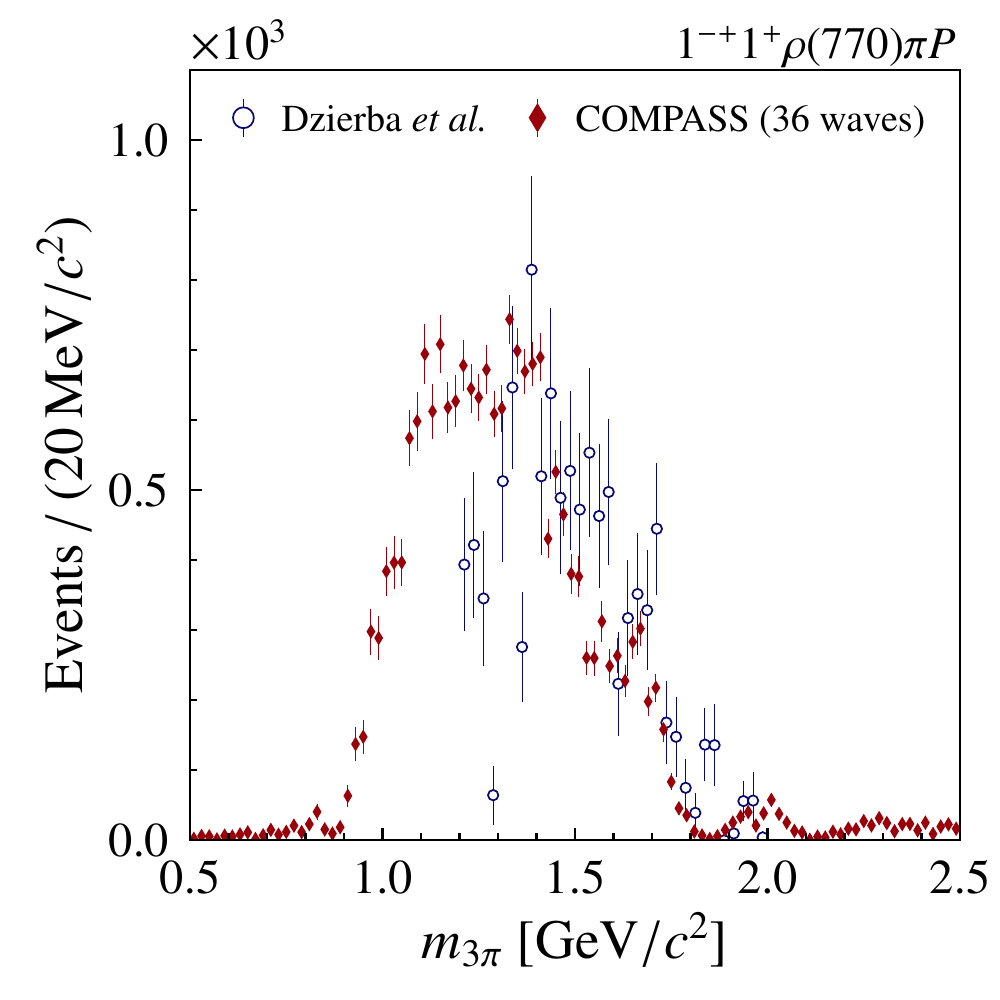}%
    \label{fig:exotic_comparison_bnlB}%
  }%
  \hspace*{\twoPlotSpacing}%
  \subfloat[][]{%
    \includegraphics[width=\twoPlotWidth]{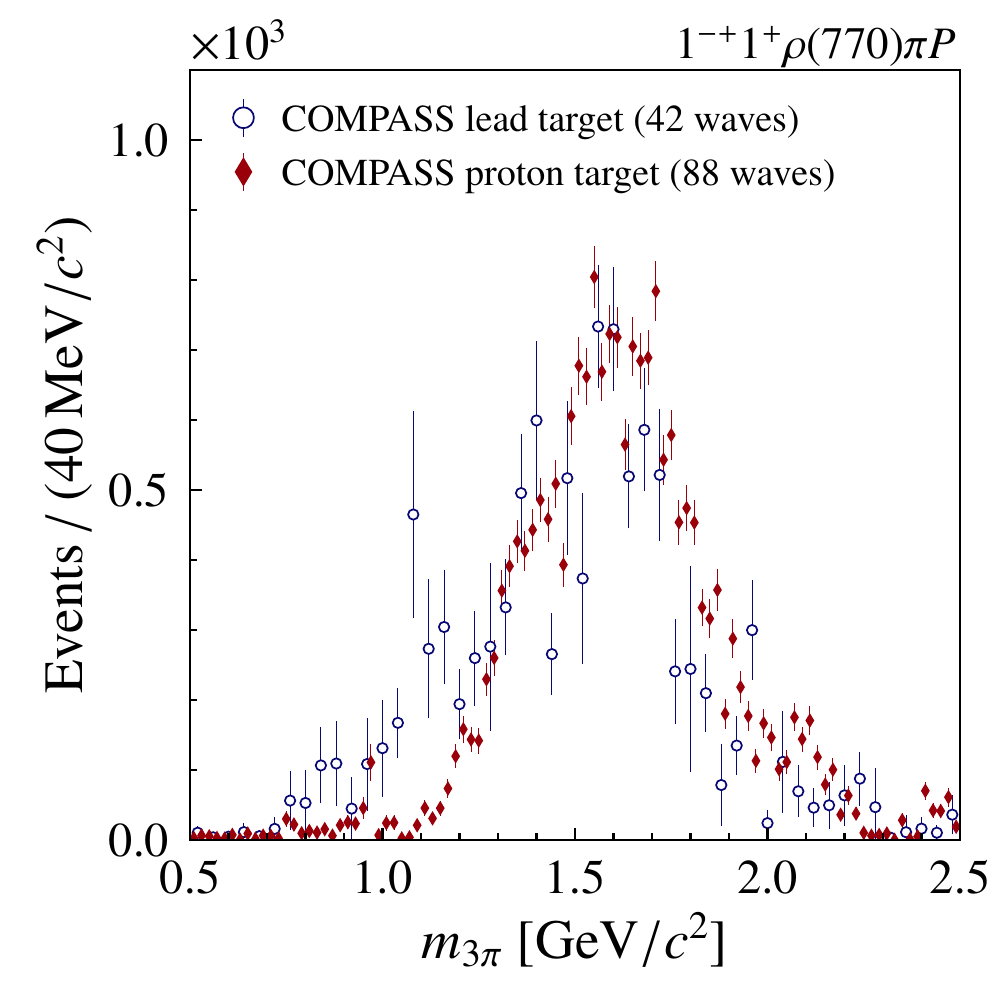}%
    \label{fig:exotic_comparison_compass_2004}%
  }%
  \caption{Comparison of intensity distributions of the spin-exotic
    \wave{1}{-+}{1}{+}{\Prho}{P} wave as obtained by different
    experiments measuring diffractive dissociation of a pion beam into
    $3\pi$.  Blue data points:
    \subfloatLabel{fig:exotic_comparison_bnlA}~21-wave fit of BNL~E852
    data in the range \SIvalRange{0.05}{\tpr}{1.0}{\GeVcsq},
    \subfloatLabel{fig:exotic_comparison_VES}~44-wave fit of VES data
    in the range \SIvalRange{0.03}{\tpr}{1.0}{\GeVcsq},
    \subfloatLabel{fig:exotic_comparison_bnlB}~36-wave fit of BNL~E852
    data in the range \SIvalRange{0.18}{\tpr}{0.23}{\GeVcsq}, and
    \subfloatLabel{fig:exotic_comparison_compass_2004}~42-wave fit of
    COMPASS lead-target data in the range
    \SIvalRange{0.1}{\tpr}{1.0}{\GeVcsq}.  The gray shaded area in
    panel~\subfloatLabel{fig:exotic_comparison_bnlA} shows the result
    of a leakage study performed by the BNL~E852
    experiment~\cite{Chung:2002pu}.  The red data points show the
    results of corresponding analyses of the COMPASS proton-target
    data using 11~\tpr bins in the range
    \SIvalRange{0.1}{\tpr}{1.0}{\GeVcsq}:
    \subfloatLabel{fig:exotic_comparison_bnlA}~\tpr-summed intensity
    distribution from the 21-wave PWA,
    \subfloatLabel{fig:exotic_comparison_VES}~\tpr-summed intensity
    distribution from the 88-wave PWA,
    \subfloatLabel{fig:exotic_comparison_bnlB}~intensity distribution
    from the 36-wave PWA in the range
    \SIvalRange{0.189}{\tpr}{0.220}{\GeVcsq}, and
    \subfloatLabel{fig:exotic_comparison_compass_2004}~intensity
    distribution from the 88-wave PWA in the range
    \SIvalRange{0.449}{\tpr}{0.724}{\GeVcsq}.  The red data points are
    scaled such that the intensity integrals of the blue and red data
    points in the region where they overlap are equal.  The blue data
    points are taken from
    \subfloatLabel{fig:exotic_comparison_bnlA}~Fig.~18(b) in
    \refCite{Chung:2002pu},
    \subfloatLabel{fig:exotic_comparison_VES}~Fig.~4(a) in
    \refCite{Zaitsev:2000rc},
    \subfloatLabel{fig:exotic_comparison_bnlB}~Fig.~25(a) in
    \refCite{Dzierba:2005jg}, and
    \subfloatLabel{fig:exotic_comparison_compass_2004}~Fig.~2(d) in
    \refCite{Alekseev:2009aa}.  The red data points in
    \subfloatLabel{fig:exotic_comparison_compass_2004}~are taken from
    Fig.~43(j) in \refCite{Akhunzyanov:2018lqa}.}%
  \label{fig:exotic_different_fits}
\end{wideFigureOrNot}

\begin{figure}[tbp]
  \centering%
  \includegraphics[width=\twoPlotWidth]{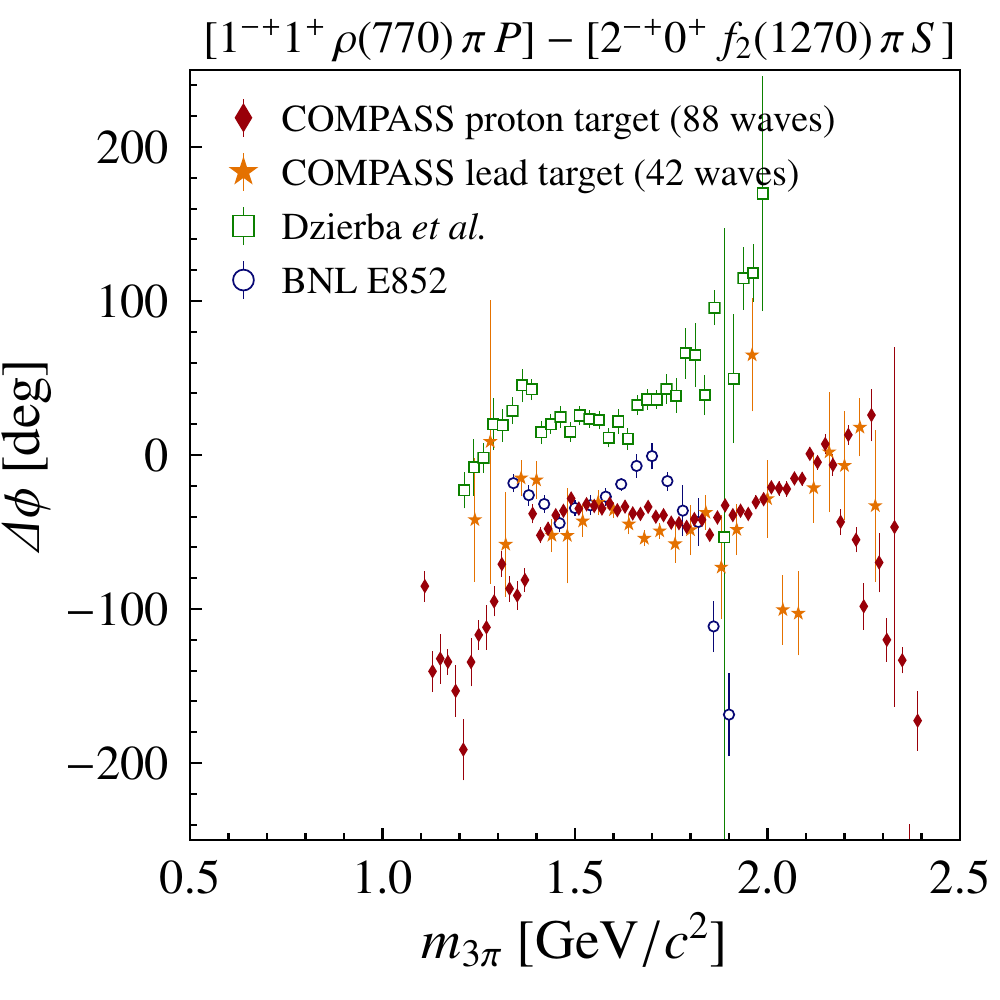}%
  \caption{Comparison of the phases of the spin-exotic
    \wave{1}{-+}{1}{+}{\Prho}{P} wave \wrt the
    \wave{2}{-+}{0}{+}{\PfTwo}{S} wave as obtained by different
    experiments measuring diffractive dissociation of a pion beam into
    $3\pi$.  Blue data points: 21-wave fit of BNL~E852 data in the
    range \SIvalRange{0.05}{\tpr}{1.0}{\GeVcsq} (shifted by
    \ang{-180}); green data points: 36-wave fit of BNL~E852 data in
    the range \SIvalRange{0.18}{\tpr}{0.23}{\GeVcsq} (shifted by
    \ang{+180}); orange data points: 42-wave fit of COMPASS
    lead-target data in the range
    \SIvalRange{0.1}{\tpr}{1.0}{\GeVcsq}; red data points: 88-wave fit
    of COMPASS proton-target data in the range
    \SIvalRange{0.449}{\tpr}{0.742}{\GeVcsq} (values for $\mThreePi <
    \SI{1.1}{\GeVcc}$ not shown).  Note that phase shifts of
    \ang{\pm180} may be caused, \eg by different choices of the
    analyzers in the definition of the coordinate systems or by
    different conventions used for the Wigner $D$~functions.  The
    data points were taken from Fig.~19(i) in \refCite{Chung:2002pu}
    (blue), Fig.~33 in \refCite{Dzierba:2005jg} (green), Fig.~3(b) in
    \refCite{Alekseev:2009aa} (orange), and Fig.~121 in
    \refCite{Akhunzyanov:2018lqa_suppl} (red).}%
  \label{fig:exotic_phases}%
\end{figure}

The BNL~E852 experiment was the first to claim a signal for
$\PpiOne[1600] \to \Prho\,\pi$ based on a PWA performed on
\num{250e3}~events obtained using an \SI{18.3}{\GeVc} pion beam
incident on a proton target in the kinematic range
\SIvalRange{0.05}{\tpr}{1.0}{\GeVcsq}~\cite{Adams:1998ff,Chung:2002pu}.
The employed PWA model included 21~waves (see \cref{tab:wave_sets} in
\cref{sec:wave_sets}) and a rank-1 spin-density matrix.  The different
\tpr~dependences of the partial-wave amplitudes were not taken into
account.  The blue data points in \cref{fig:exotic_comparison_bnlA}
show the resulting intensity distribution of the
\wave{1}{-+}{1}{+}{\Prho}{P} wave.  This distribution has two broad
enhancements.  The one in the \SIrange{1.1}{1.4}{\GeVcc} region was
attributed to wrongly assigned intensity leaking from the dominant
\wave{1}{++}{0}{+}{\Prho}{S} wave into the
\wave{1}{-+}{1}{+}{\Prho}{P} wave.  This leakage was caused by the
finite instrumental resolution in combination with a nonuniform
detector acceptance.  An estimate of this leakage obtained using Monte
Carlo techniques is shown by the gray-shaded histogram in
\cref{fig:exotic_comparison_bnlA}.  The second peak at
\SI{1.6}{\GeVcc} is accompanied by phase motions \wrt many waves (see,
\eg blue data points in \cref{fig:exotic_phases} shown in this
paper and Fig.~19 in \refCite{Chung:2002pu}) and was hence
interpreted as the \PpiOne[1600].  A simultaneous fit of the
\wave{1}{-+}{1}{+}{\Prho}{P} and \wave{2}{-+}{0}{+}{\PfTwo}{S}
amplitudes and their relative phase (see Fig.~24 in
\refCite{Chung:2002pu}) yielded Breit-Wigner parameters of
$m_{\PpiOne[1600]} = \SIsaerrs{1593}{8}{29}{47}{\MeVcc}$ and
$\Gamma_{\PpiOne[1600]} = \SIsaerrs{168}{20}{150}{12}{\MeVcc}$.  It is
noteworthy that the \PpiOne[1600] peak remained when the PWA was
performed in a low-\tpr region around \SI{0.1}{\GeVcsq}.  However, a
strong dependence of the shape and magnitude of the \PpiOne[1600]
signal on the PWA model was observed.

The VES experiment at IHEP used a \SI{36.6}{\GeVc} pion beam on a
solid-beryllium target and performed a PWA on \num{3.0e6}~events in
the kinematic range
\SIvalRange{0.03}{\tpr}{1.0}{\GeVcsq}~\cite{Zaitsev:2000rc}.  The PWA
model contained 44~waves (see \cref{tab:wave_sets} in
\cref{sec:wave_sets}) and the spin-density matrix used in the PWA fit
had maximum allowed rank.  To search for resonances, they extracted
from this spin-density matrix a rank-1 spin-density matrix of fully
coherent partial-wave amplitudes.  In the PWA model, the partial-wave
amplitudes were multiplied by an additional factor of $f_a^\refl(\tpr)
= (\tpr)^{\abs{M} / 2}\, \sqrt{e^{-b_1\, \tpr} + A\, e^{-b_2\,
\tpr}}$~\cite{Kachaev:private:2020} [see \cref{eq:trans_amp_t_model}]
to take into account the fact that the partial-wave intensity is
proportional to $(\tpr)^{\abs{M}}$.  Using this approach, also the VES
experiment observed significant intensity in the
\wave{1}{-+}{1}{+}{\Prho}{P} wave [see blue data points in
\cref{fig:exotic_comparison_VES}].  However, they did not observe a
peak at \SI{1.6}{\GeVcc} comparable to the one found in the BNL~E852
analysis in \refsCite{Adams:1998ff,Chung:2002pu}.  Instead, they found
a very broad intensity distribution with a slow phase motion of about
\ang{60} in the \SI{1.6}{\GeVcc} region (see Fig.~4 in
\refCite{Zaitsev:2000rc}).  From this they concluded that the
$\Prho\pi$ data alone are inconclusive concerning the existence of a
\PpiOne[1600] signal.  However, in a combined fit of the intensity
distributions of the $1^{-+}$ wave in the $\PbOne\pi$, $\eta'\pi$, and
$\Prho\pi$ channels, they found a satisfactory description of the data
using a \PpiOne[1600] resonance with $m_{\PpiOne[1600]} =
\SI{1560(60)}{\MeVcc}$ and $\Gamma_{\PpiOne[1600]} =
\SI{340(50)}{\MeVcc}$ (see Fig.~6 in \refCite{Zaitsev:2000rc}).

Dzierba~\etal~\cite{Dzierba:2005jg} performed a PWA of a second
BNL~E852 data sample of in total \num{5.6e6}~$3\pi$ events,
which is a factor 20~larger than the one used in the first analysis in
\refsCite{Adams:1998ff,Chung:2002pu}.  The analysis was performed
independently in 12~\tpr~bins in the range from
\SIrange{0.08}{0.53}{\GeVcsq}.  The PWA model employed a rank-1
spin-density matrix and a set of 36~partial waves (see
\cref{tab:wave_sets} in \cref{sec:wave_sets}).  This wave set was
derived from a larger parent wave set.  The resulting intensity
distribution of the \wave{1}{-+}{1}{+}{\Prho}{P} wave is broad and
structureless and shows no peak at \SI{1.6}{\GeVcc}.  As an example,
the blue data points in \cref{fig:exotic_comparison_bnlB} show the
intensity distribution in the \tpr~bin from
\SIrange{0.18}{0.23}{\GeVcsq}.  The shape of the intensity
distribution was found to change strongly with~\tpr (see Fig.~31 in
\refCite{Dzierba:2005jg}).  With increasing~\tpr, intensity moves from
the \SI{1.2}{\GeVcc} region to higher masses.  Applying the 21-wave
set from \refsCite{Adams:1998ff,Chung:2002pu} yielded a peak at
\SI{1.6}{\GeVcc} in the \wave{1}{-+}{1}{+}{\Prho}{P} intensity
distribution consistent with the first analysis of BNL~E852 data (see
Figs.~24 and~25 in \refCite{Dzierba:2005jg}).  The authors of
\refCite{Dzierba:2005jg} showed that leakage from the \PpiTwo causes
this peak, if the \wave{2}{-+}{0}{+}{\Prho}{P}, the
\wave{2}{-+}{0}{+}{\Prho}{F}, and the \wave{2}{-+}{1}{+}{\Prho}{F}
wave are omitted from the 36-wave model (see Figs.~27 and~28 in
\refCite{Dzierba:2005jg}); the latter two waves were missing in the
21-wave model used in \refsCite{Adams:1998ff,Chung:2002pu}.  Using
moments of the Wigner $D$~functions, Dzierba~\etal\ demonstrated that
the 36-wave model describes the data significantly better than the
21-wave model.  Based on these observations, they concluded that the
BNL~E852 data provide no evidence for the existence of a \PpiOne[1600]
in the $\Prho\,\pi$ channel.  For the discussion in
\cref{sec:comparison} below it is important to note that this
conclusion was based only on data in the range $\tpr <
\SI{0.53}{\GeVcsq}$ and that it was not corroborated by any kind of
resonance-model fit.  In the 36-wave PWA, Dzierba~\etal observed an
enhancement around \SI{1.6}{\GeVcc} in the higher \tpr~bins (see
Fig.~31 in \refCite{Dzierba:2005jg}) and an approximately constant
phase of the \wave{1}{-+}{1}{+}{\Prho}{P} wave \wrt the
\wave{2}{-+}{0}{+}{\PfTwo}{S} wave around \SI{1.6}{\GeVcc} [see green
data points in \cref{fig:exotic_phases} shown in this paper and
Figs.~25(b) and~33 in \refCite{Dzierba:2005jg}].  These effects could
be a sign for a $1^{-+}$ resonance with similar parameters as the
\PpiTwo, but they were both ascribed to remaining leakage from the
\PpiTwo into the $1^{-+}$ wave.

In contrast, the first analysis of a much smaller data sample of
\num{420e3}~events obtained by the COMPASS experiment using a
\SI{190}{\GeVc} pion beam on a solid-lead target showed clear evidence
for a \PpiOne[1600] signal in the \wave{1}{-+}{1}{+}{\Prho}{P}
wave~\cite{Alekseev:2009aa}.  We performed the PWA employing a rank-2
spin-density matrix and a set of 42~waves (see \cref{tab:wave_sets} in
\cref{sec:wave_sets}) in the range
\SIvalRange{0.1}{\tpr}{1.0}{\GeVcsq} using a parametrization for the
\tpr~dependence of the partial-wave amplitudes like in
\cref{eq:trans_amp_t_model} with different parameters for each wave.
The 42-wave set is similar to the 36-wave set used by Dzierba~\etal in
\refCite{Dzierba:2005jg}having 29~waves in common.  In particular, it
contains those three $2^{-+}$ waves that were found to make the peak
at \SI{1.6}{\GeVcc} disappear (see discussion above).  The resulting
intensity distribution of the \wave{1}{-+}{1}{+}{\Prho}{P} wave is
shown as blue data points in
\cref{fig:exotic_comparison_compass_2004}.  By performing a
resonance-model fit of six partial-wave amplitudes simultaneously, we
obtained Breit-Wigner parameters of $m_{\PpiOne[1600]} =
\SIsaerrs{1660}{10}{0}{64}{\MeVcc}$ and $\Gamma_{\PpiOne[1600]} =
\SIsaerrs{269}{21}{42}{64}{\MeVcc}$.  The \PpiOne[1600] parameters are
similar to the ones found for the \PpiTwo, which are $m_{\PpiTwo} =
\SIsaerrs{1658}{3}{24}{8}{\MeVcc}$ and $\Gamma_{\PpiTwo} =
\SIsaerrs{271}{9}{22}{24}{\MeVcc}$.  This explains the approximately
constant phase observed between the \wave{1}{-+}{1}{+}{\Prho}{P} and
\wave{2}{-+}{0}{+}{\PfTwo}{S} waves (see orange data points in
\cref{fig:exotic_phases}).

\subsection{Comparison of previous results with COMPASS proton-target data}%
\label{sec:comparison}

The COMPASS collaboration has recently published a detailed PWA of the
\threePi final state using a PWA model with 88~waves (see
\cref{tab:exotic_diff_models,tab:wave_sets}).  Here, we focus on the
\wave{1}{-+}{1}{+}{\Prho}{P} wave.  The red data points in
\cref{fig:exotic_comparison_VES} show the intensity distribution
summed over the 11~\tpr bins.  It is similar to the one found by the
VES experiment in a similar \tpr~range~\cite{Zaitsev:2000rc} (blue
data points).  We do not observe a peak at \SI{1.6}{\GeVcc} like in
the BNL~E852 21-wave PWA~\cite{Adams:1998ff,Chung:2002pu} [\confer
blue data points in \cref{fig:exotic_comparison_bnlA}].  Surprisingly,
the \tpr-summed intensity distributions in
\cref{fig:exotic_comparison_VES} are different from the one obtained
in the analysis of the COMPASS lead-target data~\cite{Alekseev:2009aa}
[blue points in \cref{fig:exotic_comparison_compass_2004}].
Na\"ively, one could expect these intensity distributions to be
similar because $\tpr > \SI{0.1}{\GeVcsq}$, \ie far above the region
corresponding to coherent scattering off the lead
nucleus.  Hence the beam pion
scatters off quasifree nucleons inside the nucleus.  In the COMPASS
proton-target data, the shape of the intensity distribution of the
\wave{1}{-+}{1}{+}{\Prho}{P} wave exhibits a surprisingly strong
dependence on~\tpr (see \cref{fig:exotic_deck_comparison_tbins} shown
in this paper and Fig.~43 in \refCite{Akhunzyanov:2018lqa}).
This confirms a similar observation made by
Dzierba~\etal~\cite{Dzierba:2005jg}.  At low~\tpr, the intensity
distribution is dominated by a broad structure that extends from about
\SIrange{1.0}{1.7}{\GeVcc}.  With increasing~\tpr, the structure
becomes narrower and its maximum moves to about \SI{1.6}{\GeVcc}.
Interestingly, for $\tpr \gtrsim \SI{0.5}{\GeVcsq}$ (\ie above the
kinematic range considered by Dzierba~\etal~\cite{Dzierba:2005jg}) the
intensity distribution actually resembles the one that we obtained in
the analysis of the COMPASS lead-target data in the range
\SIvalRange{0.1}{\tpr}{1.0}{\GeVcsq} [see
\cref{fig:exotic_comparison_compass_2004}].

In the COMPASS proton-target data, we observe slow phase motions of
the \wave{1}{-+}{1}{+}{\Prho}{P} wave \wrt other waves in the
\SI{1.6}{\GeVcc} region (see Fig.~44 in
\refCite{Akhunzyanov:2018lqa}).  As an example, the red data points in
\cref{fig:exotic_phases} show the phase \wrt the
\wave{2}{-+}{0}{+}{\PfTwo}{S} wave.  Compared to the rather large
differences in the intensity distributions (see
\cref{fig:exotic_different_fits}), the \mThreePi~dependence of the
phases of the \wave{1}{-+}{1}{+}{\Prho}{P} wave relative to other
waves is more robust \wrt changes of the analysis model.  The phase
motion from the COMPASS proton-target data is less pronounced than the
one observed by the BNL~E852 collaboration but agrees qualitatively
with the phase motions observed by Dzierba~\etal and in the analysis
of the COMPASS lead-target data.  We have no explanation for the
approximately \ang{+60} offset of the phase motion reported by
Dzierba~\etal~\cite{Dzierba:2005jg} \wrt the other
analyses.\footnote{The panels in Fig.~33 in \refCite{Dzierba:2005jg}
that correspond to the \tpr~bins numbered~6 and~7 show the same data
points.  Thus it is unclear whether the shown phase motion is that of
bin~6 or bin~7.  However, this probably does not explain the phase
offset \wrt the other analyses since the phase in the $\mThreePi =
\SI{1.6}{\GeVcc}$ region depends only weakly on~\tpr.}

The strong \tpr~dependence of the shape of the intensity distribution
hints at large contributions from nonresonant processes related, \eg
to the Deck effect~\cite{Deck:1964hm}.  This was confirmed by our
resonance-model fit of the COMPASS proton-target data, which describes
the partial-wave intensities and interference terms of 14~selected
partial waves simultaneously~\cite{Akhunzyanov:2018lqa}.  The
resonance-model fit was performed for the first time simultaneously in
all \tpr~bins with the resonance parameters, \ie masses and widths,
forced to be the same across the \tpr~bins.  In this \tpr-resolved
approach, we exploit the in general different \tpr~dependences of the
resonant and nonresonant amplitudes to better disentangle the two
contributions.  This eventually yields more realistic estimates for
the resonance parameters.  The model reproduces the
\wave{1}{-+}{1}{+}{\Prho}{P} intensities and phase motions well by a
\tpr-dependent interference between the \PpiOne[1600] and a
nonresonant component.\footnote{We parametrize the nonresonant
amplitude using Eqs.~(27) and~(28) in \refCite{Akhunzyanov:2018lqa}.
This is an empirical parametrization in the form of a Gaussian in the
two-body breakup momentum of the isobar-pion decay that was inspired
by \refCite{tornqvist:1995kr}.}  The latter strongly changes shape,
strength, and phase with \tpr.  At low~\tpr, the intensity is
dominated by the large nonresonant component, which interferes
constructively with the \PpiOne[1600] at low masses.  With
increasing~\tpr, the strength of the nonresonant component decreases
more quickly than that of the \PpiOne[1600] so that the latter becomes
the dominant component.  For $\tpr \gtrsim \SI{0.5}{\GeVcsq}$, \ie in
the two highest \tpr~bins, the nonresonant component is small or even
vanishing in the \SI{1.6}{\GeVcc} mass region, and the broad peak in
the data is nearly entirely described by the \PpiOne[1600].  The
resonance model is not able to reproduce a narrow enhancement at about
\SI{1.1}{\GeVcc}, which appears at low~\tpr.  This structure is not
accompanied by any phase motion and the intensity in this mass region
is sensitive to details of the PWA model.  This makes a resonance
interpretation unlikely and we hence suspect this structure to be an
artifact induced by imperfections in the analysis method.  A similar
observation has been made in the VES analysis~\cite{Zaitsev:2000rc}.
A similar structure also appears in a more advanced PWA (see
\cref{sec:freed_isobar,fig:coherent}), where we significantly reduce
the model bias introduced by the chosen parametrizations for the
dynamic isobar amplitudes.  Hence its appearance does not seem to be
tightly related to how well the isobar amplitudes are described by the
PWA model.

From the resonance-model fit, we obtain Breit-Wigner parameters of
$m_{\PpiOne[1600]} = \ifMultiColumnLayout{}{\linebreak}
\SIaerr{1600}{110}{60}{\MeVcc}$ and $\Gamma_{\PpiOne[1600]} =
\SIaerr{580}{100}{230}{\MeVcc}$.  The quoted uncertainties are
systematic only (see \refCite{Akhunzyanov:2018lqa} for details on the
performed studies); the statistical uncertainties are more than an
order of magnitude smaller and hence negligible.  Although the mass
value agrees well with the one found in our analysis of the COMPASS
lead-target data~\cite{Alekseev:2009aa} (see \cref{sec:other_models}),
the width found in the proton-target data is considerably larger.  The
reason for this discrepancy is not understood.  However, it could be
related to the fact that relative to the \PpiOne[1600] the
contribution from the nonresonant components is much larger in the
proton-target data than in the lead-target data.  Also, our resonance
models, which we use to decompose the partial-wave amplitudes into
coherent sums of Breit-Wigner resonances and nonresonant amplitudes,
might render the resonance parameters
process dependent~\cite{pdg_resonances:2018}.  In addition, due to the
much smaller data sample, the analysis of the lead-target data was
performed by integrating over~\tpr and by modeling the \tpr~dependence
of the partial-wave amplitudes according to
\cref{eq:trans_amp_t_model}.  Therefore, a potential \tpr~dependence
of the shape of the \wave{1}{-+}{1}{+}{\Prho}{P} amplitude was not
taken into account.  A recent coupled-channel analysis of COMPASS data
on diffractively produced $\eta \pi^-$ and $\eta' \pi^-$ final states
performed by the JPAC collaboration finds a resonance pole with
parameters of $m_{\PpiOne[1600]} = \SIerrs{1564}{24}{86}{\MeVcc}$ and
$\Gamma_{\PpiOne[1600]} =
\SIerrs{492}{54}{102}{\MeVcc}$~\cite{Rodas:2018owy} that are more
consistent with the Breit-Wigner parameters we find in the COMPASS
proton-target data.

Are the different results from previous analyses, in particular the
two analyses based on BNL~E852 data, caused by inconsistencies of the
data or by the different PWA models?  In order to answer this
question, we investigate the impact of the different analysis models
used for the BNL~E852 data, by applying the 21-wave set from
\refsCite{Adams:1998ff,Chung:2002pu} and the 36-wave set from
\refCite{Dzierba:2005jg} (see \cref{tab:wave_sets} in
\cref{sec:wave_sets}) to the high-precision COMPASS proton-target data
sample keeping the subdivision into 11~\tpr bins.  The red data points
in \cref{fig:exotic_comparison_bnlA} show the \tpr-summed intensity
distribution of the \wave{1}{-+}{1}{+}{\Prho}{P} wave as obtained from
the PWA using the 21-wave set.  The intensity distribution exhibits a
clear peak slightly above \SI{1.6}{\GeVcc}, similar to the signal
found in the BNL~E852 analysis in \refsCite{Adams:1998ff,Chung:2002pu}
(blue data points).  In the low-mass region, the intensities from the
two analyses shown in \cref{fig:exotic_comparison_bnlA} are not
directly comparable.  The COMPASS acceptance is much more uniform than
the acceptance of the BNL~E852 experiment and hence leakage induced by
the experimental acceptance is much suppressed.  We also confirm the
finding of Dzierba~\etal that the \SI{1.6}{\GeVcc} peak vanishes by
applying the 36-wave set to the COMPASS proton-target data.  As an
example, we compare in \cref{fig:exotic_comparison_bnlB} the intensity
distributions around $\tpr = \SI{0.2}{\GeVcsq}$.  Consequently, our
data support the conclusion from \refCite{Dzierba:2005jg} that the
\SI{1.6}{\GeVcc} peak observed in \refsCite{Adams:1998ff,Chung:2002pu}
is an artificial structure caused by using a wave set that misses
important waves.  This conclusion is further supported by the fact
that using the 21-wave set we find contrary to the expected dominance
of natural-parity exchange a peak of similar height in the same mass
region in the \wave{1}{-+}{1}{-}{\Prho}{P} wave, which has negative
reflectivity corresponding to unnatural-parity exchange.  This has
also been pointed out by VES~\cite{Zaitsev:2000rc}.

Our \tpr-resolved analysis using the 88-wave set also confirms the
finding of Dzierba~\etal that the \PpiOne[1600] signal is weak
compared to the nonresonant component in the range $\tpr \lesssim
\SI{0.5}{\GeVcsq}$.  In the range $\tpr < \SI{0.53}{\GeVcsq}$ analyzed
in \refCite{Dzierba:2005jg}, we find that the \PpiOne[1600] signal is
masked by the dominant contributions from nonresonant processes.
However, our analysis contradicts the conclusion from
\refCite{Dzierba:2005jg} that there is no evidence for the
\PpiOne[1600] in $3\pi$.  The COMPASS proton-target data require a
\PpiOne[1600] resonance in the range $\tpr \gtrsim \SI{0.5}{\GeVcsq}$
[see, \eg\
\cref{fig:piOne1600_resModelFit_int_t11,fig:piOne1600_resModelFit_phase_t11}]
and also the COMPASS lead-target data cannot be described without a
\PpiOne[1600].

It is not yet understood why the \PpiOne[1600] signal is enhanced \wrt
the nonresonant component in the lead-target data as compared to our
proton-target data.  However, we do observe a general enhancement of
the intensity of waves with spin projection $M = 1$ over those with $M
= 0$ in the lead-target data~\cite{Haas:2014bzm}.

\subsection{Summary: Previous results and comparison with COMPASS data}%
\label{sec:other_models_summary}

Using our highly precise COMPASS proton-target data we reproduce the
key PWA results of all previous analyses of the
\wave{1}{-+}{1}{+}{\Prho}{P} wave by applying their analysis models.
We conclude that this wave contains a \PpiOne[1600] signal and that
the discrepancies and mutual inconsistencies observed in previous
analyses originate either from model artifacts or from studying too
restricted \tpr~ranges.  The PWA model with 21~waves used in
\refsCite{Adams:1998ff,Chung:2002pu} contained too few waves leading
to an artificial peak being misinterpreted as the \PpiOne[1600].  The
analysis in \refCite{Dzierba:2005jg} excluded the region $\tpr >
\SI{0.53}{\GeVcsq}$ and hence missed the region, in which the
\PpiOne[1600] signal rises above the nonresonant background.  Since
the VES analysis was not performed in \tpr~bins, their \PpiOne[1600]
signal was also diluted by large nonresonant contributions.

A remaining puzzle is that in $\gamma + \pi^\pm \to \pi^\pm \pi^-
\pi^+$ reactions the production of the \PpiOne[1600] seems to be much
less prominent than expected considering vector-meson dominance and
the observation of the $\Prho \pi$ decay.\footnote{The absolute
partial width $\PpiOne[1600] \to \Prho \pi$ is currently unknown.
However, our results suggest that the branching fraction might be in
the percent region.}  The CLAS
experiment~\cite{Nozar:2008aa,Eugenio:2013xua} and the COMPASS
Primakoff experiment~\cite{Ketzer:2012vn,Grabmuller:2012oja} find
nearly vanishing intensities of the $1^{-+}$ wave in the
\SI{1.6}{\GeVcc} mass region.  This, however, could in principle be
due to destructive interference of a \PpiOne[1600] with a nonresonant
component---a hypothesis that could be verified by resonance-model
fits.  In the future, much more precise photoproduction data from
Jefferson Laboratory will help to clarify the situation.
\section{Study of dynamic isobar amplitudes}%
\label{sec:freed_isobar}

The partial-wave analyses of the $3\pi$ final state performed so far
(see \cref{sec:other_models}) used the conventional isobar model where
isobar resonances are described using fixed parametrizations for their
dynamic amplitude $\Delta_a(m_\xi)$ (see \cref{sec:pwa}) with
resonance parameters taken from previous
experiments~\cite{Tanabashi:2018zz}.\footnote{Here, $m_\xi
\equiv \mTwoPi$.}  Even though this approach is quite common, it
might introduce a model bias in the analysis because the fixed dynamic
amplitudes might deviate from the true ones present in real data.  The
differences could be due to distortions of the
\twoPi dynamic amplitudes, caused by the
presence of the third pion, or due to contributions from excited
isobar resonances or nonresonant processes.

To study this possible bias in our PWA model, we reanalyze our dataset
using the freed-isobar PWA method presented in detail in
\refsCite{Krinner:2017dba,Krinner:2018bwg}.  This analysis technique
no longer relies on fixed parametrizations for the dynamic isobar
amplitudes, but allows us to extract these amplitudes from the data
themselves with much reduced model dependence.  In this approach, the
fixed parametrization for the dynamic amplitude $\Delta_a(m_\xi)$ of
an isobar~$\xi$ in wave~$a$ [see \cref{eq:wave_index}] is replaced by
a set of piecewise constant amplitudes defined over a contiguous set
of intervals in the \twoPi mass~$m_\xi$ that are indexed by~$k$,
\ie\footnote{\label{footnote:model}In the following, we discuss PWA
models with rank~1 and waves with positive reflectivity.  We hence
omit the \refl~and $r$~indices from here onwards.}
\begin{equation}
  \label{eq:dyn_amp_freed}
  \begin{splitOrNot}
    \alignOrNot
    \Delta_a(m_\xi)
    = \sum_k \mathscr{T}_{a, k}\, \Pi_{k, \xi}(m_\xi)
    \newLineOrNot
    \ifMultiColumnLayout{&}{\quad}
    \text{with}\quad
    \Pi_{k, \xi}(m_\xi)
    = \begin{cases}
        1, & \text{if $m_{k, \xi} \leq m_\xi < m_{k + 1, \xi}$,} \\
        0, & \text{otherwise}.
      \end{cases}
  \end{splitOrNot}
\end{equation}
This way, the dynamic amplitude for isobar~$\xi$ is approximated by
the set $\{\mathscr{T}_{a, k}\}$ of complex-valued constants.  This
method allows us not only to estimate the model bias caused by the
fixed dynamic isobar amplitudes in our PWA model, but also to study
the dynamic isobar amplitudes themselves.

In our PWA model, we factorize the decay amplitude~$\decayAmp_a$ of a
partial wave~$a$ in \cref{eq:intensity_bin} into the dynamic isobar
amplitude~$\Delta_a$ and an angular amplitude~$\angularPart_a$.
Including the Bose symmetrization \wrt the two
indistinguishable~$\pi^-$ in the $\pi^-_1 \pi^-_2 \pi^+_3$ final
state, we write for a given $(\mThreePi, \tpr)$ cell [see Eq.~(47) in
\refCite{Adolph:2015tqa}]:\footnote{The angular amplitude is that part
of the decay amplitude, which depends only on the decay angles and not
on~\mThreePi or~\mTwoPi.  It is given by Eqs.~(11) and~(7) in
\refCite{Adolph:2015tqa} without the dynamic parts $f_{\lambda\,0}^J$
and $f_{0\,0}^{J_\xi}$.}
\begin{equation}
  \label{eq:decay_amp_factorization}
  \decayAmp_a(\tau_{13}, \tau_{23})
  = \angularPart_a(\tau_{13})\, \Delta_a(m_{13})
    + \angularPart_a(\tau_{23})\, \Delta_a(m_{23}).
\end{equation}
It is important that $\Delta_a$ depends only on the invariant
mass~$m_{ij}$ of the $\pi^-_i \pi^+_j$ subsystems forming the isobar
and that $\angularPart_a$ depends only on the four angular variables
in the set of five phase-space variables of the three-body system
represented by~$\tau_{ij}$ (see Sec.~III~A in \refCite{Adolph:2015tqa}
for details on the definition of the coordinate systems).  Inserting
\cref{eq:dyn_amp_freed} into \cref{eq:decay_amp_factorization} and
defining separate transition and decay amplitudes for every \twoPi
mass interval~$k$ via
\begin{equation}
  \label{eq:prod_amp_freed}
  \prodAmp_{a, k}
  \equiv \prodAmp_a\, \mathscr{T}_{a, k}
\end{equation}
and
\begin{multlineOrEq}
  \label{eq:decay_amp_freed}
  \decayAmp_{a, k}(\tau_{13}, \tau_{23})
  \equiv \angularPart_a(\tau_{13})\, \Pi_{k, \xi}(m_{13}) \newLineOrNot
    + \angularPart_a(\tau_{23})\, \Pi_{k, \xi}(m_{23}),
\end{multlineOrEq}
the expression for the intensity distribution in
\cref{eq:intensity_bin} can be written as\footnote{See
\cref{footnote:model}.}
\begin{equation}
  \label{eq:intensity_freed}
  \intensity(\tau_{13}, \tau_{23})
  = \Abs[3]{\sum_a \sum_k \prodAmp_{a, k}\, \decayAmp_{a, k}(\tau_{13}, \tau_{23})}^2
    + \intensity_\text{flat}.
\end{equation}
Note that although \cref{eq:intensity_freed} contains an additional
sum over the two-pion mass intervals~$k$, the mathematical structure
is exactly the same as in \cref{eq:intensity_bin}.  We can thus use
the same extended maximum likelihood approach to determine the set
$\{\prodAmp_{a, k}\}$ of the unknown fit parameters from the data.

Performing a freed-isobar PWA in $(\mThreePi, \tpr)$ cells, yields
transition amplitudes $\prodAmp_{a, k}(\mThreePi, \tpr) =
\prodAmp_a(\mThreePi, m_\xi, \tpr)$ that now depend not only
on~$\mThreePi$ and~$\tpr$ but also on~$m_\xi$ via the index~$k$.
According to \cref{eq:prod_amp_freed}, a freed-isobar transition
amplitude contains information on both the $3\pi$ system and the
\twoPi subsystem.  For each freed-isobar wave in the PWA model and
each $(\mThreePi, \tpr)$ cell, the method yields an \Argand ranging
in~$m_\xi$ from $2 m_\pi$ to $\mThreePi - m_\pi$.  It is important to
note that in the freed-isobar approach, we do not make any assumptions
on the resonance content of the \twoPi subsystem.  The freed-isobar
PWA thus allows us to determine from the data the overall amplitude of
all \twoPi intermediate states with given \JPC quantum numbers in the
$3\pi$ partial wave defined by~$a$.  This amplitude hence includes in
principle all contributing \twoPi resonances, potential nonresonant
contributions, as well as distortions due to final-state interactions.
Note that a \twoPi system with even relative orbital angular momentum,
\ie even total spin~$J$, has \IGJPC quantum numbers $0^+\,J^{++}$, which
correspond to $f_J$~states, or $2^+\,J^{++}$, which would be
flavor exotic.  A \twoPi system with odd relative orbital angular
momentum, \ie odd~$J$, has \IGJPC quantum numbers $1^+\,J^{--}$, which
correspond to $\rho_J$~states.

In the ansatz in \cref{eq:intensity_freed}, we sum coherently over the
index~$k$ of the \mTwoPi~intervals.  This takes into account the
interference of the amplitudes in different \mTwoPi~intervals due to
Bose symmetrization of the final-state particles.  This is
conceptually different from the binning in~\mThreePi and~\tpr, where
all kinematic bins are independent.

The obtained dynamic isobar amplitudes can be different for every
wave~$a$, even though they might describe \twoPi subsystems with the
same relative orbital angular momentum.  The reduced model dependence
of the freed-isobar method and the additional information on the
\twoPi subsystems come at the price of a considerably larger number of
fit parameters compared to the conventional fixed-isobar PWA\@.  Thus
even for large datasets, the freed-isobar approach is feasible only
when it is applied to a selected subset of partial waves in the PWA
model, while for the remaining partial waves the conventional fixed
isobar parametrizations are used.

Based on the COMPASS proton-target data, we have performed a first
freed-isobar PWA already in \refCite{Adolph:2015tqa} to extract the
dynamic \twoPi $S$-wave amplitudes in three different $3\pi$ partial
waves.

\subsection{Freed-isobar analysis model}%
\label{sec:freed_isobar_pwa_model}

In the following, we apply the freed-isobar method to the
\wave{1}{-+}{1}{+}{\Prho}{P} wave.  Since this wave has a low relative
intensity of only \SI{0.8}{\percent} it is prone to potential leakage
effects.  Therefore, it does not suffice to free the dynamic isobar
amplitude only in the $1^{-+}$ wave.  Small imperfections in the
description of the dynamic isobar amplitudes of waves with much higher
relative intensity could create tensions between model and data, which
in turn could induce leakage into the freed $1^{-+}$ wave due to its
high flexibility.

Therefore, we free those 12~waves of our 88-wave PWA
model (see \cref{tab:wave_sets} in \cref{sec:wave_sets}), that
obtained a relative intensity of more than \SI{1}{\percent} in the
conventional PWA\@.  In addition to these 12~waves, we free the
\wave{1}{-+}{1}{+}{\Prho}{P} wave to study its $1^{--}$ dynamic isobar
amplitude.  The 88-wave PWA model contains subsets of waves with
identical quantum numbers but different $\IGJPC = 0^+\, 0^{++}$ isobar
resonances.  Such waves are absorbed into a single freed-isobar wave
with $\JPC = 0^{++}$ of the \twoPi subsystem (indicated by the
brackets in \cref{tab:freedIsobarList}).  As a consequence, three
additional waves with a relative intensity below \SI{1}{\percent} are
also freed.  In total, we replace the dynamic isobar amplitudes of~16
of the original 88~fixed-isobar waves by 12~waves with freed-isobar
amplitudes (see \cref{tab:freedIsobarList,tab:wave_sets}); 72~waves
with fixed dynamic isobar amplitudes remain in the freed-isobar PWA
model.  In the conventional fixed-isobar PWA, the intensity sum of the
16~freed waves accounts for \SI{83.3}{\percent} of the total
intensity.

\begin{table}[tbp]
  \begin{minipage}{\linewidth}
    \setcounter{mpfootnote}{\value{footnote}}%
    \refstepcounter{mpfootnote}%
    \renewcommand{\thempfootnote}{\ifMultiColumnLayout{\arabic{mpfootnote}}{[\alph{mpfootnote}]}}%
    \ifMultiColumnLayout{}{\centering}%
    \renewcommand{\arraystretch}{1.2}%
    \caption{Waves in the freed-isobar PWA model with dynamic isobar
      amplitudes parametrized according to \cref{eq:dyn_amp_freed}.
      The notation \pipiJF represents a \twoPi subsystem with
      well-defined \JPC quantum
      numbers.\protect\footnotemark[\value{mpfootnote}]  The center
      column lists the corresponding waves in the conventional 88-wave
      fixed-isobar PWA (see \cref{tab:wave_sets} in
      \cref{sec:wave_sets}) and the right column their relative
      intensity as obtained in \refCite{Adolph:2015tqa}.}%
    \label{tab:freedIsobarList}
    \begin{tabular}{l@{\,}lr}
      \toprule
      \textbf{Freed wave}            & \textbf{Fixed wave(s)}               & \textbf{Relative} \\
                                     &                                      & \textbf{intensity} \\
      \midrule
      \ldelim\{{3}{*}[\wave{0}{-+}{0}{+}{\pipiSF}{S}~]
                                     & \wave{0}{-+}{0}{+}{\pipiS}{S}        &  \SI{8.0}{\percent} \\
                                     & \wave{0}{-+}{0}{+}{\PfZero[980]}{S}  &  \SI{2.4}{\percent} \\
                                     & \wave{0}{-+}{0}{+}{\PfZero[1500]}{S} &  \SI{0.1}{\percent} \\[1.2ex]

      \wave{0}{-+}{0}{+}{\pipiPF}{P} & \wave{0}{-+}{0}{+}{\Prho}{P}         &  \SI{3.5}{\percent} \\[1.2ex]

      \ldelim\{{2}{*}[\wave{1}{++}{0}{+}{\pipiSF}{P}~]
                                     & \wave{1}{++}{0}{+}{\pipiS}{P}        &  \SI{4.1}{\percent} \\
                                     & \wave{1}{++}{0}{+}{\PfZero[980]}{P}  &  \SI{0.3}{\percent} \\[1.2ex]

      \wave{1}{++}{0}{+}{\pipiPF}{S} & \wave{1}{++}{0}{+}{\Prho}{S}         & \SI{32.7}{\percent} \\[1.2ex]

      \wave{1}{++}{1}{+}{\pipiPF}{S} & \wave{1}{++}{1}{+}{\Prho}{S}         &  \SI{4.1}{\percent} \\[1.2ex]

      \wave{1}{-+}{1}{+}{\pipiPF}{P} & \wave{1}{-+}{1}{+}{\Prho}{P}         &  \SI{0.8}{\percent} \\[1.2ex]

      \wave{2}{++}{1}{+}{\pipiPF}{D} & \wave{2}{++}{1}{+}{\Prho}{D}         &  \SI{7.7}{\percent} \\[1.2ex]

      \ldelim\{{2}{*}[\wave{2}{-+}{0}{+}{\pipiSF}{D}~]
                                     & \wave{2}{-+}{0}{+}{\pipiS}{D}        &  \SI{3.0}{\percent} \\
                                     & \wave{2}{-+}{0}{+}{\PfZero[980]}{D}  &  \SI{0.6}{\percent} \\[1.2ex]

      \wave{2}{-+}{0}{+}{\pipiPF}{P} & \wave{2}{-+}{0}{+}{\Prho}{P}         &  \SI{3.8}{\percent} \\[1.2ex]

      \wave{2}{-+}{1}{+}{\pipiPF}{P} & \wave{2}{-+}{1}{+}{\Prho}{P}         &  \SI{3.3}{\percent} \\[1.2ex]

      \wave{2}{-+}{0}{+}{\pipiPF}{F} & \wave{2}{-+}{0}{+}{\Prho}{F}         &  \SI{2.2}{\percent} \\[1.2ex]

      \wave{2}{-+}{0}{+}{\pipiDF}{S} & \wave{2}{-+}{0}{+}{\PfTwo[1270]}{S}  &  \SI{6.7}{\percent} \\
      \midrule
                                     & \textbf{Intensity sum}               & \textbf{\SI{83.3}{\percent}} \\
      \bottomrule
    \end{tabular}
    \footnotetext[\value{mpfootnote}]{In the fixed-isobar PWA, \pipiS
    represents a parametrization for the broad component of the \twoPi
    $S$-wave amplitude based on~\refCite{Au:1986vs} (see Sec.~III~A in
    \refCite{Adolph:2015tqa} for details).}%
    \setcounter{footnote}{\value{mpfootnote}}%
  \end{minipage}
\end{table}

For the freed-isobar waves, we choose \mTwoPi~intervals with a width
of \SI{40}{\MeVcc} except in the regions of the known \Prho,
\PfZero[980], and \PfTwo resonances, where we use a finer binning.
For the waves with $\JPC = 1^{--}$ isobars, we use an interval width
of \SI{20}{\MeVcc} in the range from
\SIrange{0.64}{0.92}{\GeVcc}.\footnote{For $\JPC = 0^{++}$ isobars, we
use an interval width of \SI{10}{\MeVcc} in the range from
\SIrange{0.92}{1.08}{\GeVcc}; for $\JPC = 2^{++}$, we use an interval
width of \SI{20}{\MeVcc} from \SIrange{1.18}{1.40}{\GeVcc}.}  Our
freed-isobar PWA model (see \cref{tab:freedIsobarList}) has a much
larger number of fit parameters than the conventional fixed-isobar
PWA\@.\footnote{In the highest \mThreePi~bin at \SI{2.48}{\GeVcc}, the
number of free real-valued parameters in the freed-isobar PWA is
\num{1520}.  This number decreases with decreasing~\mThreePi because
fewer \mTwoPi~intervals are kinematically allowed.  In the same
\mThreePi~bin, the 88-wave fixed-isobar PWA model (see
\cref{tab:wave_sets} in \cref{sec:wave_sets}) has \num{184}~free
real-valued parameters.}  In order to sufficiently constrain the fit
parameters by data, we increase the \mThreePi~bin width from
\SI{20}{\MeVcc} in the fixed-isobar PWA to \SI{40}{\MeVcc} in the
freed-isobar PWA and reduce in addition the number of \tpr~bins
from~11 to~4 (see \cref{tab:t-bins}).  We thus decrease the total
number of kinematic $(\mThreePi, \tpr)$ cells in the analyzed range
from~\num{1100} in the fixed-isobar PWA to~200 in the freed-isobar
PWA\@.  All other parameters of the PWA remain as described in
\refCite{Adolph:2015tqa}.  For the \mThreePi~bins below
\SI{0.98}{\GeVcc}, the results from the freed-isobar PWA turn out to
be not well determined by the data.  This is probably related to the
fact that this \mThreePi~region corresponds to the range $\mTwoPi
\lesssim \SI{0.8}{\GeVcc}$ where most isobar resonances, which
otherwise stabilize the fit, are absent.  Therefore, we exclude this
\mThreePi~range from the following analysis.

\begin{table*}[tbp]
  \caption{Borders of the four nonequidistant \tpr~bins, in which the
    freed-isobar PWA is performed.  The intervals are chosen such that
    each bin contains approximately \num{11.5e6} events.}%
  \label{tab:t-bins}
  \renewcommand{\arraystretch}{1.2}
  \newcolumntype{Z}{%
    >{\Makebox[0pt][c]\bgroup}%
    c%
    <{\egroup}%
  }
  \setlength{\tabcolsep}{0pt}  %
  \begin{tabular}{l@{\extracolsep{12pt}}c@{\extracolsep{6pt}}Z*{10}{cZ}c}
    \toprule
    \textbf{Bin} && 1 && 2 && 3 && 4 & \\
    \midrule
    \textbf{\tpr [\si{\GeVcsq}]} &
    \num{0.100} &&
    \num{0.141} &&
    \num{0.194} &&
    \num{0.326} &&
    \num{1.000} \\
    \bottomrule
  \end{tabular}%
\end{table*}

In a freed-isobar PWA mathematical ambiguities, so-called zero modes,
may arise at the level of the decay amplitudes leading to ambiguous
solutions for the transition amplitudes $\{\prodAmp_{a, k}\}$. These
ambiguities can be resolved by imposing conditions on the
\mTwoPi~dependence of the dynamic isobar
amplitudes~\cite{Krinner:2017dba,Krinner:2018bwg}.  We give details on
the zero mode in the \wave{1}{-+}{1}{+}{\pipiPF}{P} wave and its
resolution in \cref{app:zeroMode}.  The zero modes are confined to
sectors with the same \JPCMrefl quantum numbers of the $3\pi$ system.
Therefore, similar ambiguities present in other waves have no
influence on the results extracted for the spin-exotic wave.  In the
following, we will discuss only zero-mode corrected results.

\subsection{Freed-isobar results for the $\JPC = 1^{-+}$ wave}%
\label{sec::fipwa_exotic}

In the following, we will present results for the
\wave{1}{-+}{1}{+}{\pipiPF}{P} wave obtained from the freed-isobar PWA
with 12~freed waves as listed in
\cref{tab:freedIsobarList,tab:wave_sets}.  The corresponding
sets $\cbrk{\prodAmp_{a, k}}$ of transition amplitudes [see
\cref{eq:dyn_amp_freed,eq:prod_amp_freed}] for all \mTwoPi~intervals
and all $(\mThreePi, \tpr)$ cells are provided in computer-readable
format at~\cite{paper4_hepdata}. \Cref{fig:twoDplots} shows the
partial-wave intensities $\abs{\overline{\prodAmp}_{\!\!a, k}}^2$ for
the four~\tpr~bins listed in \cref{tab:t-bins}.  Here,
\begin{equation}
  \label{eq:prod_amp_freed_norm}
  \overline{\prodAmp}_{\!\!a, k}
  = \frac{\prodAmp_{a, k}}{\sqrt{w_k}}
\end{equation}
is the transition amplitude normalized by the width~$w_k$ of the
\mTwoPi interval~$k$.  We observe a clear correlation of the
\mThreePi~distribution of the $3\pi$ system with $\IGJPCMrefl = 1^-\,
1^{-+}1^+$ with the \mTwoPi~distribution of the \twoPi subsystem with
$\IGJPC = 1^-\, 1^{--}$.  The \mTwoPi~spectra are dominated by a peak
in the \Prho region.  The shape of the \mThreePi~spectrum in the \Prho
region depends strongly on~\tpr.  At low~\tpr, it is characterized by
a broad structure peaking at low values of~\mThreePi around
\SI{1.1}{\GeVcc}.  A similar enhancement is observed in the
conventional fixed-isobar PWA [see
\cref{fig:piOne1600_resModelFit_int_t1,sec:comparison}].  The
freed-isobar PWA shows that this enhancement indeed contains mainly
$\Prho \pi$ (see discussion below).  With increasing~\tpr, the
intensity in the low-mass region decreases quickly and in the highest
\tpr~bin, a peak emerges in the $\mThreePi = \SI{1.6}{\GeVcc}$ region
[see \cref{fig:twoDplots_tbin4}].

\begin{wideFigureOrNot}[tbp]
  \centering%
  \subfloat[][]{%
    \includegraphics[width=\twoPlotWidth]{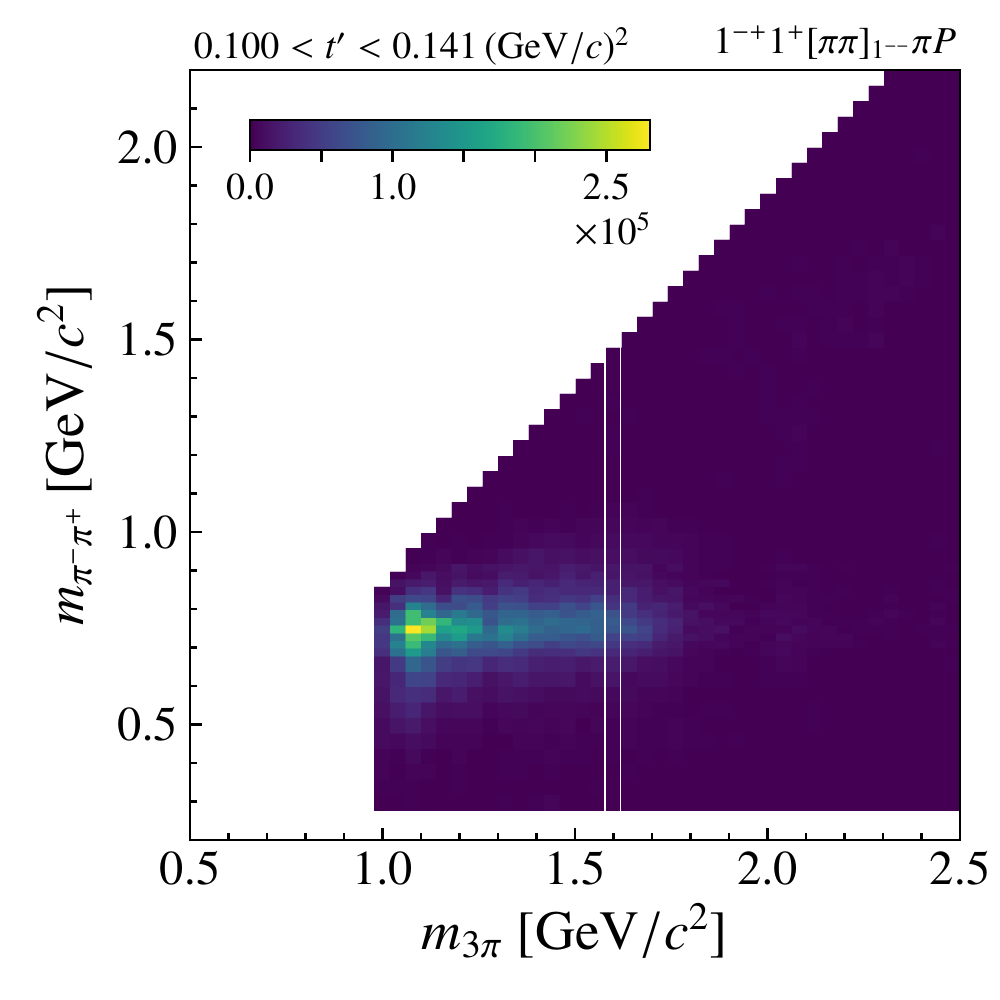}%
  }%
  \hspace*{\twoPlotSpacing}%
  \subfloat[][]{%
    \includegraphics[width=\twoPlotWidth]{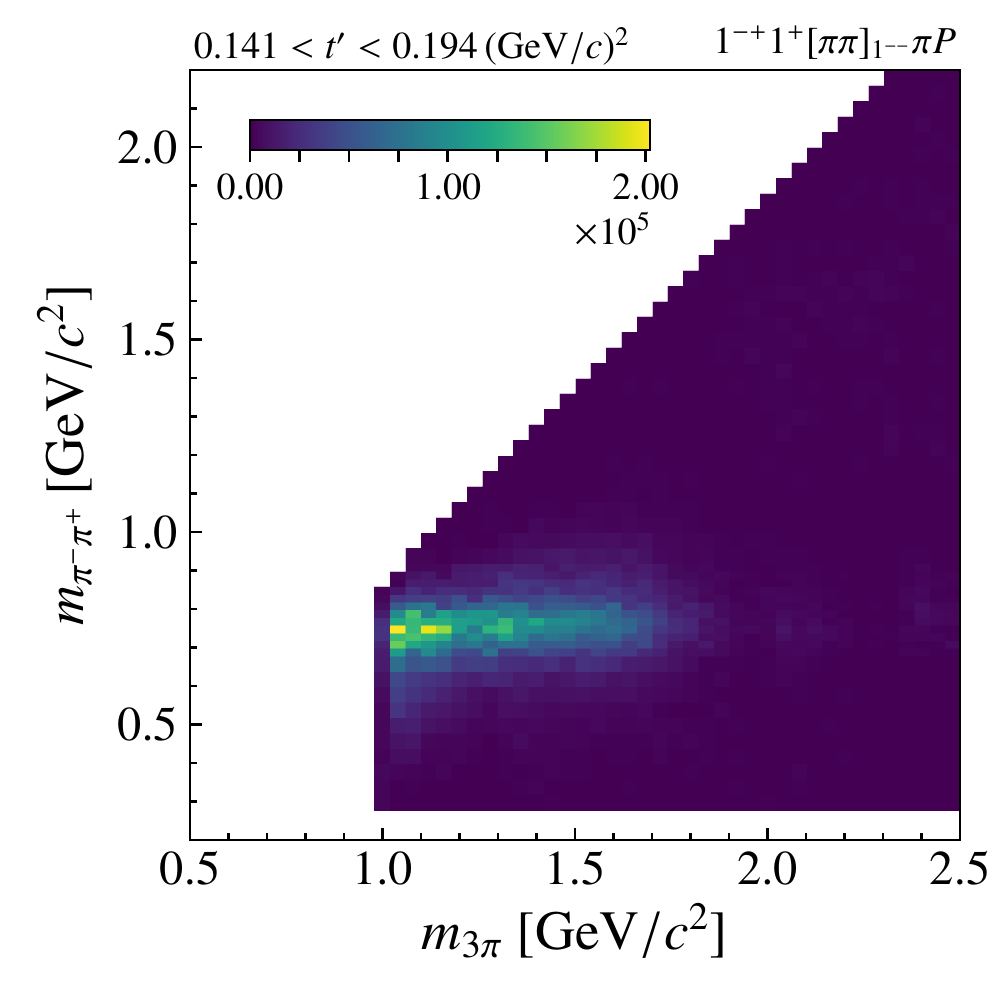}%
  }%
  \\%
  \subfloat[][]{%
    \includegraphics[width=\twoPlotWidth]{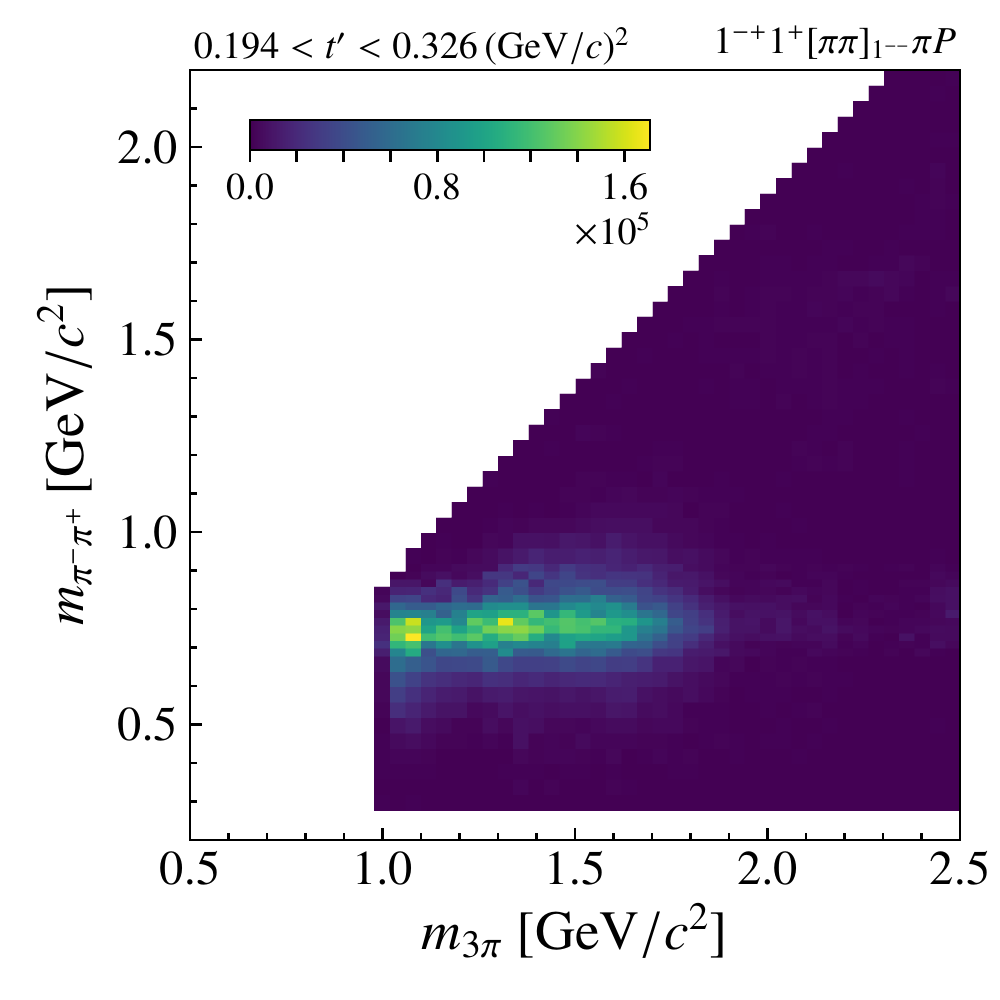}%
  }%
  \hspace*{\twoPlotSpacing}%
  \subfloat[][]{%
    \includegraphics[width=\twoPlotWidth]{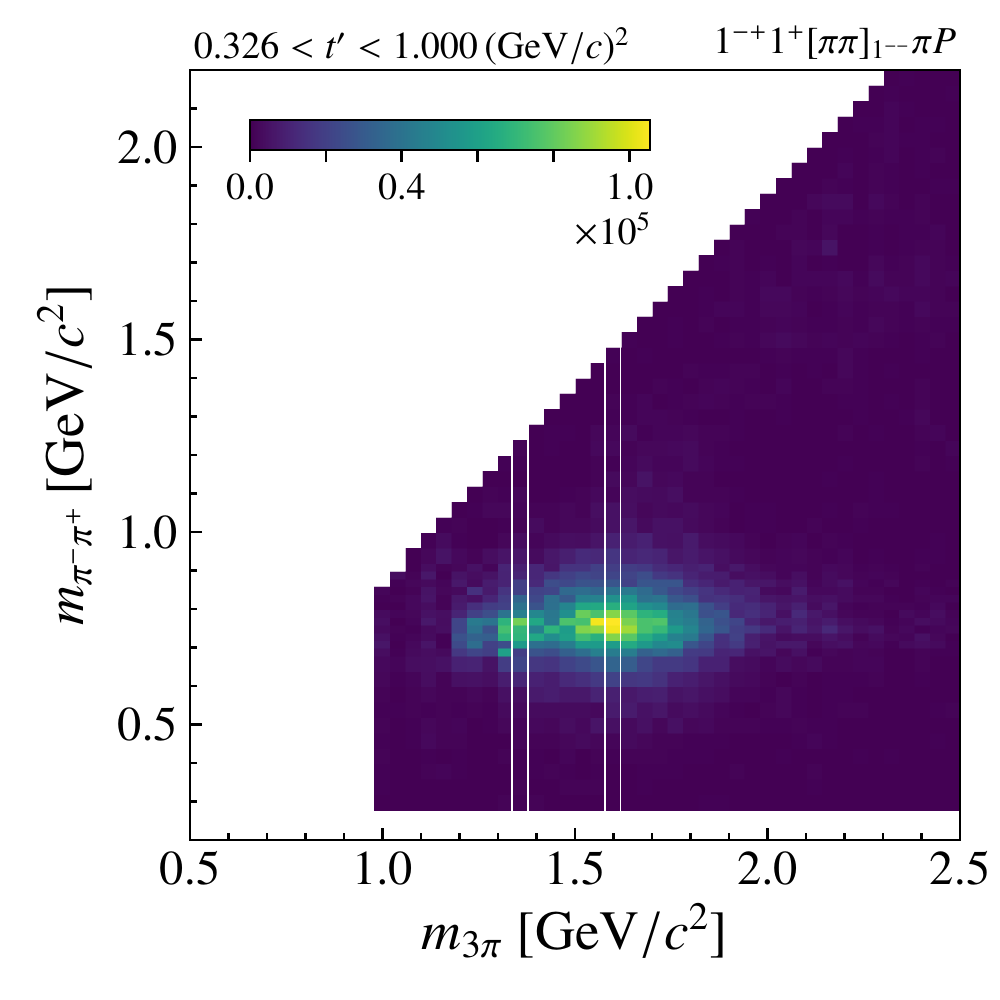}%
    \label{fig:twoDplots_tbin4}%
  }%
  \caption{Two-dimensional intensity distribution of the
    \wave{1}{-+}{1}{+}{\pipiPF}{P} wave obtained in the freed-isobar
    PWA (after correction for the zero mode) as a function
    of~\mThreePi and~\mTwoPi for all four~\tpr~bins.  The color scale
    represents the intensity in units of number of events per
    \SI{40}{\MeVcc} interval in~\mTwoPi and in~\mThreePi.  The white
    vertical lines indicate the \mThreePi~bins shown in
    \cref{fig:slices}.}%
  \label{fig:twoDplots}
\end{wideFigureOrNot}

The left column of \cref{fig:slices} shows the intensity distributions
as a function of \mTwoPi for selected \mThreePi~bins that are
indicated by vertical lines in
\cref{fig:twoDplots}.\footnote{The \mTwoPi intensity
distributions for all $(\mThreePi, \tpr)$ cells are shown in
\ifMultiColumnLayout{Appendix~F of the Supplemental Material of this
paper~\cite{paper4_supplemental_material}}{the Supplemental Material
in \cref{suppl:freed_isobar_dyn_amp}}.}  The mass bin
\SIvalRange{1.34}{\mThreePi}{1.38}{\GeVcc} is dominated by nonresonant
contributions, whereas the bin
\SIvalRange{1.58}{\mThreePi}{1.62}{\GeVcc} lies in the \PpiOne[1600]
resonance region.  For the low-mass region, we show, as an example,
only the data in the highest \tpr~bin, while for the \PpiOne[1600]
resonance region, we present the results for the lowest and the
highest of the four~\tpr~bins.

\begin{wideFigureOrNot}[tbp]
  \centering%
  \subfloat[][]{%
    \includegraphics[width=\ifMultiColumnLayout{\twoPlotWidth}{0.4\textwidth}]{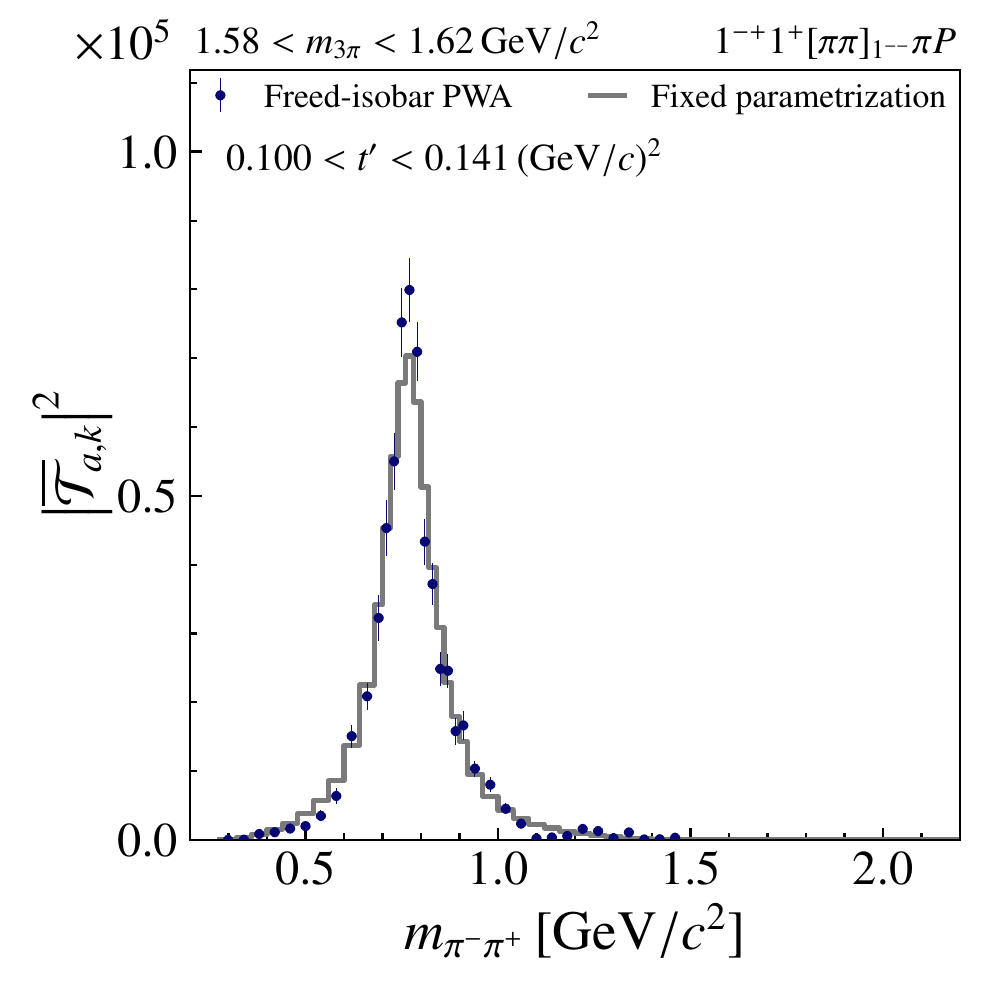}%
  }%
  \hspace*{\twoPlotSpacing}%
  \subfloat[][]{%
    \includegraphics[width=\ifMultiColumnLayout{\twoPlotWidth}{0.4\textwidth}]{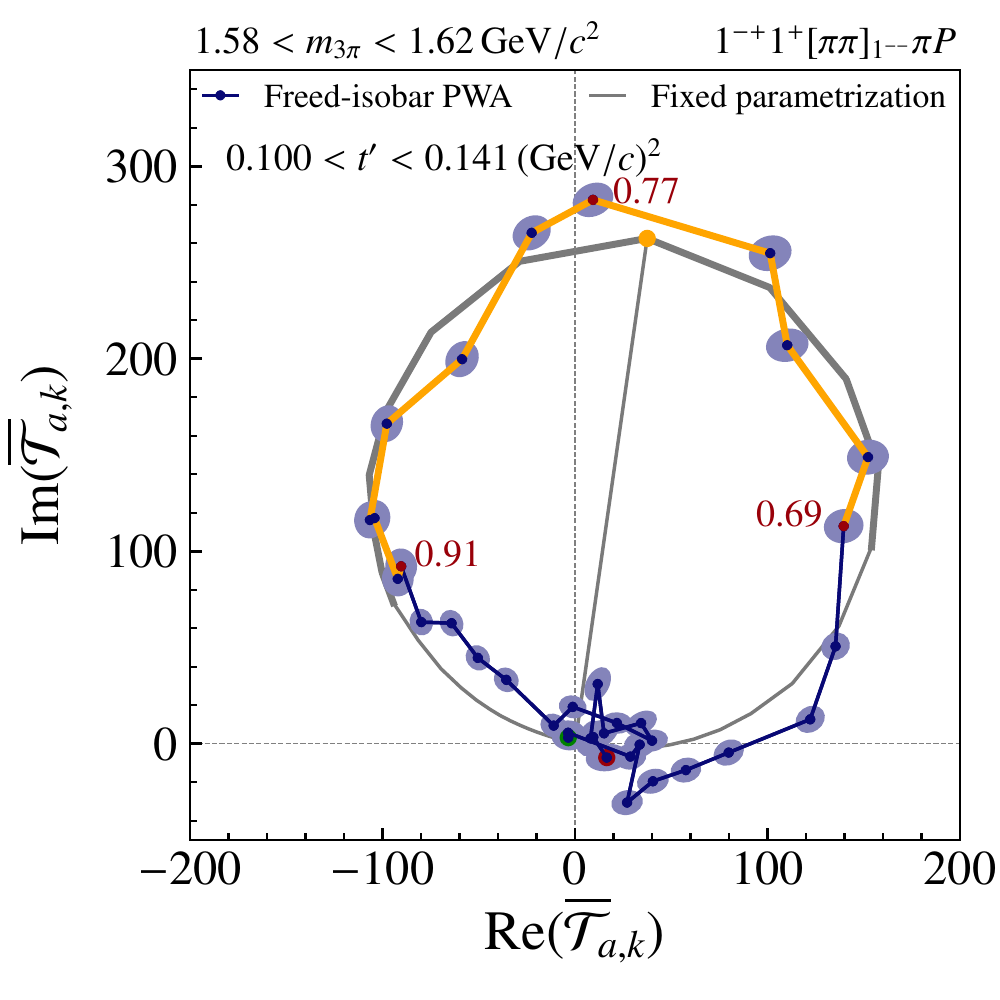}%
  }%
  \\%
  \subfloat[][]{%
    \includegraphics[width=\ifMultiColumnLayout{\twoPlotWidth}{0.4\textwidth}]{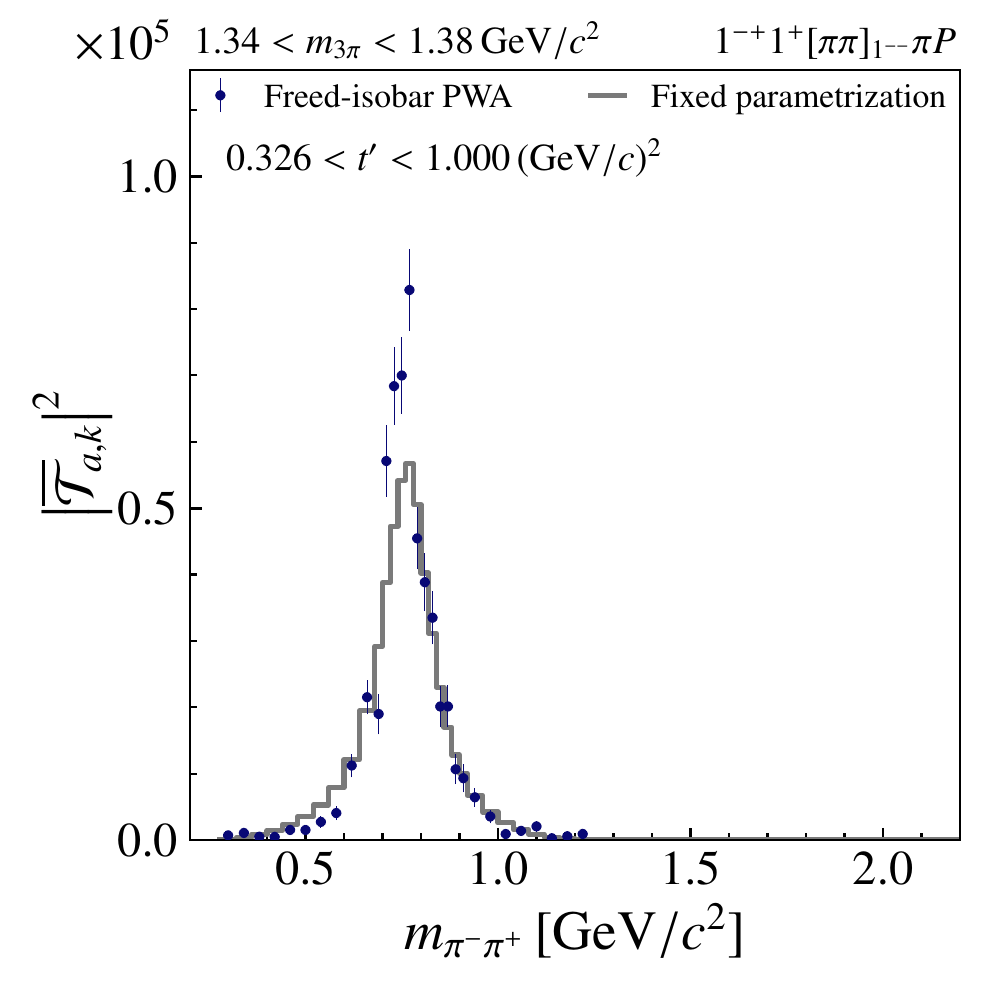}%
  }%
  \hspace*{\twoPlotSpacing}%
  \subfloat[][]{%
    \includegraphics[width=\ifMultiColumnLayout{\twoPlotWidth}{0.4\textwidth}]{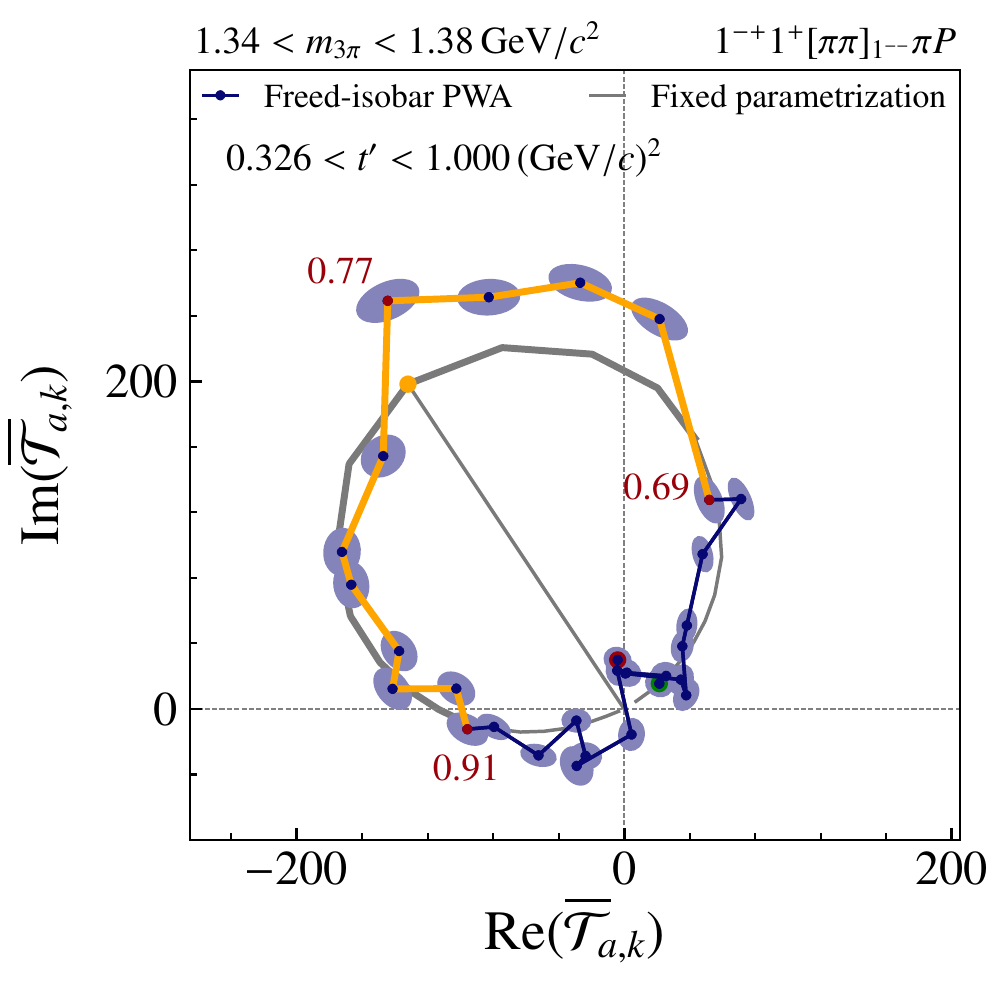}%
  }%
  \\%
  \subfloat[][]{%
    \includegraphics[width=\ifMultiColumnLayout{\twoPlotWidth}{0.4\textwidth}]{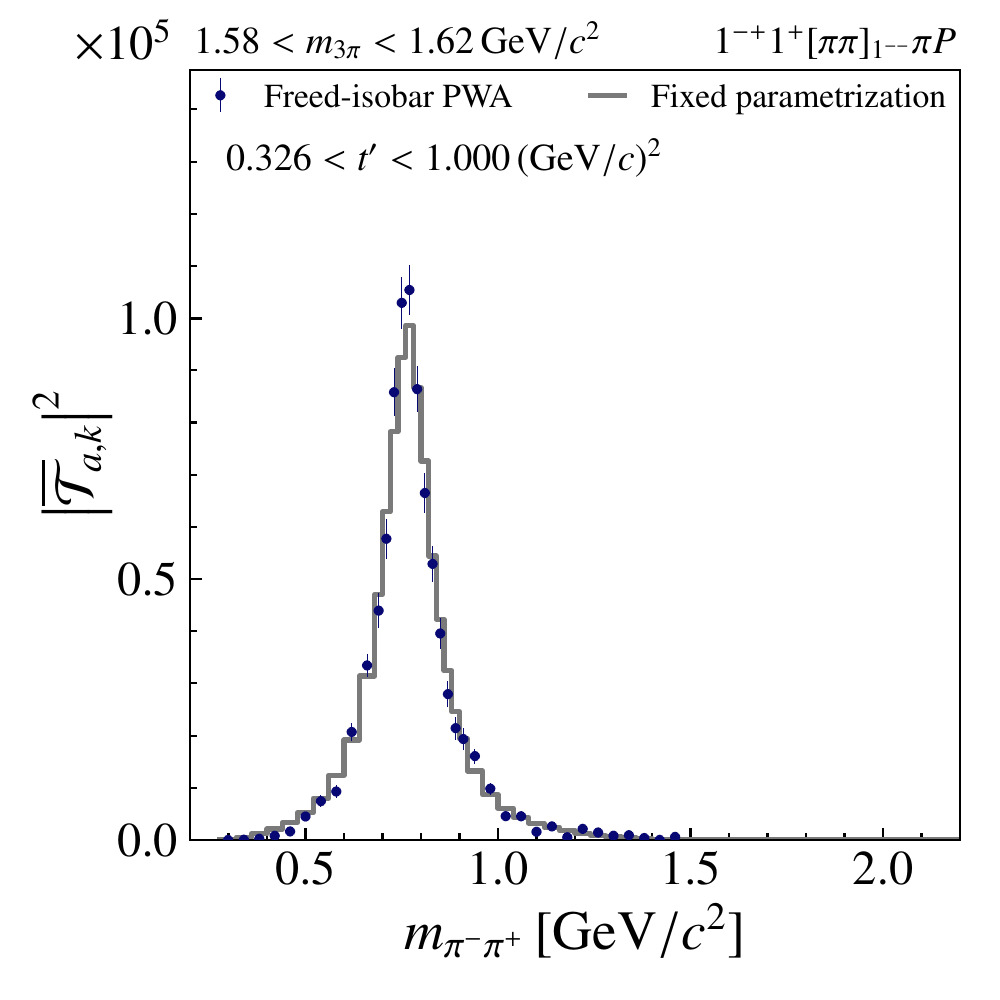}%
  }%
  \hspace*{\twoPlotSpacing}%
  \subfloat[][]{%
    \includegraphics[width=\ifMultiColumnLayout{\twoPlotWidth}{0.4\textwidth}]{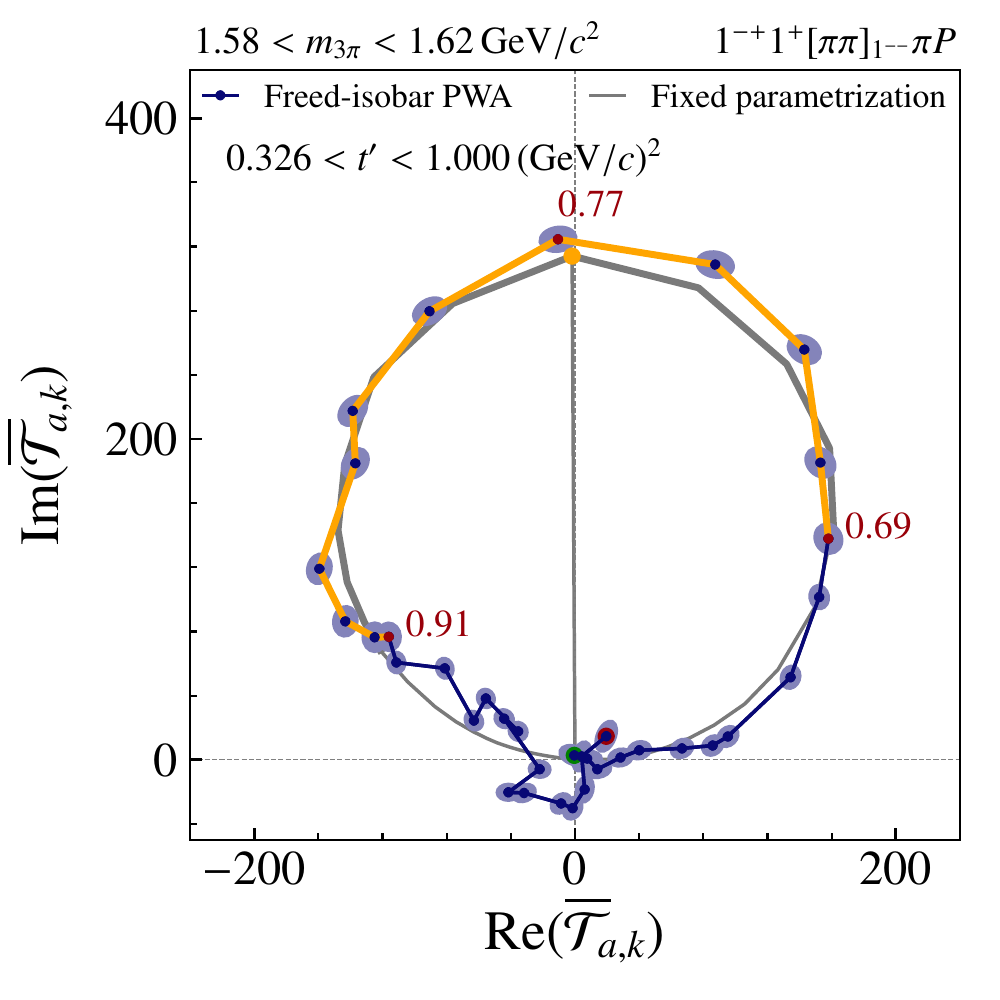}%
  }%
  \caption{The \pipiPF dynamic isobar amplitude in the
    \wave{1}{-+}{1}{+}{\pipiPF}{P} wave as a function of~\mTwoPi for
    selected \mThreePi~and \tpr~bins.  Left column: intensities; right
    column: \Argands.  The blue data points with error bars or error
    ellipses, respectively, are the result of the freed-isobar PWA
    corrected for the zero mode.  In the \Argands, the data points are
    connected by lines to indicate the order and the red numbers
    correspond to \mTwoPi~values in \si{\GeVcc}.  The line segments
    highlighted in orange correspond to the \mTwoPi~range from
    \SIrange{0.64}{0.92}{\GeVcc} around the \Prho.  The overall phase
    of the \Argands is fixed by the \wave{4}{++}{1}{+}{\Prho}{G} wave.
    For comparison, the fixed parametrization of the dynamic isobar
    amplitude for the \Prho as used in the conventional PWA is shown
    by the gray lines with the \Prho region indicated by thicker
    lines.  In the \Argands, the orange point indicates the nominal
    \Prho mass and the green- and red-circled points indicate the
    lowest and the highest \mTwoPi~interval, respectively.}%
  \label{fig:slices}
\end{wideFigureOrNot}

Since the freed-isobar PWA extracts the amplitude as a function
of~\mThreePi and~\mTwoPi, we have also information about the phase as
a function of~\mTwoPi.  This is shown in the right column of
\cref{fig:slices} in the form of \Argands for the selected
\mThreePi~bins.\footnote{The \Argands for all $(\mThreePi,
\tpr)$ cells are shown in \ifMultiColumnLayout{Appendix~F of the
Supplemental Material of this
paper~\cite{paper4_supplemental_material}}{the Supplemental Material
in \cref{suppl:freed_isobar_dyn_amp}}.}  The dominant \Prho peak in
the intensity spectra corresponds to a clear circular structure in the
\Argands\ with a phase motion by about \ang{180}.  This confirms, that
the presence of the \Prho has not been artificially enforced by the
fixed parametrizations of the dynamic isobar amplitudes as used in
previous analyses.  Just as the \mTwoPi~spectra, also the \Argands
exhibit no strong dependence on~\mThreePi or~\tpr.  The spin-exotic
wave is clearly dominated by the \Prho over the full \mThreePi~region
and in all four \tpr~bins.

We study the freed-isobar transition amplitudes that we extracted from
the data in terms of isobar resonances and possible distortions.  In a
first study, we investigate the \Prho resonance in the presence of
another pion, which together form a $3\pi$ system with $\JPC =
1^{-+}$.  Lacking an elaborate model, we perform this study by fitting
the $\JPC = 1^{--}$ dynamic isobar amplitudes with a \Prho
Breit-Wigner model of the form
\begin{wideTextOrNot}%
  \begin{equation}
    \label{eq:bw_model_isobar}
    \hat{\prodAmp}_a(\mTwoPi; \mThreePi, \tpr)
    = \mathcal{C}_a(\mThreePi, \tpr)\, \frac{\mathcal{N}_a(\mThreePi, \mTwoPi)}{m_{\Prho}^2 - \mTwoPi^2 - i\, m_{\Prho} \Gamma(\mTwoPi)}
  \end{equation}
\end{wideTextOrNot}%
in every $(\mThreePi, \tpr)$ cell independently.  Here, $a =
\wave{1}{-+}{1}{+}{\pipiPF}{P}$, $\mathcal{N}_a(\mThreePi, \mTwoPi)$
is a normalization factor, which takes into account the variation of
the \mTwoPi bin width, the self-interference of the Breit-Wigner
amplitude due to Bose symmetrization, and the angular-momentum barrier
factors $F_L(m_{3\pi}; \mTwoPi, m_\pi)$ and $F_{J_\xi}(\mTwoPi; m_\pi,
m_\pi)$ from Eqs.~(10) and~(8) of \refCite{Adolph:2015tqa}, and
$\Gamma(\mTwoPi)$ is the mass-dependent total width of the \Prho as
given by Eq.~(40) in \refCite{Adolph:2015tqa}.  In the fits, the
resonance parameters $m_{\Prho}$~and~$\Gamma_{\Prho}$ are fixed to the
values used in the conventional fixed-isobar PWA (see Table~III in
\refCite{Adolph:2015tqa}).  The only free fit parameter is the
complex-valued coupling~$\mathcal{C}_a(\mThreePi, \tpr)$, which
determines strength and phase of the \Prho signal in the given
$(\mThreePi, \tpr)$ cell, \ie radius and rotation of the resonance
circle about the origin in the \Argand.\footnote{These fits also
resolve the mathematical ambiguity discussed in
\cref{sec:freed_isobar_pwa_model}.  This is explained in detail in
\cref{app:zeroMode}.}  The model is evaluated at those \mTwoPi~values
that correspond to the centers of the \mTwoPi~intervals defined in
\cref{eq:dyn_amp_freed}.

The fits are limited to the region $\mTwoPi < \SI{1.12}{\GeVcc}$ to
avoid bias from excited \Prho* resonances at higher masses.  The
results of these fits are shown as gray curves in \cref{fig:slices}.
The resulting curves are in good agreement with the extracted dynamic
isobar amplitudes in the \mThreePi~region of the \PpiOne[1600], which
confirms the validity of the isobar model.  For the lower
\mThreePi~bin shown, the agreement is slightly worse, which could hint
at a stronger influence of nonresonant contributions in this mass
region.

In a second study, we let the \Prho resonance parameters float in the
fit and determine them independently for every $(\mThreePi, \tpr)$
cell.  We hence do not assume anymore that we can factorize the
\mTwoPi~dependence of the transition amplitudes from their~\mThreePi
and \tpr~dependence.
The weighted average of the obtained \Prho mass values is about
\SI{760}{\MeVcc}, only slightly below the PDG averages.  The weighted
average of the obtained \Prho width values is approximately
\SI{130}{\MeVcc}, which lies \SIrange{15}{20}{\MeVcc} below the PDG
averages.  We have currently no explanation why the \Prho in the
$1^{-+}$ wave appears so much narrower.\footnote{Allowing the range
parameter~$q_R$ [see below Eq.~(39) in \refCite{Adolph:2015tqa}] of
the angular-momentum barrier factor $F_{J_\xi}$ of the \Prho decay as
an additional free fit parameter, yields slightly larger width values
in some \mThreePi regions.  However, qualitatively the picture remains
unchanged.}  The \Prho parameters exhibit variations of about $\pm
\SI{10}{\percent}$ with~\mThreePi (see \cref{fig:rhoParameterFits}),
while they show little variation with~\tpr~\cite{Krinner:2018bwg}.
More advanced models are needed to study the distortion of the \Prho
line shape due to effects of final-state interaction and interfering
contributions from nonresonant processes in the \twoPi system.

\begin{figure}[tbp]
  \centering%
  \subfloat[][]{%
    \includegraphics[width=\twoPlotWidth]{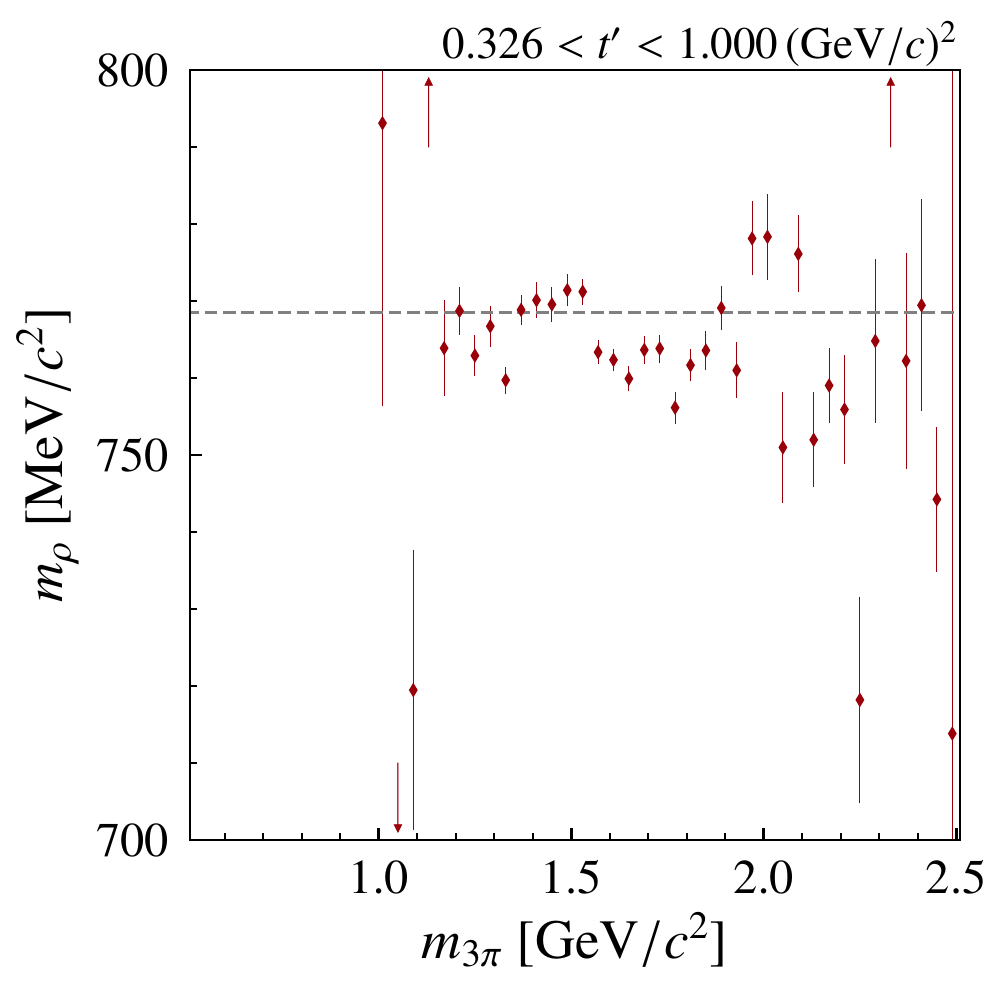}%
    \label{fig:rhoMassFit}%
  }%
  \newLineOrHspace{\twoPlotSpacing}%
  \subfloat[][]{%
    \includegraphics[width=\twoPlotWidth]{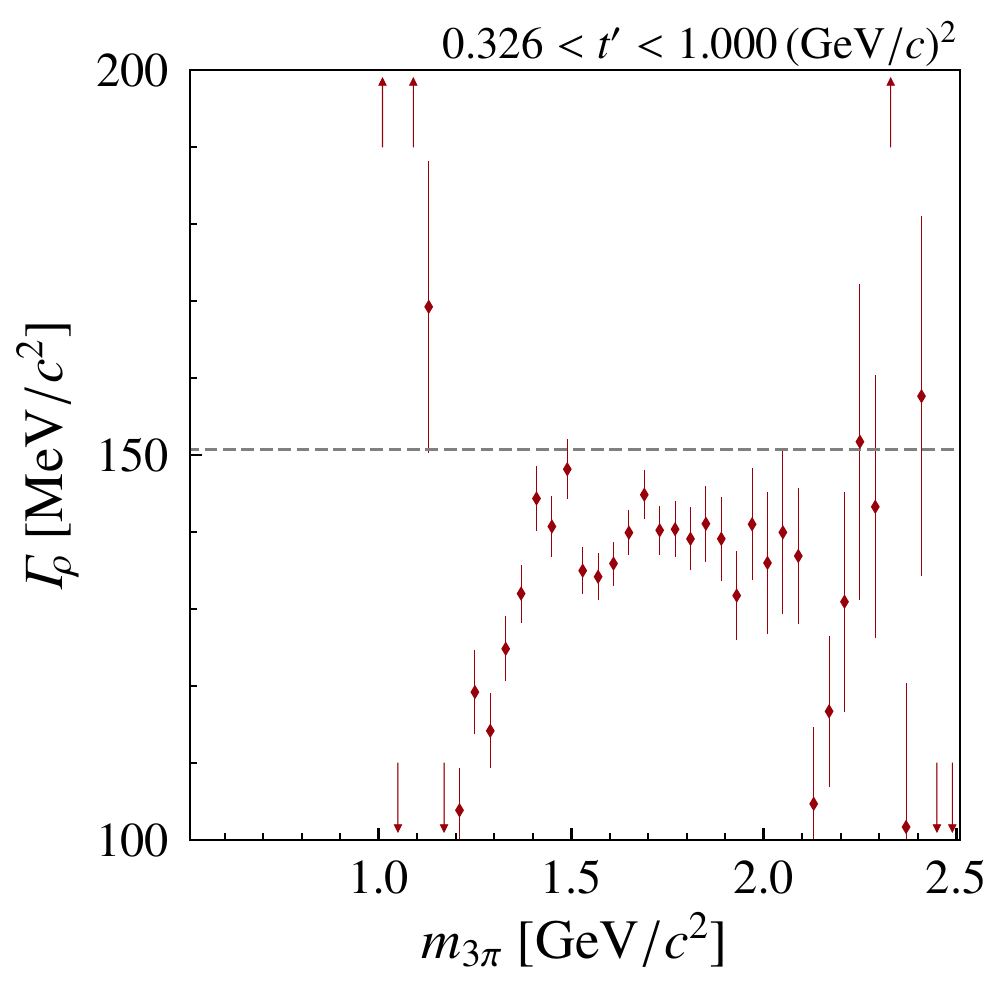}%
    \label{fig:rhoWidthFit}%
  }%
  \caption{Parameters of the \Prho resonance obtained by fitting the
    $\JPC = 1^{--}$ dynamic isobar amplitudes of the spin-exotic wave
    from the freed-isobar PWA.  The fit is performed independently in
    every $(\mThreePi, \tpr)$ cell; the results shown are for the
    highest \tpr~bin.  \subfloatLabel{fig:rhoMassFit}~shows the \Prho
    mass and \subfloatLabel{fig:rhoWidthFit}~the \Prho width. The gray
    lines indicate the corresponding parameter values used in the
    conventional PWA.}%
  \label{fig:rhoParameterFits}
\end{figure}

\subsection{Comparison with the conventional partial-wave analysis}

In order to directly compare the results from the freed-isobar PWA
with the conventional PWA with fixed parametrizations for the dynamic
isobar amplitudes, we repeated the latter with the same 88-wave PWA
model as in \refCite{Adolph:2015tqa} but applying the coarser binning
in~\mThreePi and~\tpr from the freed-isobar PWA (see
\cref{sec:freed_isobar_pwa_model,tab:t-bins}).  From the result of the
freed-isobar PWA, we obtain intensity distributions as a function
of~\mThreePi and~\tpr alone by summing the contributions of the
freed-isobar transition amplitudes $\{\prodAmp_{a, k}\}$ [see
\cref{eq:dyn_amp_freed,eq:prod_amp_freed}] from all \mTwoPi~intervals
coherently.  Doing so, we take into account the interference of
amplitudes in different \mTwoPi~intervals, \ie the so-called overlaps,
that arise due to Bose symmetrization of the final-state particles.
The intensity of these coherent sums is by definition not affected by
the zero-mode ambiguity mentioned in \cref{sec:freed_isobar_pwa_model}
(see also \cref{app:zeroMode}).  The intensity distributions of
the coherent sums are provided in computer-readable format
at~\cite{paper4_hepdata}.

In \cref{fig:coherent}, we overlay the intensity distributions from
the freed-isobar PWA obtained as described above (orange data points)
with the corresponding distributions from the conventional
fixed-isobar PWA (blue data points).\footnote{The intensity
distributions for the two intermediate \tpr~bins are shown in
\ifMultiColumnLayout{Appendix~E of the Supplemental Material of this
paper~\cite{paper4_supplemental_material}}{the Supplemental Material
in \cref{suppl:freed_isobar_comp}}.}  Although the $1^{-+}$ wave
contributes only about \SI{1}{\percent} to the total intensity, the
distributions are surprisingly similar.  The shapes of the intensity
distributions are consistent in both approaches, regardless of the
\tpr~bin.  However, the intensity of the $1^{-+}$ wave is higher for
the freed-isobar PWA\@.  This intensity increase is not caused by
freeing the dynamic isobar amplitude of the $1^{-+}$ wave itself, but
rather by the other 11~freed-isobar waves (see
\cref{tab:freedIsobarList}).  Keeping these 11~freed waves but fixing
the dynamic isobar amplitude in the $1^{-+}$ wave to the \Prho, like
in the conventional PWA, yields basically the same $1^{-+}$ intensity
distribution as in the PWA with 12~freed waves.  Further systematic
studies show that no single freed wave causes the increase of the
$1^{-+}$ intensity, but that this is the result of the interplay of
all 11~freed waves.  This suggests that---unlike for the $1^{-+}$
wave---the fixed-isobar amplitudes, which in the conventional 88-wave
PWA correspond to the these 11~freed-isobar waves, do not match the
data completely.  Deviations could be caused, for example, by
unsuitable parametrizations and/or parameters used for the dynamic
isobar amplitudes or by neglecting higher excited isobar resonances
that may become relevant at higher values of~\mThreePi.

\begin{figure}[tbp]
  \centering%
  \subfloat[][]{%
    \includegraphics[width=\twoPlotWidth]{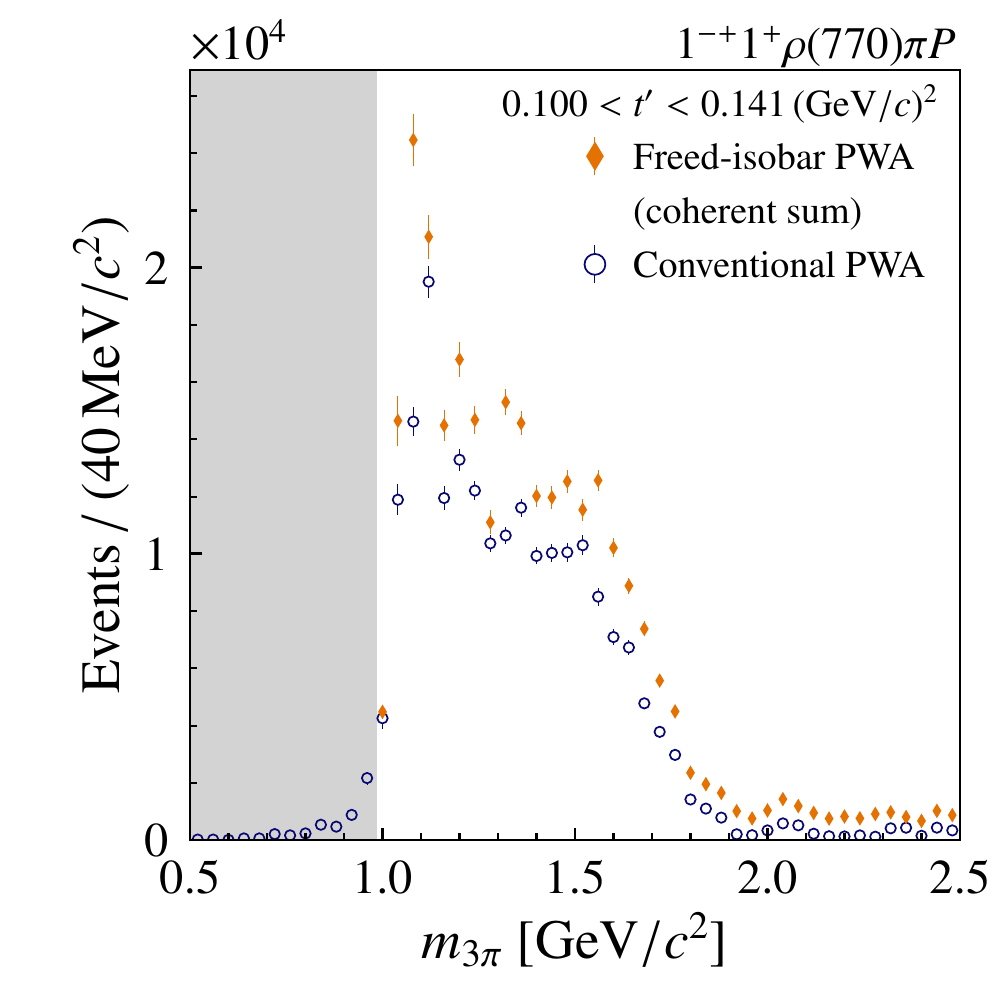}%
    \label{fig:coherent_lowt}%
  }%
  \newLineOrHspace{\twoPlotSpacing}%
  \subfloat[][]{%
    \includegraphics[width=\twoPlotWidth]{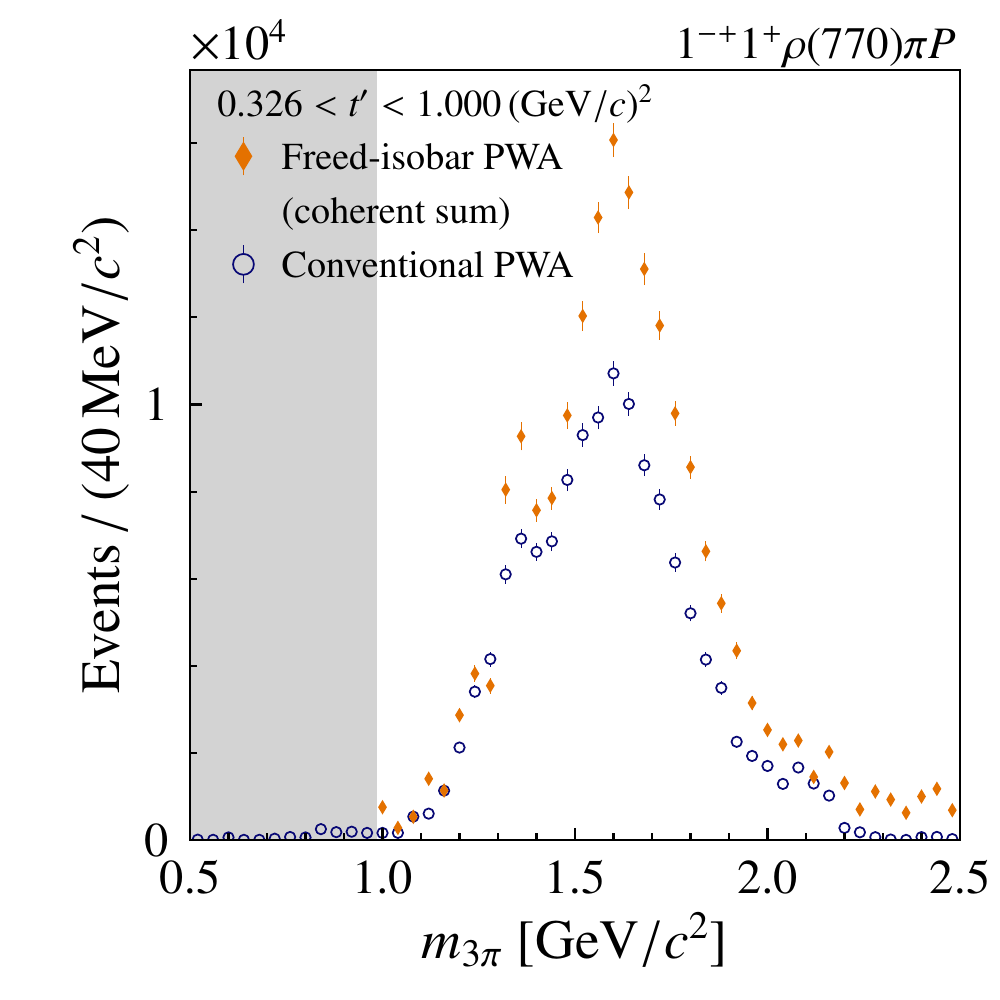}%
    \label{fig:coherent_hight}%
  }%
  \caption{Comparison of the \mThreePi~intensity distributions of the
    \wave{1}{-+}{1}{+}{\pipiPF}{P} wave from the freed-isobar PWA
    (orange data points) and of the \wave{1}{-+}{1}{+}{\Prho}{P} wave
    from the conventional PWA (blue data points).  For the former, the
    intensities are calculated by coherently summing the contributions
    from all \mTwoPi~intervals.
    \subfloatLabel{fig:coherent_lowt}~shows the lowest and
    \subfloatLabel{fig:coherent_hight}~the highest \tpr~bin.  The
    shaded \mThreePi~range is excluded from the freed-isobar PWA\@.}%
  \label{fig:coherent}
\end{figure}

Another way to compare the two PWA methods is to use the information
that we obtain by fitting the \mTwoPi~dependence of the amplitudes
extracted by the freed-isobar PWA with \cref{eq:bw_model_isobar} using
the same fixed \Prho parameters as in \refCite{Adolph:2015tqa} (gray
curves in \cref{fig:slices}).  The interesting information is
contained in the complex-valued quantity
\begin{equation}
  \label{eq:bw_coupling}
  \mathfrak{T}_a(\mThreePi, \tpr)
  \equiv \mathcal{C}_a(\mThreePi, \tpr)\, \mathcal{N}_a(\mThreePi)
\end{equation}
that we determine for every $(\mThreePi, \tpr)$ cell and that is
directly comparable to the transition amplitude $\prodAmp_a(\mThreePi,
\tpr)$ obtained in the fixed-isobar PWA.\footnote{Note that
$\mathcal{N}_a$ contains a normalization factor we choose such that
$\abs{\mathfrak{T}_a}^2$ gives the number of events per interval
in~\mThreePi and~\tpr.}  In \cref{fig:modeledComparison_int}, we
compare the intensity distribution $\abs{\mathfrak{T}_a(\mThreePi)}^2$
from the freed-isobar PWA (red data points) to the intensity
distribution from the fixed-isobar PWA (blue data points) for the
$1^{-+}$ wave in the highest \tpr~bin.\footnote{The intensity
distributions for the other \tpr~bins are shown in
\ifMultiColumnLayout{Appendix~E of the Supplemental Material of this
paper~\cite{paper4_supplemental_material}}{the Supplemental Material
in \cref{suppl:freed_isobar_comp}}.}  The red data points in
\cref{fig:modeledComparison_int} are very similar to the orange ones
in \cref{fig:coherent_hight}.  This confirms that the $1^{-+}$ wave is
well described a dynamic isobar amplitude containing only the \Prho.

\begin{figure}[tbp]
  \centering%
  \subfloat[][]{%
    \includegraphics[width=\twoPlotWidth]{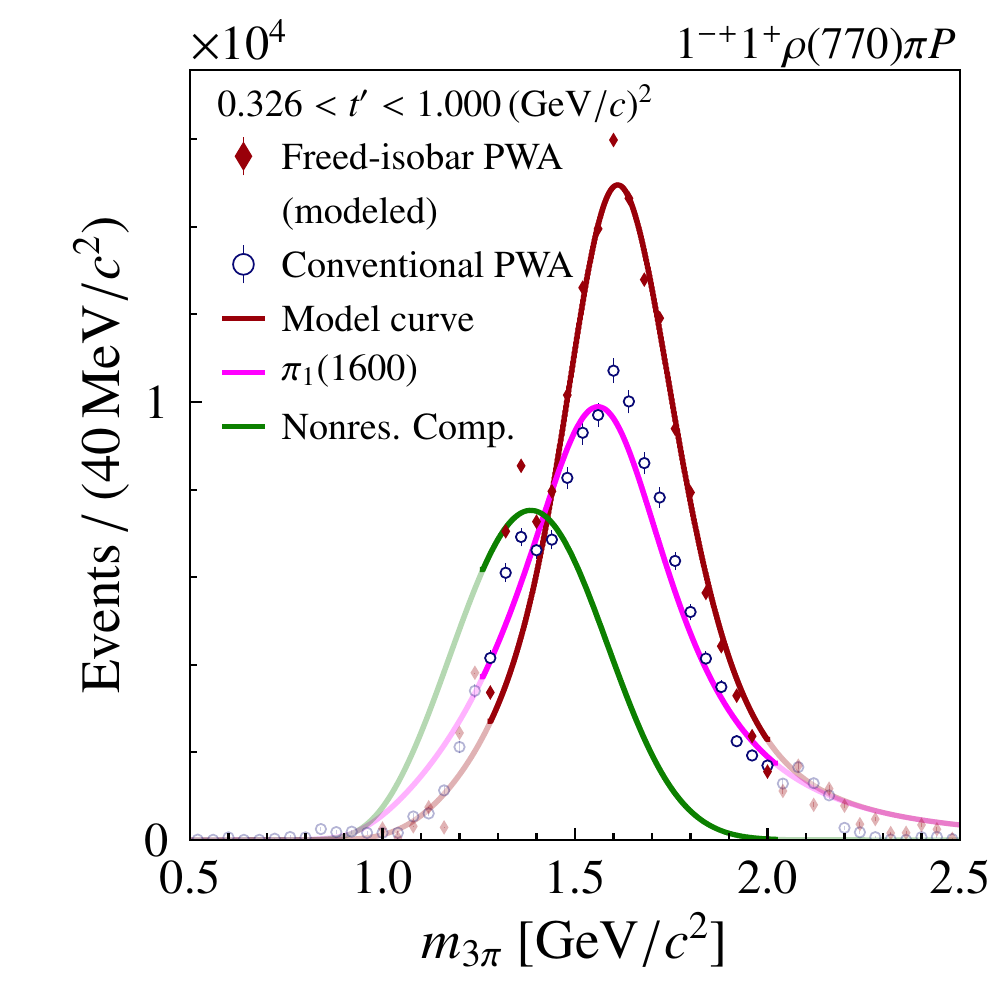}%
    \label{fig:modeledComparison_int}%
  }%
  \newLineOrHspace{\twoPlotSpacing}%
  \subfloat[][]{%
    \includegraphics[width=\twoPlotWidth]{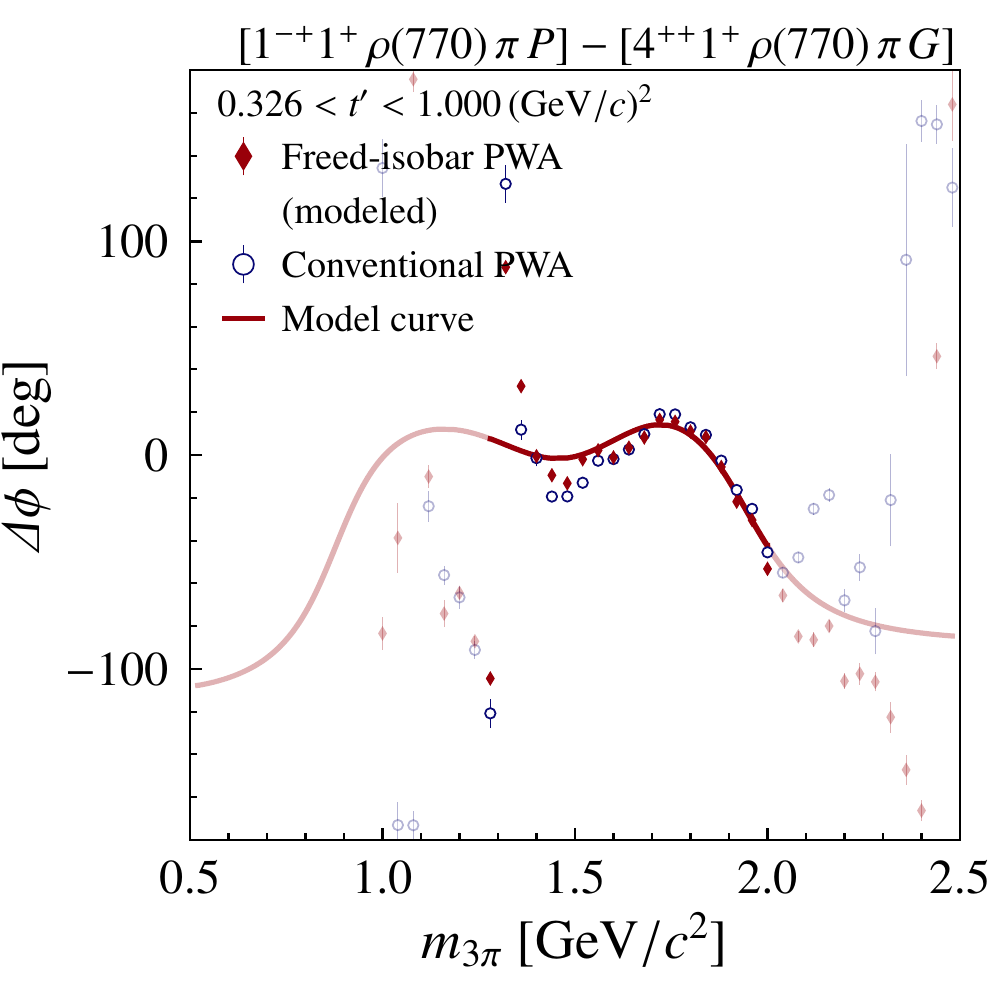}%
    \label{fig:modeledComparison_phase}%
  }%
  \caption{\subfloatLabel{fig:modeledComparison_int}~intensity
    distribution and \subfloatLabel{fig:modeledComparison_phase}~phase
    of the spin-exotic wave \wrt the \wave{4}{++}{1}{+}{\Prho}{G} wave
    in the highest \tpr~bin.  The red data points represent the
    \wave{1}{-+}{1}{+}{\pipiPF}{P} amplitude from the freed-isobar PWA
    after modeling the \mTwoPi~dependence using the \Prho Breit-Wigner
    amplitude in \cref{eq:bw_model_isobar}; the blue data points
    represent the \wave{1}{-+}{1}{+}{\Prho}{P} amplitude from the
    conventional fixed-isobar PWA.  The red curve represents the
    result of a fit of a resonance model to the red data points, which
    is the coherent sum of a resonant amplitude for the \PpiOne[1600]
    (magenta curve) and a nonresonant term (green curve).}%
  \label{fig:modeledComparison}
\end{figure}

As explained in \cref{sec:freed_isobar}, the amplitudes
$\{\prodAmp_{a, k}\}$ determined by the freed-isobar PWA contain
information on both the $3\pi$ system and the \twoPi subsystem [see
\cref{eq:dyn_amp_freed,eq:prod_amp_freed}].  In order to consistently
extract the \mThreePi~dependence of the phase of a freed wave, we thus
need to model the \mTwoPi~dependence of the freed-isobar transition
amplitude.  This is accomplished by the introduction of the amplitude
$\mathfrak{T}_a(\mThreePi, \tpr)$ via
\cref{eq:bw_model_isobar,eq:bw_coupling}, which defines the phase of
the $1^{-+}$ wave that can be compared to the phase obtained in the
fixed-isobar PWA\@.  We use the fixed-isobar
\wave{4}{++}{1}{+}{\Prho}{G} wave as a reference wave since it
exhibits a nonzero intensity distribution over a broad \mThreePi~range
and a clear signal of the \PaFour.\footnote{Usually, the largest waves
in the PWA model are used as reference waves.  However, in the
freed-isobar PWA model these waves use freed dynamic isobar amplitudes
(see \cref{tab:freedIsobarList,tab:wave_sets}) and therefore do not
offer a consistent reference phase.}  From the fits of
\cref{eq:bw_model_isobar} with fixed \Prho parameters we obtain
\cref{fig:modeledComparison_phase}.  In this figure, we compare the
phase of the $1^{-+}$ wave \wrt the \wave{4}{++}{1}{+}{\Prho}{G} wave
obtained from the freed-isobar PWA in the way described above (red
data points) with the corresponding phase from the fixed-isobar PWA
(blue data points).\footnote{The phases for the other \tpr~bins
are shown in \ifMultiColumnLayout{Appendix~E of the Supplemental
Material of this paper~\cite{paper4_supplemental_material}}{the
Supplemental Material in \cref{suppl:freed_isobar_comp}}.}  The two
phase motions are in qualitative agreement in the \mThreePi~range from
about \SIrange{1.4}{2.0}{\GeVcc}.  For $\mThreePi \lesssim
\SI{1.2}{\GeVcc}$, the phase is not well determined because the
intensities of the two waves are small.  The rapid phase motion at
\SI{1.2}{\GeVcc} is caused by the nearly vanishing intensity of the
$4^{++}$ wave.  The rising phase motion in the range from
\SIrange{1.4}{1.7}{\GeVcc} indicates the presence of the \PpiOne[1600]
in the $1^{-+}$ wave, whereas the falling phase motion from
\SIrange{1.7}{2.0}{\GeVcc} is caused by the \PaFour in the $4^{++}$
wave.  Similar rising phase motions are also observed \wrt other
waves, \eg\ \wrt the \wave{4}{-+}{0}{+}{\Prho}{F} and
\wave{6}{-+}{0}{+}{\Prho}{H} waves discussed in \cref{sec:deck_model}.

To further check the consistency between the conventional and the
freed-isobar PWA, we fit the \mThreePi~dependence of the
amplitude~$\mathfrak{T}_a$ defined in
\cref{eq:bw_model_isobar,eq:bw_coupling} simultaneously for all
four~\tpr~bins using the same Breit-Wigner model as in
\refCite{Akhunzyanov:2018lqa}.  However, we cannot perform the same
14-wave fit as given in Table~II of \refCite{Akhunzyanov:2018lqa}
because most of the selected 14~waves are in the set of 12~freed waves
(see \cref{tab:freedIsobarList,tab:wave_sets}) and hence do not
provide well-defined phases.  Here, we perform a much simpler fit that
only includes the $1^{-+}$ intensity distribution, \ie
$\abs{\mathfrak{T}_a(\mThreePi)}^2$, and the phase
of~$\mathfrak{T}_a(\mThreePi)$ \wrt the amplitude of the fixed-isobar
\wave{4}{++}{1}{+}{\Prho}{G} wave.  The \wave{4}{++}{1}{+}{\Prho}{G}
wave was included in the 14-wave resonance-model fit in
\refCite{Akhunzyanov:2018lqa}, which was another reason to choose it
as the reference wave.  The \mThreePi~fit range is restricted to the
overlap region from \SIrange{1.26}{2.02}{\GeVcc} of the fit ranges of
the $1^{-+}$ and the $4^{++}$ waves in the resonance-model fit in
\refCite{Akhunzyanov:2018lqa}.  We take the parametrizations for the
$1^{-+}$ and the $4^{++}$ partial-wave amplitudes from
\refCite{Akhunzyanov:2018lqa} and use them to model the real and
imaginary part of $\mathfrak{T}_a(\mThreePi)$.  In the fit, we let the
parameters of the \PpiOne[1600] and the nonresonant component in the
$1^{-+}$ wave float.  We describe the phase of the $4^{++}$ wave
without any free parameters by using the fit result from
\refCite{Akhunzyanov:2018lqa}.  Since the phase of the $4^{++}$ wave
depends on the \tpr~bin, we have to translate the 11~\tpr~bins used in
\refCite{Akhunzyanov:2018lqa} to the four bins used here.  Thus, for
our four \tpr~bins we construct a linear combination of the phases in
the 11~\tpr~bins, using the overlap of the corresponding \tpr~bins
weighted with the \tpr~distribution of all events as coefficients.
This is possible, since the phase of the $4^{++}$ wave changes
smoothly with~\tpr.  In an alternate approach, we replace these linear
combinations of the phases in \tpr~bins by a single phase taken from
the closest of the 11~\tpr~bins yielding a very similar result.  The
found \PpiOne[1600] resonance parameters are
\begin{alignOrEq}
  \label{eq:resonanceFitResult}
  m_{\PpiOne[1600]}
  \alignOrNot= \SI{1550}{\MeVcc} \newLineOrNot
  \ifMultiColumnLayout{\intertext{and}}{\quad\text{and}\quad}
  \Gamma_{\PpiOne[1600]}
  \alignOrNot= \SI{500}{\MeVcc}.
\end{alignOrEq}
These values are compatible with those found in
\refCite{Akhunzyanov:2018lqa} and are based on the same data.  We do
not give any uncertainties in \cref{eq:resonanceFitResult}, since we
did not perform systematic studies and the statistical uncertainties
are negligible compared to the systematic ones.  The latter are
expected to be in the same order of magnitude as those quoted in
\refCite{Akhunzyanov:2018lqa}.  Increasing the lower \mThreePi~limit
of the fit range to \SI{1.34}{\GeVcc}, for example, yields a
\PpiOne[1600] that is \SI{30}{\MeVcc} heavier and \SI{180}{\MeVcc}
narrower.

\subsection{Summary: Dynamic isobar amplitude in the $\JPC = 1^{-+}$ wave}%
\label{sec:freed_isobar_summary}

In conclusion, the results for the spin-exotic $\JPC = 1^{-+}$ wave
from the freed-isobar PWA confirm the findings from the conventional
PWA with fixed parametrizations of the dynamic isobar amplitudes
presented in \refsCite{Adolph:2015tqa,Akhunzyanov:2018lqa} in several
important aspects:  \one~the emergence of the \Prho resonance in the
\twoPi subsystem of the $1^{-+}$ wave, as shown in
\cref{fig:twoDplots,fig:slices}, confirms that the assumption of the
$1^{-+}$ wave decaying via a \Prho isobar is indeed valid.  \two~The
observed agreement of the extracted dynamic isobar amplitude of the
$\JPC = 1^{--}$ \twoPi subsystem with the \Prho amplitude used in the
conventional PWA validates the chosen \Prho parametrization and the
parameter values within about \SI{10}{\percent}.  \three~We observe
good agreement between the results from the freed-isobar PWA with
those from the fixed-isobar PWA in terms of the phase motions and the
shape of the intensity distributions as function of~\mThreePi, as
shown in \cref{fig:coherent,fig:modeledComparison}.  Thus the
structures observed in the $1^{-+}$ amplitude in the conventional PWA
are not an artifact due to the employed parametrizations for the
dynamic isobar amplitudes.  This is supported by the similarity of the
\PpiOne[1600] resonance parameters from the freed-isobar PWA with
those from the 14-wave resonance-model fit from
\refCite{Akhunzyanov:2018lqa}.
\section{The Deck process and its projection into the $\JPC = 1^{-+}$ wave}%
\label{sec:deck_model}

Most partial-wave amplitudes contained in the 88-wave set used to
analyze the COMPASS proton-target data (see
\cref{sec:pwa,sec:comparison}) contain coherent contributions from
resonant and nonresonant processes.  Aiming at extracting the resonant
components through fits of resonance models to the~\mThreePi and~\tpr
dependence of the spin-density matrix, COMPASS has used a simple
empirical description for the amplitude of the nonresonant
processes~(see Sec.~IV~A~2 in \refCite{Akhunzyanov:2018lqa_suppl}).
Our resonance-model fits reveal contributions of nonresonant processes
that are very different for the various partial waves.  The intensity
of the nonresonant contributions shows a strong dependence on~\tpr
that is often more pronounced than that of the resonances.  In the
analyzed~\mThreePi and \tpr~range, the nonresonant components are
expected to originate predominantly from double-Regge exchange
processes, of which the so-called Deck effect is the most prominent
one.  In \cref{fig:Deck_mechanism_Mandelstam}, we show the diagram of
the Deck process for the \threePi final state.  In this process, a
quasi-on-shell pion is exchanged between the vertices~$a$ and~$b$
becoming real by scattering off the target proton via Pomeron or
Reggeon exchange.  The \twoPi state produced at vertex~$a$, originally
taken to be the \Prho, is the only appearing resonance.  The described
process was proposed by R.~T.~Deck in \refCite{Deck:1964hm} as an
alternative explanation to \PaOne resonance production in the
$\Prho\,\pi$ $S$-wave channel~\cite{Goldhaber:1964zz,Chung:1964zza}.

\begin{figure}[tbp]
  \centering
  \includegraphics[width=\linewidthOr{0.6\textwidth}]{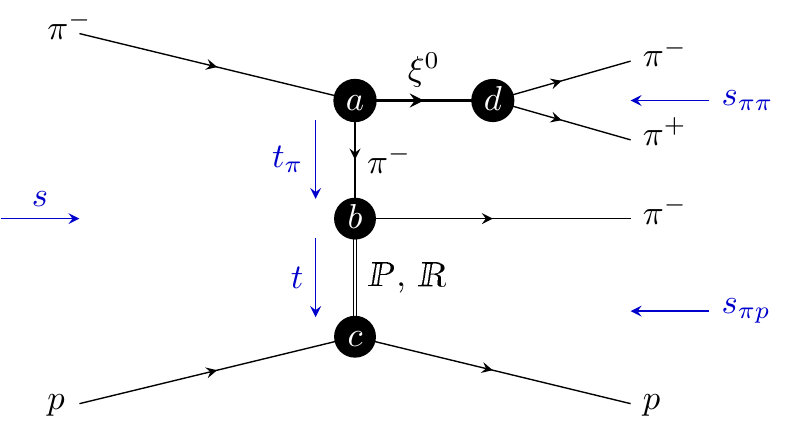}
  \caption{Schematic diagram of the Deck process with the relevant
    kinematic variables.}%
  \label{fig:Deck_mechanism_Mandelstam}
\end{figure}

In the COMPASS proton-target data, we found that the shape of the
intensity distribution of the spin-exotic \wave{1}{-+}{1}{+}{\Prho}{P}
wave changes strongly with~\tpr~\cite{Akhunzyanov:2018lqa}, which is
indicative of large nonresonant contributions that seem to contribute
particularly at low~\tpr as already discussed in
\cref{sec:other_models}.  This is consistent with the result of our
resonance-model fit, where we found that the $1^{-+}$ wave is strongly
dominated by nonresonant amplitudes at low~\tpr.  In this paper, we
investigate the role of the Deck process in this wave by determining
the intensity distribution of a Deck model in the
\wave{1}{-+}{1}{+}{\Prho}{P} wave and by comparing it to the
analytical description of the nonresonant component that we obtained
in our resonance-model fit.  We also study projections of the Deck
amplitude into waves with higher spin for which no confirmed
resonances exist~\cite{Tanabashi:2018zz} and compare with the
corresponding intensity distributions obtained from real data.  To
perform these studies we use pseudodata generated according to a model
of the Deck process using Monte Carlo techniques.  Since the
pseudodata contain only nonresonant contributions, the direct
comparison with real data neglects the interference between the
resonant and nonresonant wave components.  However, similar intensity
distributions in real and pseudodata would point to dominant
contributions from nonresonant Deck-like processes in the real data.

For our first attempt to model the Deck process, we use the simplified
model from \refsCite{Ascoli:1974sp,Ascoli:1974hi,Daum:1980ay} to
construct the Deck amplitude that we use to generate the pseudodata.
In the model, the Deck amplitude is factorized into three terms
(\confer\ \cref{fig:Deck_mechanism_Mandelstam}): \one~an
amplitude~$\mathcal{A}_\pipi$ that describes the $\twoPi \to \twoPi$
scattering including the vertices~$a$ and~$d$, \two~a stable-particle
propagator that describes the pion exchange, and \three~an
amplitude~$\mathcal{A}_{\pi p}$ that describes the $\pi^- p \to \pi^-
p$ scattering including the vertices~$b$ and~$c$.  We hence write the
Deck amplitude as\footnote{Since we do not use the Deck amplitude to
calculate absolute cross sections, the normalization of
\cref{eq:deck_ampl} is irrelevant.}
\begin{multlineOrEq}
  \label{eq:deck_ampl}
  \mathcal{A}_\text{Deck}(s_\pipi, s_{\pi p}, t_\pi, t) \newLineOrNot
  = \mathcal{A}_\pipi(s_\pipi, t_\pi)\, \frac{e^{\frac{b}{2}\, t_\pi}}{m_\pi^2 - t_\pi}\, \mathcal{A}_{\pi p}(s_{\pi p}, t).
\end{multlineOrEq}
It is important to note that we Bose-symmetrize the amplitude in
\cref{eq:deck_ampl} \wrt the two indistinguishable~$\pi^-$.  The
kinematic variables are defined in
\cref{fig:Deck_mechanism_Mandelstam} with $s_\pipi$~being the squared
center-of-momentum energy of the \twoPi system between vertices~$a$
and~$d$, $t_\pi$~the squared four-momentum transferred by the exchange
pion, $s_{\pi p}$~the squared center-of-momentum energy of the $\pi^-
p$ system including the vertices~$b$ and~$c$, and $t$~the squared
four-momentum transferred to the target.  Note, that both
$t$~and~$t_\pi$ are negative.  Like in the original Deck model in
\refCite{Deck:1964hm}, we only take into account pion exchange between
vertices~$a$ and~$b$ in \cref{fig:Deck_mechanism_Mandelstam}, while
possible additional processes like \Prho meson exchange are neglected.

The $\twoPi \to \twoPi$ amplitude in \cref{eq:deck_ampl} is taken from
\refCite{Hyams:1973zf} using the result of a so-called
energy-dependent analysis based on data for the reaction $\pi^- p \to
\pi^- \pi^+ n$ measured at \SI{17.2}{\GeVc} pion-beam momentum.  This
analysis yielded the \pipi partial-wave
amplitudes~$\mathcal{T}_\ell^I$ for orbital angular momenta $\ell = 0,
1, 2, 3$ between the two pions.  The model included amplitudes with
isospin~$I = 0$ of the \pipi system for even~$\ell$ and $I = 1$~for
odd~$\ell$.  In addition, an $I = 2$ amplitude was included for $\ell
= 0$.  The $P$-, $D$-, and $F$-wave amplitudes are dominated by \Prho,
\PfTwo, and \PrhoThree, respectively.  The parametrization of the
$S$-, $P$-, and $D$-wave amplitudes~$\mathcal{T}_0^0$,
$\mathcal{T}_1^1$, and~$\mathcal{T}_2^0$ is based on $K$~matrices that
take into account the \pipi and \KKbar channels.  The $F$-wave
amplitude is parametrized by a dynamic-width Breit-Wigner amplitude
for the \PrhoThree\ [see Eq.~(12d) in \refCite{Hyams:1973zf}].  The
$S$- and $P$-wave amplitudes use a $K$~matrix containing two poles and
a constant background term [see Eqs.~(12a), (12b) and (13a) in
\refCite{Hyams:1973zf}], where the \Prho pole in the $P$~wave includes
the angular-momentum barrier factor.  The $K$~matrix for the $D$-wave
amplitude contains a single pole, which includes the angular-momentum
barrier factor, and a constant background term [see Eqs.~(12c) and
(13b) in \refCite{Hyams:1973zf}].  For the $I = 2$ $S$-wave amplitude,
a scattering-length formula is used [see Eq.~(14)  in
\refCite{Hyams:1973zf}].  We use the parameters given in Table~1 of
\refCite{Hyams:1973zf}.

For the propagator of the exchanged pion, we use in
\cref{eq:deck_ampl} the nonreggeized form containing only the pion
pole $1 / (m_\pi^2 - t_\pi)$ and a form factor $e^{b\, t_\pi}$ [see
Eq.~(2.1) in \refCite{Ascoli:1974hi}].  We use a slope parameter of $b
= \SI{3.4}{\perGeVcsq}$, which provides a reasonable description of
the angular distributions of the COMPASS data in the \mThreePi range
from \SIrange{2.3}{2.5}{\GeVcc} assuming that the Deck process
dominates in this mass range.

For the $\pi^- p \to \pi^- p$ amplitude in \cref{eq:deck_ampl} we
employ the simple parametrization from Eq.~(3.1) in
\refCite{Ascoli:1974sp}:
\begin{equation}
  \label{eq:pi_p_scattering_amp}
  \mathcal{A}_{\pi p}(s_{\pi p}, t)
  = i\, s_{\pi p}\, e^{\frac{a}{2}\, t},
\end{equation}
with an exponential slope of $a = \SI{8}{\perGeVcsq}$.

\Cref{fig:MC-distributions} shows the~\mThreePi and~\mTwoPi
distributions of the \num{75e6}~Deck Monte Carlo events.
The various \twoPi resonances that are included in the model are
reflected in the \mTwoPi~distribution.  In order to roughly estimate
the Deck-like contributions to the real-data intensity distributions,
we perform a PWA of the Deck Monte Carlo data using the same 88-wave
PWA model that was applied to the COMPASS proton-target data in
\refCite{Akhunzyanov:2018lqa} (see \cref{tab:wave_sets} in
\cref{sec:wave_sets}).  We show in \cref{fig:deck_high_spin} as an
example the intensity distributions of the Deck model in the
\wave{4}{-+}{0}{+}{\Prho}{F} and \wave{6}{-+}{0}{+}{\Prho}{H} waves
for low and high values of~\tpr superimposed with the real-data
distributions.  Owing to the absence of confirmed resonances in these
waves, we expect the measured intensities to be dominated by
nonresonant contributions.  This hypothesis is supported by the fact
that the shapes of the Deck intensity distributions are in good
qualitative agreement with the real data over the full \tpr~range.
Note that the Deck Monte Carlo data are normalized using only one
common factor for all waves that is determined from the $1^{-+}$ wave
as described further below.

\begin{figure}[tbp]
  \centering%
  \subfloat[]{%
    \includegraphics[width=\twoPlotWidth]{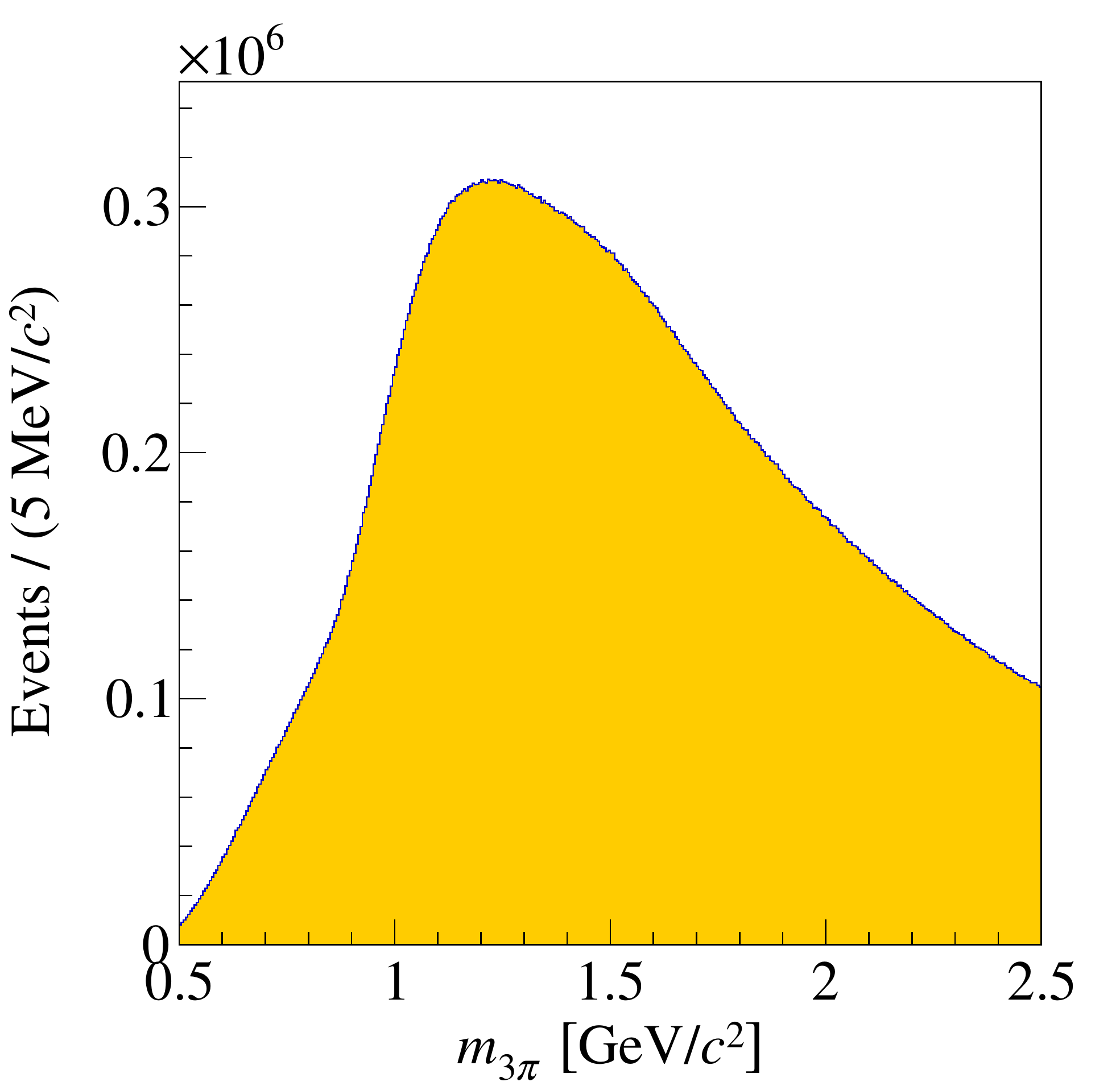}%
    \label{fig:MC-distributions_3pi}%
  }%
  \newLineOrHspace{\twoPlotSpacing}%
  \subfloat[]{%
    \includegraphics[width=\twoPlotWidth]{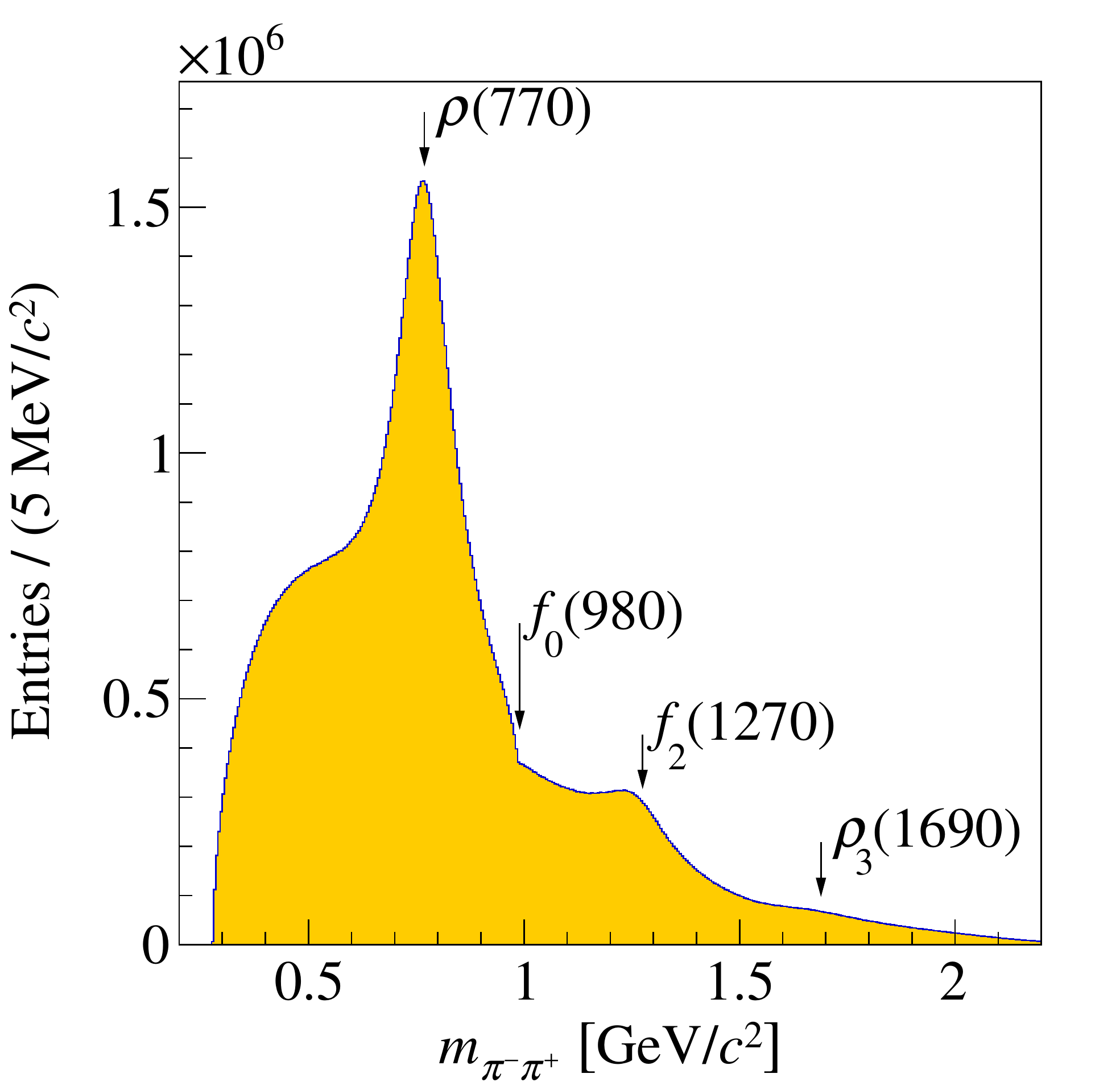}%
    \label{fig:MC-distributions_2pi}%
  }%
  \caption{\subfloatLabel{fig:MC-distributions_3pi}~\threePi invariant
    mass spectrum for the Deck Monte Carlo sample.
    \subfloatLabel{fig:MC-distributions_2pi}~invariant mass
    distribution of the \twoPi subsystem (two entries per event).  The
    arrows indicate $2\pi$ resonances included in the Deck model.}%
  \label{fig:MC-distributions}
\end{figure}

\begin{wideFigureOrNot}[tbp]
  \centering%
  \subfloat[]{%
    \includegraphics[width=\twoPlotWidth]{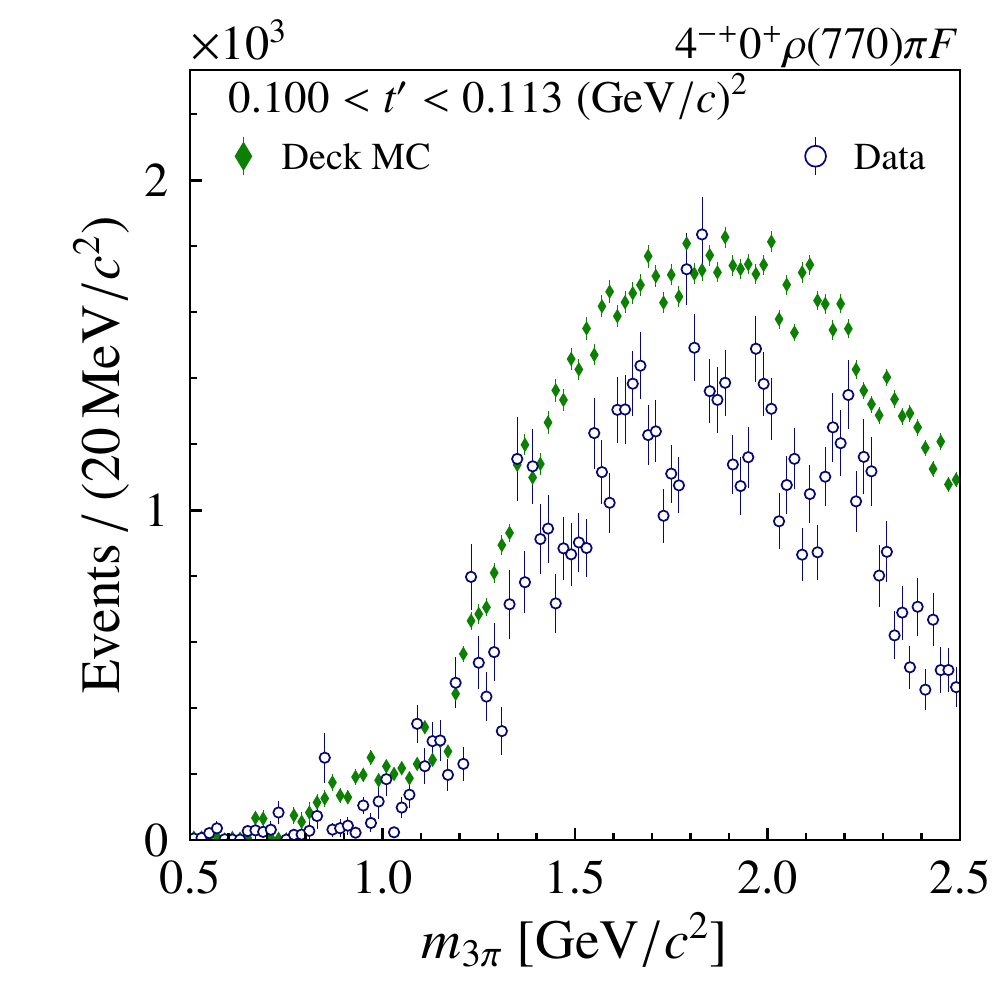}%
  }%
  \hspace*{\twoPlotSpacing}%
  \subfloat[]{%
    \includegraphics[width=\twoPlotWidth]{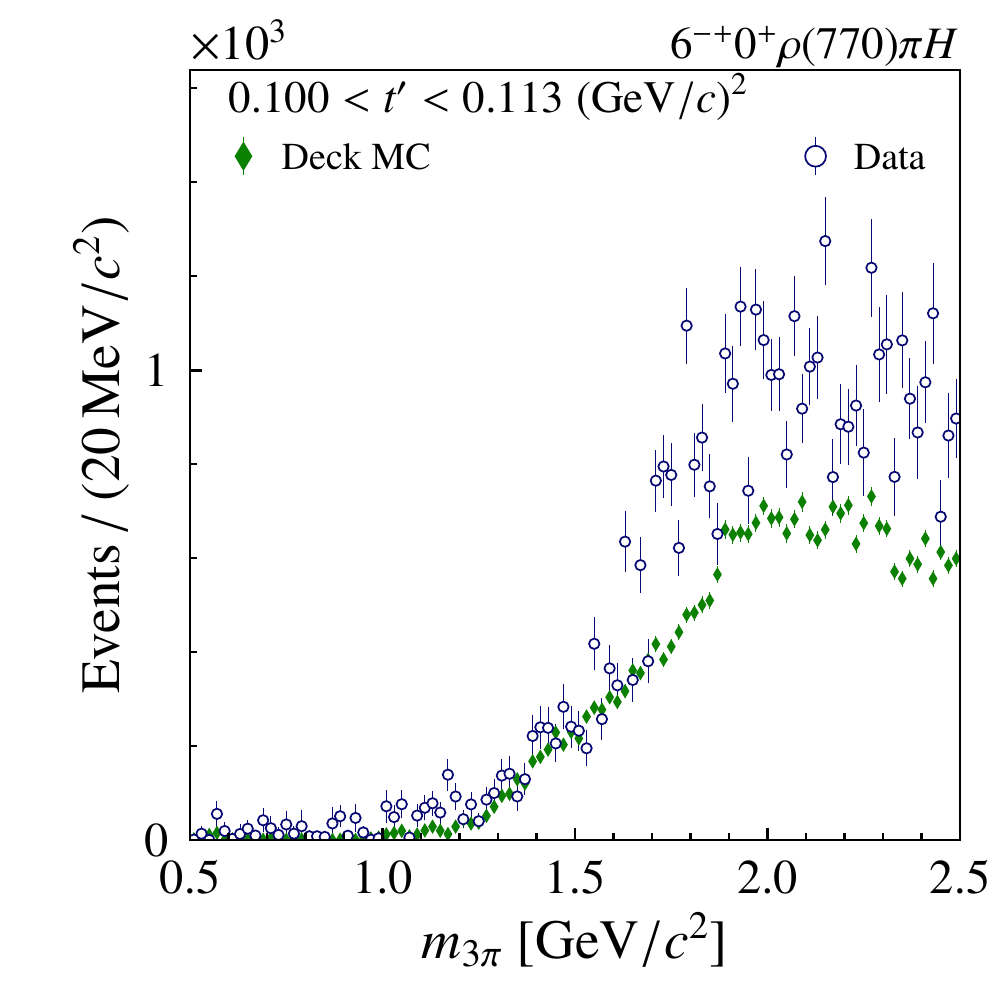}%
  }%
  \\%
  \subfloat[]{%
    \includegraphics[width=\twoPlotWidth]{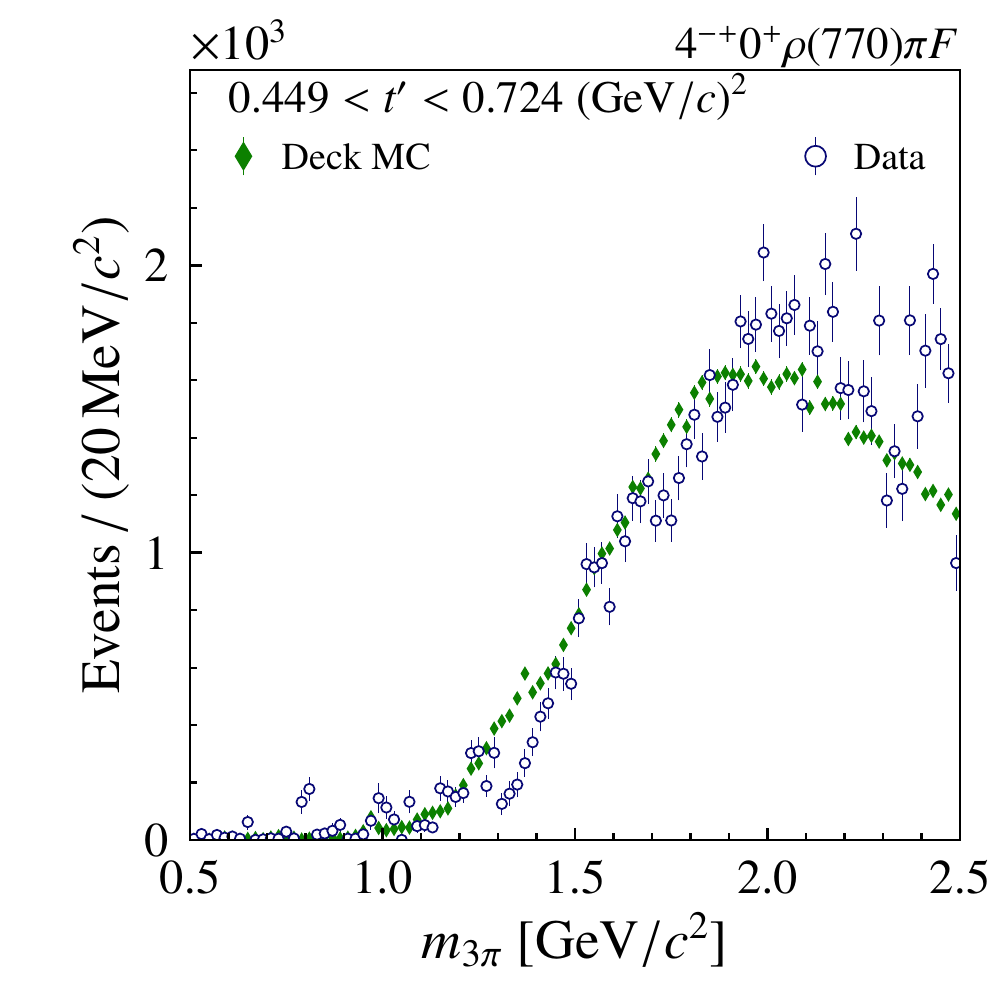}%
  }%
  \hspace*{\twoPlotSpacing}%
  \subfloat[]{%
    \includegraphics[width=\twoPlotWidth]{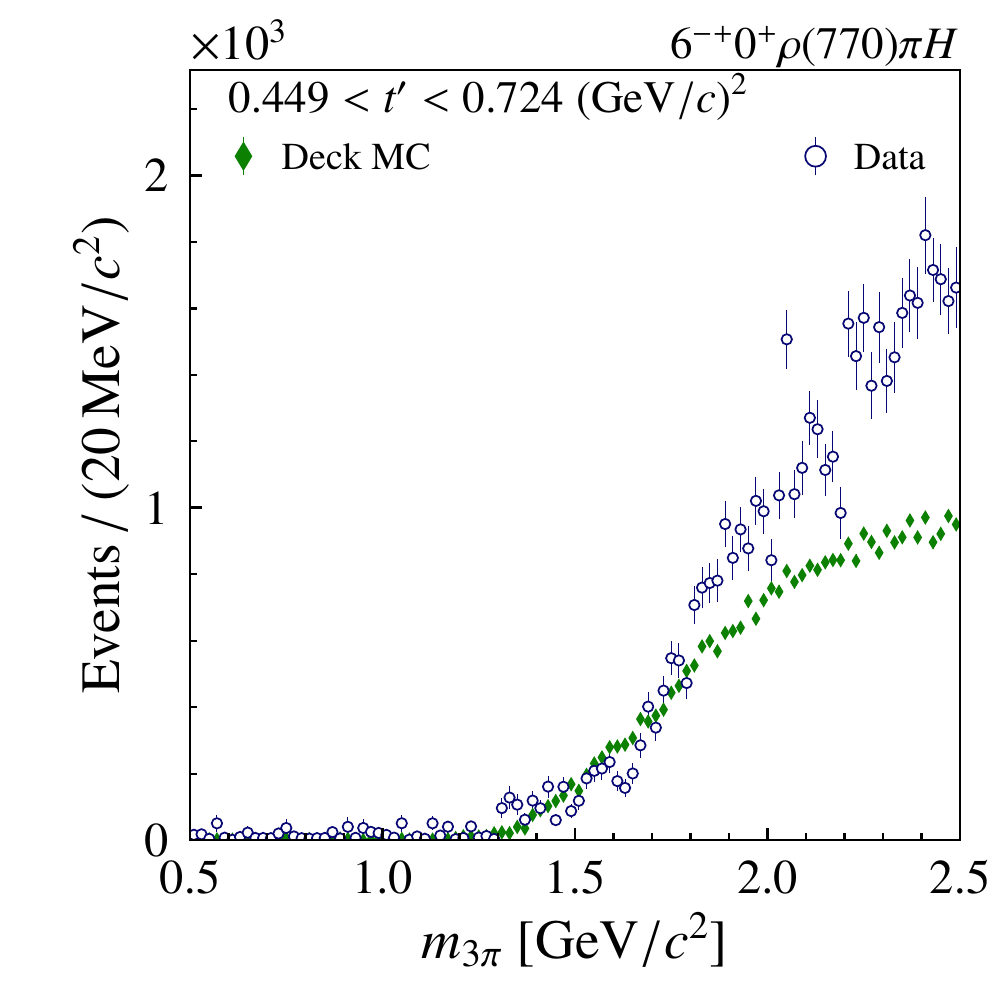}%
  }%
  \caption{Intensity distributions of the \wave{4}{-+}{0}{+}{\Prho}{F}
    wave (left column) and the \wave{6}{-+}{0}{+}{\Prho}{H} wave
    (right column) for two \tpr~bins as obtained from the 88-wave
    PWA\@.  The real data are represented by the blue data points; the
    Deck pseudodata by the green data points.  The pseudodata are
    normalized using a common factor for all waves (see text for
    details).}%
  \label{fig:deck_high_spin}
\end{wideFigureOrNot}

In \cref{fig:exotic_deck_comparison_tbins}, we compare the intensity
distributions of the \wave{1}{-+}{1}{+}{\Prho}{P} wave for Deck
pseudodata and real data, where the latter contains contributions from
resonances as well as nonresonant processes.  In addition, we show
curves that correspond to the nonresonant component found in the
resonance-model fit in \refCite{Akhunzyanov:2018lqa}.  We normalize
the Deck intensity in the $1^{-+}$ wave to the nonresonant curves from
the resonance-model fit by matching their \mThreePi-integrated
intensities summed over the lowest 9~of the 11~\tpr bins.  The two
highest \tpr~bins are excluded from the calculation of the
normalization factor because in this region the \PpiOne[1600]
resonance dominates the $1^{-+}$ wave and hence the nonresonant
component has a large systematic uncertainty.  The same normalization
factor is also used for the other waves shown in
\cref{fig:deck_high_spin}.  For the first 9~\tpr~bins, \ie for $\tpr
\lesssim \SI{0.5}{\GeVcsq}$, the shapes of the Deck intensity
distributions are in qualitative agreement with the nonresonant curves
from the resonance-model fit [see
\cref{fig:exotic_deck_comparison_tbin_1,fig:exotic_deck_comparison_tbin_5,fig:exotic_deck_comparison_tbin_9}].
This shows that the empirical parametrization used for the nonresonant
component in the resonance-model fit is able to capture the gross
features of the Deck amplitudes.  We find that the \tpr~dependence of
the Deck intensity is shallower than that of the nonresonant curve
leading to an undershoot of the Deck intensity at low~\tpr [see
\cref{fig:exotic_deck_comparison_tbin_1}] and an overshoot at
high~\tpr [see \cref{fig:exotic_deck_comparison_tbin_9}].  For the two
highest \tpr~bins, \ie for $\tpr \gtrsim \SI{0.5}{\GeVcsq}$, the
shapes of the Deck intensity distribution and the nonresonant curve
start to deviate [see \cref{fig:exotic_deck_comparison_tbin_10}].  The
observed behavior of the Deck model is consistent with our finding in
\refCite{Akhunzyanov:2018lqa} that at low~\tpr the broad structure in
the $1^{-+}$ intensity distribution is mostly due to nonresonant
contributions masking the small \PpiOne[1600] signal.

\begin{wideFigureOrNot}[tbp]
  \centering%
  \subfloat[]{%
    \includegraphics[width=\twoPlotWidth]{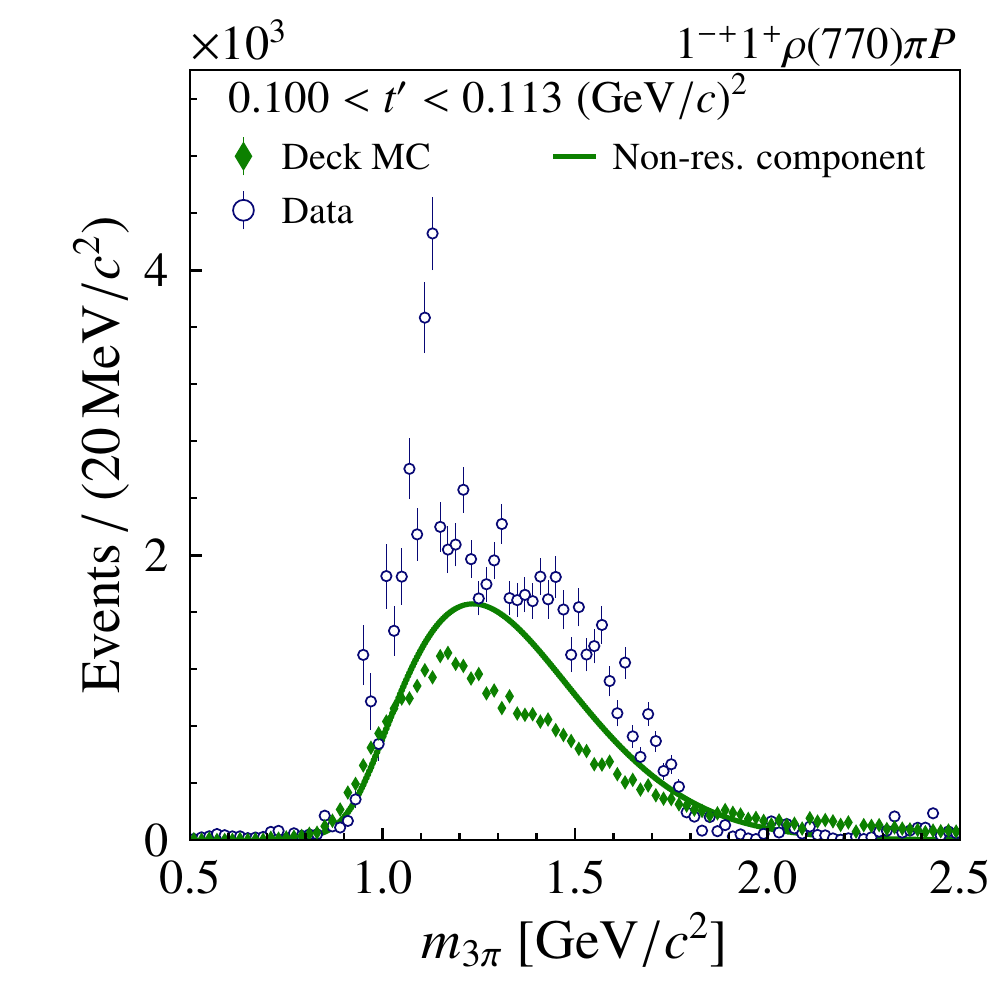}%
    \label{fig:exotic_deck_comparison_tbin_1}%
  }%
  \hspace*{\twoPlotSpacing}%
  \subfloat[]{%
    \includegraphics[width=\twoPlotWidth]{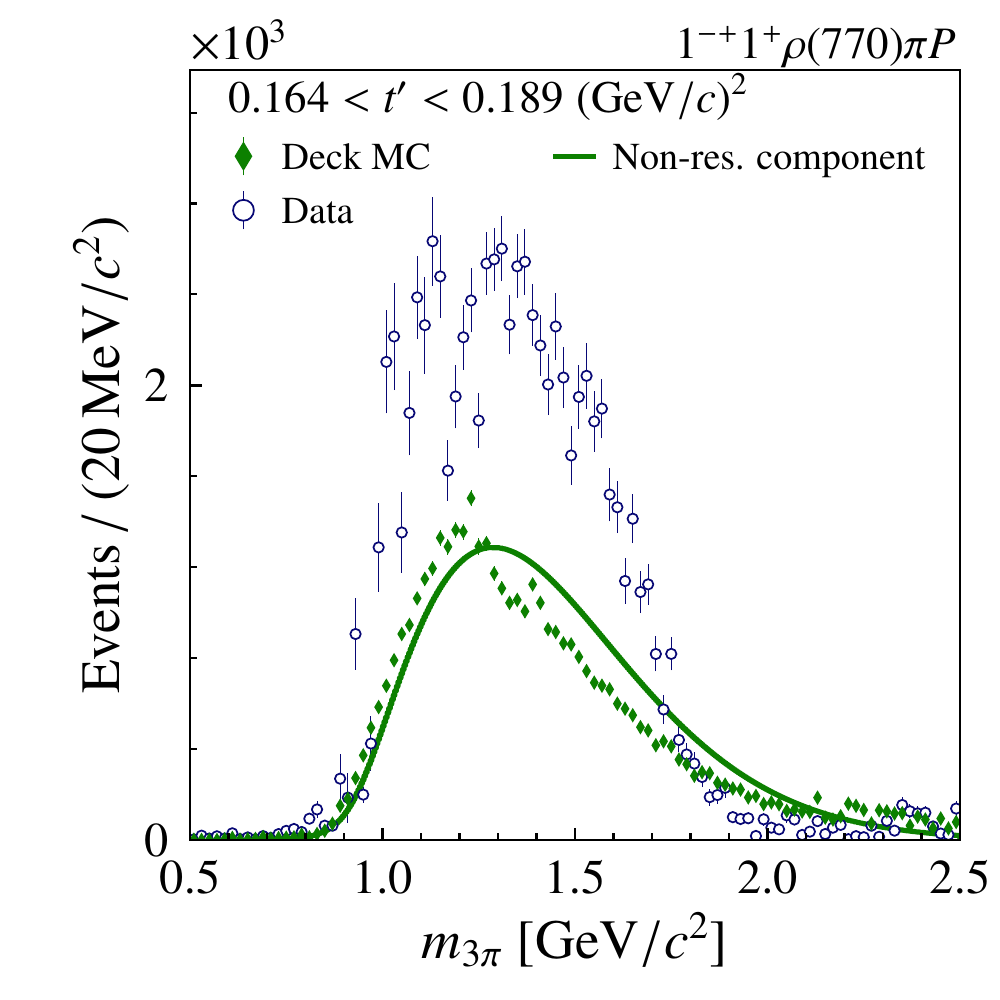}%
    \label{fig:exotic_deck_comparison_tbin_5}%
  }%
  \\%
  \subfloat[]{%
    \includegraphics[width=\twoPlotWidth]{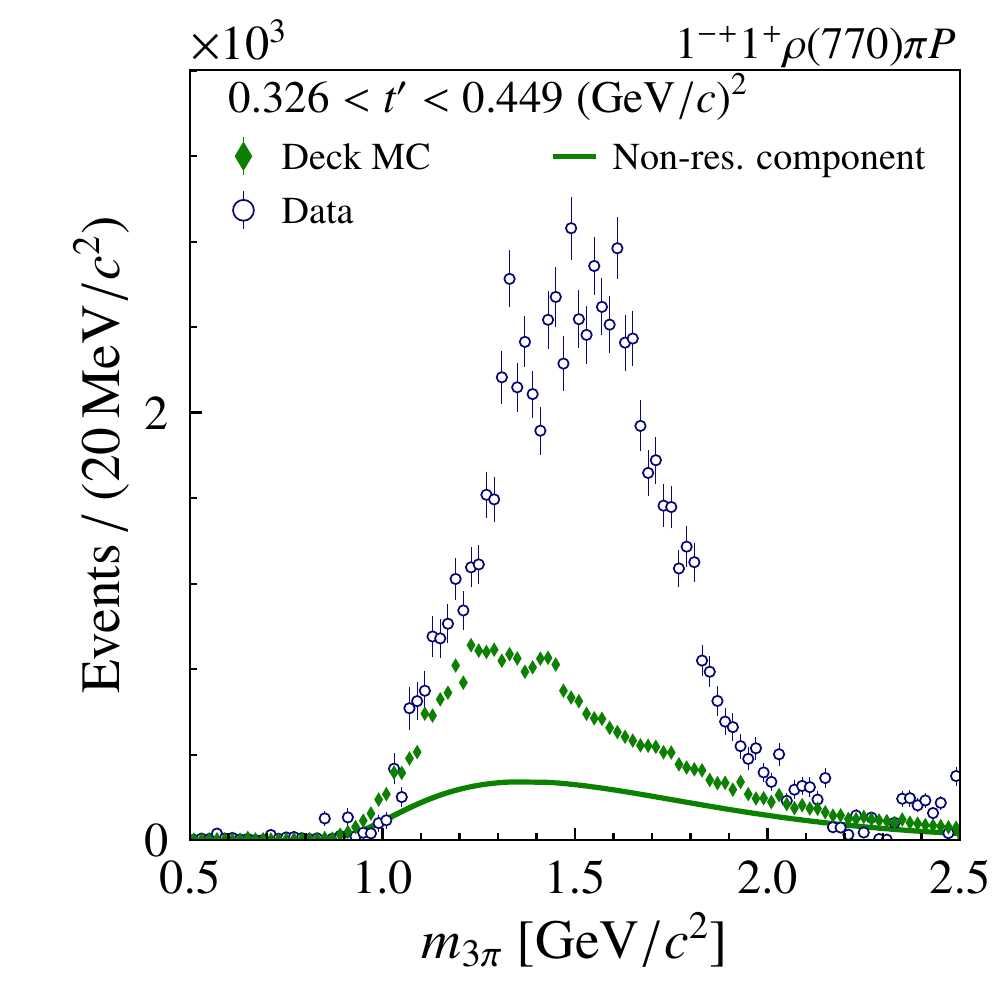}%
    \label{fig:exotic_deck_comparison_tbin_9}%
  }%
  \hspace*{\twoPlotSpacing}%
  \subfloat[]{%
    \includegraphics[width=\twoPlotWidth]{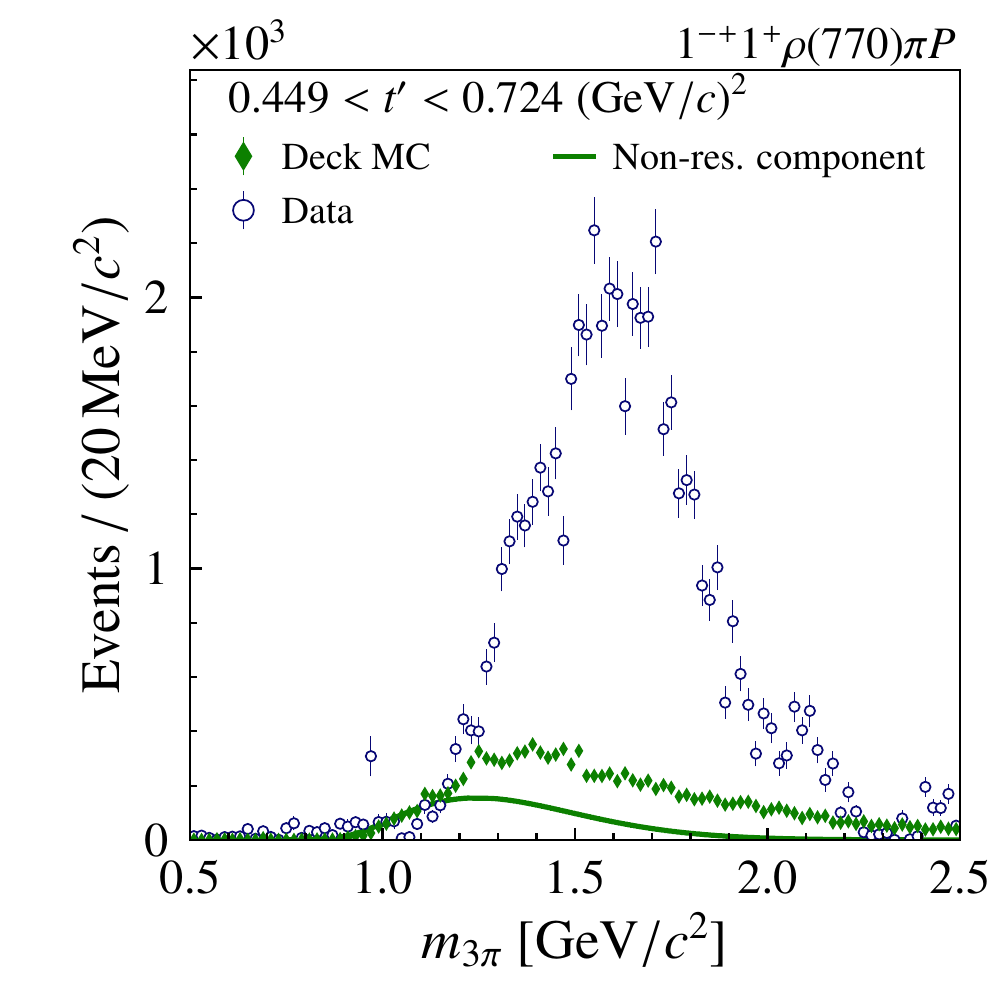}%
    \label{fig:exotic_deck_comparison_tbin_10}%
  }%
  \caption{Intensity distributions of the \wave{1}{-+}{1}{+}{\Prho}{P}
    wave in different \tpr~bins as obtained from the 88-wave PWA\@.
    The real data are represented by the blue data points; the Deck
    pseudo data by the green data points.  The green curve represents
    the nonresonant component determined in the resonance-model fit in
    \refCite{Akhunzyanov:2018lqa}.  The pseudodata are normalized
    using a common factor for all waves (see text for details).}%
  \label{fig:exotic_deck_comparison_tbins}
\end{wideFigureOrNot}

\subsection{Summary: The Deck process and the $\JPC = 1^{-+}$ wave}%
\label{sec:deck_model_summary}

We have studied a simple model for the Deck process [see
\cref{eq:deck_ampl,fig:Deck_mechanism_Mandelstam}] and have compared
its intensity distribution in selected waves with the ones obtained
from real data.  With only one common normalization factor, we find
that the Deck intensity is in qualitative agreement with the measured
intensity distributions of the \wave{4}{-+}{0}{+}{\Prho}{F} and
\wave{6}{-+}{0}{+}{\Prho}{H} waves in the analyzed \tpr~range.  This
is consistent with the expectation that these waves are dominated by
nonresonant components because there are no confirmed $\pi_4$~or
$\pi_6$~resonances~\cite{Tanabashi:2018zz}.

We find that the Deck intensity distribution in the spin-exotic
\wave{1}{-+}{1}{+}{\Prho}{P} wave qualitatively reproduces the strong
\tpr~dependence of intensity and shape of the nonresonant component
that we extracted in our resonance-model fit in
\refCite{Akhunzyanov:2018lqa} in the range $\tpr \lesssim
\SI{0.5}{\GeVcsq}$.  In this \tpr~range, the intensity of the
nonresonant contribution is similar to or larger than the intensity of
the \PpiOne[1600] component.  However, with regard to the
\tpr~dependence it must be recognized that the present simple version
of the Deck model does not adequately describe the background yield in
the high-\tpr range where the \PpiOne[1600] resonance dominates the
$1{-+}$ wave (see \cref{fig:exotic_deck_comparison_tbins}).

Nuclear effects such as absorption seem to play an important role in
the scattering process in the \tpr~range between
\SIlist{0.1;1.0}{\GeVcsq}, so that the process cannot be described
simply as incoherent scattering off quasifree nucleons.  There are
currently no models available that describe Deck-like processes on
nuclear targets.  Such models could help to better understand the
enhancement of the \PpiOne[1600] signal relative to the nonresonant
component that we observe in the lead-target data as compared to the
proton-target data (see \cref{sec:other_models}).

\clearpage
\appendix
\makeatletter
\@addtoreset{equation}{section}  %
\makeatother
\section*{Appendix}
\ifMultiColumnLayout{\clearpage\onecolumngrid}{}
\section{Compilation of wave sets used in partial-wave analyses of the $3\pi$ system}%
\label{sec:wave_sets}

\Cref{tab:wave_sets} lists the wave sets used in the partial-wave
analyses that are summarized in \cref{tab:exotic_diff_models}.

\ifMultiColumnLayout{}{\begin{scriptsize}}
\begin{wideLongTableOrNot}{ccccccccc}
  \caption{Comparison of the 88-wave set used for the COMPASS
    proton-target data with the wave sets of BNL~E852, VES,
    Dzierba~\etal, and the one used for the COMPASS lead-target data.
    Entries that are not in the COMPASS 88-wave set are marked
    with~$*$.  Entries marked with~$\dagger$ indicate waves that are
    replaced in the freed-isobar PWA by waves with dynamic isobar
    amplitudes parametrized according to \cref{eq:dyn_amp_freed} (see
    also \cref{tab:freedIsobarList}).%
  }%
  \label{tab:wave_sets}%
  \\
  \toprule
  & \textbf{\JPCMrefl} & \textbf{Isobar} & \textbf{$L$} & \textbf{COMPASS}  & \textbf{BNL E852} & \textbf{VES}      & \textbf{Dzierba~\etal} & \textbf{COMPASS} \\
  &                    &                 &              & \textbf{88 waves} & \textbf{21 waves} & \textbf{44 waves} & \textbf{36 waves} & \textbf{42 waves} \\
  &                    &                 &              & Table~IX in~\cite{Adolph:2015tqa} & Table~I in~\cite{Chung:2002pu} & {\cite{Zaitsev:2000rc}} & Table~IV in~\cite{Dzierba:2005jg} & {\cite{Alekseev:2009aa}} \\
  \midrule
  \endfirsthead
  \caption{\textit{(Continued)}}
  \\
  \toprule
  & \textbf{\JPCMrefl} & \textbf{Isobar} & \textbf{$L$} & \textbf{COMPASS}  & \textbf{BNL E852} & \textbf{VES}      & \textbf{Dzierba~\etal} & \textbf{COMPASS} \\
  &                    &                 &              & \textbf{88 waves} & \textbf{21 waves} & \textbf{44 waves} & \textbf{36 waves} & \textbf{42 waves} \\
  \midrule
  \endhead
  \bottomrule
  \addlinespace[1ex]
  \multicolumn{9}{r}{\textit{(Table continued)}} \\
  \endfoot
  \bottomrule
  \endlastfoot

  $\dagger$ & $0^{-+}\,0^+$ & \pipiS        & $S$ & \checkmark & \checkmark & \checkmark & \checkmark & \checkmark \\
  $\dagger$ & $0^{-+}\,0^+$ & \Prho         & $P$ & \checkmark & \checkmark & \checkmark & \checkmark & \checkmark \\
  $\dagger$ & $0^{-+}\,0^+$ & \PfZero[980]  & $S$ & \checkmark & \checkmark & \checkmark & \checkmark & \checkmark \\
            & $0^{-+}\,0^+$ & \PfTwo        & $D$ & \checkmark &            & \checkmark &            &            \\
  $\dagger$ & $0^{-+}\,0^+$ & \PfZero[1500] & $S$ & \checkmark &            &            &            &            \\
  \midrule
  $\dagger$ & $1^{++}\,0^+$ & \pipiS       & $P$ & \checkmark &            & \checkmark & \checkmark & \checkmark \\
            & $1^{++}\,1^+$ & \pipiS       & $P$ & \checkmark &            & \checkmark &            & \checkmark \\
  $\dagger$ & $1^{++}\,0^+$ & \Prho        & $S$ & \checkmark & \checkmark & \checkmark & \checkmark & \checkmark \\
  $\dagger$ & $1^{++}\,1^+$ & \Prho        & $S$ & \checkmark & \checkmark & \checkmark & \checkmark & \checkmark \\
            & $1^{++}\,0^+$ & \Prho        & $D$ & \checkmark & \checkmark & \checkmark &            & \checkmark \\
            & $1^{++}\,1^+$ & \Prho        & $D$ & \checkmark &            & \checkmark &            & \checkmark \\
  $\dagger$ & $1^{++}\,0^+$ & \PfZero[980] & $P$ & \checkmark &            & \checkmark & \checkmark &            \\
            & $1^{++}\,1^+$ & \PfZero[980] & $P$ & \checkmark &            &            &            &            \\
            & $1^{++}\,0^+$ & \PfTwo       & $P$ & \checkmark &            & \checkmark &            & \checkmark \\
            & $1^{++}\,1^+$ & \PfTwo       & $P$ & \checkmark &            & \checkmark & \checkmark & \checkmark \\
            & $1^{++}\,0^+$ & \PfTwo       & $F$ & \checkmark &            &            &            &            \\
            & $1^{++}\,0^+$ & \PrhoThree   & $D$ & \checkmark &            &            &            &            \\
            & $1^{++}\,0^+$ & \PrhoThree   & $G$ & \checkmark &            &            &            &            \\
  \midrule
  $\dagger$ & $1^{-+}\,1^+$ & \Prho        & $P$ & \checkmark & \checkmark & \checkmark & \checkmark & \checkmark \\
  \midrule
  $\dagger$ & $2^{++}\,1^+$ & \Prho        & $D$ & \checkmark & \checkmark & \checkmark & \checkmark & \checkmark \\
            & $2^{++}\,2^+$ & \Prho        & $D$ & \checkmark &            &            &            &            \\
            & $2^{++}\,1^+$ & \PfTwo       & $P$ & \checkmark &            & \checkmark &            & \checkmark \\
            & $2^{++}\,2^+$ & \PfTwo       & $P$ & \checkmark &            &            &            &            \\
            & $2^{++}\,1^+$ & \PrhoThree   & $D$ & \checkmark &            &            &            &            \\
  \midrule
  $\dagger$ & $2^{-+}\,0^+$ & \pipiS       & $D$ & \checkmark & \checkmark & \checkmark & \checkmark & \checkmark \\
            & $2^{-+}\,1^+$ & \pipiS       & $D$ & \checkmark &            & \checkmark & \checkmark & \checkmark \\
  $\dagger$ & $2^{-+}\,0^+$ & \Prho        & $P$ & \checkmark & \checkmark & \checkmark & \checkmark & \checkmark \\
  $\dagger$ & $2^{-+}\,1^+$ & \Prho        & $P$ & \checkmark &            & \checkmark & \checkmark & \checkmark \\
            & $2^{-+}\,2^+$ & \Prho        & $P$ & \checkmark &            &            &            &            \\
  $\dagger$ & $2^{-+}\,0^+$ & \Prho        & $F$ & \checkmark &            & \checkmark & \checkmark & \checkmark \\
            & $2^{-+}\,1^+$ & \Prho        & $F$ & \checkmark &            & \checkmark & \checkmark & \checkmark \\
  $\dagger$ & $2^{-+}\,0^+$ & \PfZero[980] & $D$ & \checkmark &            & \checkmark & \checkmark &            \\
  $*$       & $2^{-+}\,1^+$ & \PfZero[980] & $D$ &            &            &            & \checkmark &            \\
  $\dagger$ & $2^{-+}\,0^+$ & \PfTwo       & $S$ & \checkmark & \checkmark & \checkmark & \checkmark & \checkmark \\
            & $2^{-+}\,1^+$ & \PfTwo       & $S$ & \checkmark & \checkmark & \checkmark & \checkmark & \checkmark \\
            & $2^{-+}\,2^+$ & \PfTwo       & $S$ & \checkmark &            &            &            &            \\
            & $2^{-+}\,0^+$ & \PfTwo       & $D$ & \checkmark & \checkmark & \checkmark & \checkmark & \checkmark \\
            & $2^{-+}\,1^+$ & \PfTwo       & $D$ & \checkmark & \checkmark & \checkmark & \checkmark & \checkmark \\
            & $2^{-+}\,2^+$ & \PfTwo       & $D$ & \checkmark &            &            &            &            \\
            & $2^{-+}\,0^+$ & \PfTwo       & $G$ & \checkmark &            &            &            &            \\
            & $2^{-+}\,0^+$ & \PrhoThree   & $P$ & \checkmark &            &            & \checkmark &            \\
            & $2^{-+}\,1^+$ & \PrhoThree   & $P$ & \checkmark &            &            & \checkmark &            \\
  \midrule
            & $3^{++}\,0^+$ & \pipiS       & $F$ & \checkmark &            &            &            &            \\
            & $3^{++}\,1^+$ & \pipiS       & $F$ & \checkmark &            &            &            &            \\
            & $3^{++}\,0^+$ & \Prho        & $D$ & \checkmark &            & \checkmark & \checkmark & \checkmark \\
            & $3^{++}\,1^+$ & \Prho        & $D$ & \checkmark &            &            &            & \checkmark \\
            & $3^{++}\,0^+$ & \Prho        & $G$ & \checkmark &            &            &            &            \\
            & $3^{++}\,1^+$ & \Prho        & $G$ & \checkmark &            &            &            &            \\
            & $3^{++}\,0^+$ & \PfTwo       & $P$ & \checkmark &            & \checkmark & \checkmark & \checkmark \\
            & $3^{++}\,1^+$ & \PfTwo       & $P$ & \checkmark &            &            &            & \checkmark \\
            & $3^{++}\,0^+$ & \PrhoThree   & $S$ & \checkmark & \checkmark & \checkmark & \checkmark & \checkmark \\
            & $3^{++}\,1^+$ & \PrhoThree   & $S$ & \checkmark &            & \checkmark &            & \checkmark \\
            & $3^{++}\,0^+$ & \PrhoThree   & $I$ & \checkmark &            &            &            &            \\
  \midrule
            & $3^{-+}\,1^+$ & \Prho        & $F$ & \checkmark &            &            &            &            \\
            & $3^{-+}\,1^+$ & \PfTwo       & $D$ & \checkmark &            &            &            &            \\
  \midrule
            & $4^{++}\,1^+$ & \Prho        & $G$ & \checkmark &            & \checkmark & \checkmark & \checkmark \\
            & $4^{++}\,2^+$ & \Prho        & $G$ & \checkmark &            &            &            &            \\
            & $4^{++}\,1^+$ & \PfTwo       & $F$ & \checkmark &            & \checkmark & \checkmark & \checkmark \\
            & $4^{++}\,2^+$ & \PfTwo       & $F$ & \checkmark &            &            &            &            \\
            & $4^{++}\,1^+$ & \PrhoThree   & $D$ & \checkmark &            &            & \checkmark &            \\
  \midrule
            & $4^{-+}\,0^+$ & \pipiS       & $G$ & \checkmark &            &            &            &            \\
            & $4^{-+}\,0^+$ & \Prho        & $F$ & \checkmark &            & \checkmark & \checkmark & \checkmark \\
            & $4^{-+}\,1^+$ & \Prho        & $F$ & \checkmark &            &            &            & \checkmark \\
            & $4^{-+}\,0^+$ & \PfTwo       & $D$ & \checkmark &            & \checkmark &            &            \\
            & $4^{-+}\,1^+$ & \PfTwo       & $D$ & \checkmark &            &            &            &            \\
            & $4^{-+}\,0^+$ & \PfTwo       & $G$ & \checkmark &            &            &            &            \\
  $*$       & $4^{-+}\,0^+$ & \PrhoThree   & $P$ &            &            & \checkmark & \checkmark &            \\
  \midrule
            & $5^{++}\,0^+$ & \pipiS       & $H$ & \checkmark &            &            &            &            \\
            & $5^{++}\,1^+$ & \pipiS       & $H$ & \checkmark &            &            &            &            \\
            & $5^{++}\,0^+$ & \Prho        & $G$ & \checkmark &            &            &            &            \\
            & $5^{++}\,0^+$ & \PfTwo       & $F$ & \checkmark &            &            &            &            \\
            & $5^{++}\,1^+$ & \PfTwo       & $F$ & \checkmark &            &            &            &            \\
            & $5^{++}\,0^+$ & \PfTwo       & $H$ & \checkmark &            &            &            &            \\
            & $5^{++}\,0^+$ & \PrhoThree   & $D$ & \checkmark &            &            &            &            \\
  \midrule
            & $6^{++}\,1^+$ & \Prho        & $I$ & \checkmark &            &            &            &            \\
            & $6^{++}\,1^+$ & \PfTwo       & $H$ & \checkmark &            &            &            &            \\
  \midrule
            & $6^{-+}\,0^+$ & \pipiS       & $I$ & \checkmark &            &            &            &            \\
            & $6^{-+}\,1^+$ & \pipiS       & $I$ & \checkmark &            &            &            &            \\
            & $6^{-+}\,0^+$ & \Prho        & $H$ & \checkmark &            &            &            &            \\
            & $6^{-+}\,1^+$ & \Prho        & $H$ & \checkmark &            &            &            &            \\
            & $6^{-+}\,0^+$ & \PfTwo       & $G$ & \checkmark &            &            &            &            \\
            & $6^{-+}\,0^+$ & \PrhoThree   & $F$ & \checkmark &            &            &            &            \\
  \midrule
            & $1^{++}\,1^-$ & \Prho        & $S$ & \checkmark & \checkmark & \checkmark &            & \checkmark \\
  \midrule
            & $1^{-+}\,0^-$ & \Prho        & $P$ & \checkmark & \checkmark & \checkmark & \checkmark & \checkmark \\
            & $1^{-+}\,1^-$ & \Prho        & $P$ & \checkmark & \checkmark & \checkmark & \checkmark & \checkmark \\
  \ifMultiColumnLayout{\midrule}{}
            & $2^{++}\,0^-$ & \Prho        & $D$ & \checkmark & \checkmark & \checkmark & \checkmark & \checkmark \\
  $*$       & $2^{++}\,1^-$ & \Prho        & $D$ &            &            & \checkmark &            &            \\
            & $2^{++}\,0^-$ & \PfTwo       & $P$ & \checkmark &            &            &            & \checkmark \\
            & $2^{++}\,1^-$ & \PfTwo       & $P$ & \checkmark &            &            &            & \checkmark \\
  \midrule
  $*$       & $2^{-+}\,1^-$ & \Prho        & $P$ &            &            & \checkmark &            &            \\
            & $2^{-+}\,1^-$ & \PfTwo       & $S$ & \checkmark & \checkmark & \checkmark &            & \checkmark \\
  \midrule
            & Flat          &              &     & \checkmark & \checkmark & \checkmark & \checkmark & \checkmark \\
\end{wideLongTableOrNot}
\ifMultiColumnLayout{}{\end{scriptsize}}
 \ifMultiColumnLayout{\twocolumngrid}{}
\section{Ambiguity in the $\JPC = 1^{-+}$ amplitude in the freed-isobar partial-wave analysis}%
\label{app:zeroMode}

As has been already mentioned in \cref{sec:freed_isobar}, continuous
mathematical ambiguities of certain decay amplitudes, called zero
modes, may appear in a freed-isobar PWA\@.  Methods to detect and
resolve zero modes are discussed in detail in
\refsCite{Krinner:2017dba,Krinner:2018bwg}.  Because of zero modes,
different values of the transition amplitudes $\{\prodAmp_{a, k}\}$
defined in \cref{eq:dyn_amp_freed,eq:prod_amp_freed} may lead to the
same total amplitude $\sum_k \prodAmp_{a, k}\, \decayAmp_{a, k}$ in
\cref{eq:intensity_freed}, where $k$~labels the \mTwoPi~intervals.
Zero modes may appear in sets of freed waves that have the same
\JPCMrefl quantum numbers but describe decays via different isobars.
In special cases, such as the one described below, zero modes may also
appear within a single freed wave due to Bose symmetrization of
final-state particles.

In the following, we focus on the zero mode present within the
\wave{1}{-+}{1}{+}{\pipiPF}{P} wave.  For the employed wave set (see
\cref{tab:freedIsobarList,tab:wave_sets}), it is mathematically well
defined and it is the only zero mode affecting this wave.  First, we
show the origin of this zero-mode ambiguity in the $1^{-+}$ wave.  To
this end, we express the angular amplitude $\angularPart_a(\tau)$ with
$a = \wave{1}{-+}{1}{+}{\pipiPF}{P}$ in \cref{eq:decay_amp_freed} in
the helicity formalism:\footnote{Also \confer with Eqs.~(11) and~(7)
in \refCite{Adolph:2015tqa}.}\footnote{To ease the notation, we omit
the wave index~$a$ in this section.}
\begin{multlineOrEq}
  \label{eq:cross_product_amplitude}
  \angularPart_a(\tau)
  \propto \sum_{\lambda = \pm 1} -\frac{\lambda}{\sqrt{2}}\,
  {}^{(\refl = +1)}\!D^{1 \text{*}}_{1, \lambda}(\phiGJ, \thetaGJ, 0)\, \newLineOrNot
  D^{1 \text{*}}_{\lambda, 0}(\phiHF, \thetaHF, 0).
\end{multlineOrEq}
Here, $\lambda$~is the helicity of the $1^{--}$ isobar.  The factor
$-\lambda/ \sqrt{2}$ is the Clebsch-Gordan coefficient
\clebsch{L}{0}{J_\xi}{\lambda}{J}{\lambda} that describes the coupling
of the relative orbital angular momentum $L = 1$ between the isobar
and the bachelor~$\pi^-$ with the spin $J_\xi = 1$ of the isobar to
the spin~$J = 1$ of~$X$.  The angular distributions of the decays $X^-
\to \xi^0 + \pi^-$ and $\xi^0 \to \pi^- + \pi^+$ are described by
Wigner $D$~functions, where the one for the $X^-$~decay is defined in
the reflectivity basis according to Eq.~(19) in
\refCite{Adolph:2015tqa}.  The subscripts \GJ\ and \HF\ of the angles
denote the Gottfried-Jackson and helicity rest frames of~$X$ and the
isobar, respectively (see Sec.~III~A in \refCite{Adolph:2015tqa} for
the definition of the coordinate systems).  Inserting the
$D$~functions
\begin{multlineOrEq}
  {}^{(\refl = +1)}\!D^1_{1, \lambda}(\phiGJ, \thetaGJ, 0)
  = \frac{1}{\sqrt{2}}\, \big( \cos\phiGJ \newLineOrNot
    - i \lambda\, \sin\phiGJ\, \cos\thetaGJ \big)
\end{multlineOrEq}
and
\begin{equation}
  D^1_{\lambda, 0}(\phiHF, \thetaHF, 0)
  = -\frac{\lambda}{\sqrt{2}}\, e^{-i \lambda\, \phiHF}\, \sin\thetaHF
\end{equation}
with $\lambda = \pm 1$ into \cref{eq:cross_product_amplitude}, we find
\begin{multlineOrEq}
  \label{eq:K-1}
  \angularPart_a(\tau)
  \propto \big( \cos\phiGJ\, \cos\phiHF \newLineOrNot
  - \cos\thetaGJ\, \sin\phiGJ\, \sin\phiHF \big)\, \sin\thetaHF.
\end{multlineOrEq}

The $\pi^-_1 \pi^-_2 \pi^+_3$ system contains two
indistinguishable~$\pi^-$, and hence \cref{eq:K-1} needs to be
Bose-symmetrized.  We choose the isobar to decay into $\pi_1^-
\pi_3^+$ and the vector~$\vec{p}_1^{\,\HF}$ to represent the momentum
of~$\pi_1^-$ in the helicity rest frame of this isobar.  We calculate
the magnitude of this vector using the two-body breakup momentum
\begin{multlineOrEq}
  q^2(m, m_1, m_2) \newLineOrNot
  = \frac{\sBrk{m^2 - (m_1 + m_2)^2}\, \sBrk{m^2 - (m_1 - m_2)^2}}{4 m^2}.
\end{multlineOrEq}
Thus $\abs{\vec{p}_1^{\,\HF}} = q_{1 3}$, where
\begin{equation}
  \label{eq:breakupTwo}
  q_{i j}
  \equiv q(m_{i j}, m_\pi, m_\pi)
\end{equation}
is the breakup momentum between pions~$i$ and~$j$ with $m_{i j}$~being
the invariant mass of the two-pion system.  Using the above equations,
we express the~$x$ and $y$~components of~$\vec{p}_1^{\,\HF}$ using the
helicity angles as spherical coordinates:
\begin{align}
  \label{eq:HF_angles_x}
  q_{1 3}\, \cos\phiHF\, \sin\thetaHF
  &= p_{1, x}^\HF \ifMultiColumnLayout{\nonumber \\}{}
  \alignOrNot= \frac{\cos\thetaGJbose}{\sin\thetaGJ}\, Q_{2 3}
  - \frac{\cos\thetaGJ}{\sin\thetaGJ}\, \frac{\vec{p}_{1 3} \cdot \vec{p}_{2 3}}{Q_{13}},
  \\
  \label{eq:HF_angles_y}
  q_{1 3}\, \sin\phiHF\, \sin\thetaHF
  &= p_{1, y}^\HF \ifMultiColumnLayout{\nonumber \\}{}
  \alignOrNot= Q_{2 3}\, \sin\thetaGJbose\, \sin(\phiGJ - \phiGJbose).
\end{align}
Here, angles with a hat~(\textquote{\^{}}) indicate the
Bose-symmetrized system, where the isobar decays into $\pi^-_2
\pi^+_3$.  The $\vec{p}_{i j}$ represent the sums of the momenta of
particles~$i$ and~$j$ in the $3\pi$ center-of-momentum system and the
$Q_{i j}$ are the two-body breakup momenta of the $3\pi$ system given
by
\begin{equation}
  \label{eq:breakupThree}
  Q_{i j}
  \equiv q(\mThreePi, m_{i j}, m_\pi).
\end{equation}
The right-hand sides of \cref{eq:HF_angles_x,eq:HF_angles_y} are
obtained from transforming the four-momentum vector of~$\pi^+_1$ from
the Gottfried-Jackson frame into the helicity frame of the isobar.
This calculation can be found in
\ifMultiColumnLayout{Appendix~D of the Supplemental Material
of this paper~\cite{paper4_supplemental_material}}{the Supplemental
Material in \cref{suppl:zeroMode}}

The angular amplitude $\angularPart_a(\tau)$ in \cref{eq:K-1} depends
on the helicity angles with the coordinate system in the helicity rest
frame depending on the particles forming the isobar.
\Cref{eq:HF_angles_x,eq:HF_angles_y} allow us to replace the
dependencies of $\angularPart_a(\tau)$ on the helicity angles by
expressions depending only on Gottfried-Jackson angles for the two
combinations of the final-state particles that correspond to Bose
symmetrization.  The coordinate system in the Gottfried-Jackson rest
frame does not depend on the particles forming the
isobar.\footnote{Therefore, $(\thetaGJ, \phiGJ)$ and $(\thetaGJbose,
\phiGJbose)$ in \cref{eq:boseMinus} are two different sets of angles
defined in the same coordinate system.}  Expressing in addition the
scalar product $\vec{p}_{1 3} \cdot \vec{p}_{2 3}$ in
\cref{eq:HF_angles_x} in terms of the Gottfried-Jackson angles, we
obtain
\begin{multlineOrEq}
  \label{eq:boseMinus}
  \angularPart_a(\tau_{1 3})
  \propto \frac{Q_{2 3}}{q_{1 3}}\,
  \big( \cos\phiGJ\, \sin\thetaGJ\, \cos\thetaGJbose \newLineOrNot
  - \cos\phiGJbose\, \sin\thetaGJbose\, \cos\thetaGJ \big),
\end{multlineOrEq}
where $\tau_{1 3}$~is the set of phase-space variables for the isobar
decaying into $\pi_1^- \pi_3^+$.  Performing the Bose symmetrization
and including the dynamic isobar amplitude~$\Delta(m_{i j})$, we find
for the decay amplitude
\begin{equation}
  \Psi_a(\tau_{1 3}, \tau_{2 3})
  = \angularPart_a(\tau_{1 3})\, \Delta(m_{1 3}) + \angularPart_a(\tau_{2 3})\, \Delta(m_{2 3}).
\end{equation}
This amplitude exactly vanishes at every point in phase space if the
dynamic amplitude $\Delta(m_{i j})$ has the form
\begin{equation}
  \label{eq:zero_mode_shape}
  \tilde{\Delta}(m_{i j})
  = Q_{i j}\, q_{i j}.
\end{equation}
This is because the two terms in the bracket in \cref{eq:boseMinus}
are Bose-symmetrized versions of each other and thus
\begin{equation}
  \angularPart_a(\tau_{1 3})
  = -\frac{Q_{2 3}}{q_{1 3}}\, \frac{q_{2 3}}{Q_{1 3}}\, \angularPart_a(\tau_{2 3}).
\end{equation}
Consequently, changing the dynamic isobar amplitude according to
\begin{equation}
  \label{eq:shift}
  \Delta(m_{i j})
  \to \Delta(m_{i j}) + \zeroModeCoefficient\, \tilde{\Delta}(m_{i j})
\end{equation}
with an arbitrary complex-valued coefficient~\zeroModeCoefficient does
not alter the decay amplitude and hence also leaves the intensity as
well as the likelihood function unchanged.  Therefore, the
coefficient~\zeroModeCoefficient represents a mathematical ambiguity,
or zero mode, in the PWA model that is defined by the real-valued
zero-mode shape $\tilde{\Delta}(m_{i j})$.  In the conventional PWA,
this ambiguity does not appear owing to the fixed parametrization of
the dynamic isobar amplitudes.  However, in a freed-isobar PWA
a shift in the direction of the zero mode is possible due to the
freedom in the dynamic isobar amplitudes.

Let~$\{\prodAmp^\text{fit}_k\}$ be the set of binned transition
amplitudes as defined in \cref{eq:dyn_amp_freed,eq:prod_amp_freed}
that are extracted in a freed-isobar PWA, with~$k$ being the index of
the \mTwoPi~interval.  The~$\{\prodAmp^\text{fit}_k\}$ might be shifted away
from their physical values~$\{\prodAmp^\text{phys}_k\}$ in the
direction of the zero mode $\tilde{\Delta}(m_{i j})$ according
to \cref{eq:shift}.  Thus for each \mTwoPi~interval,
the center of which is represented by~$m_k$, we have\footnote{Since
the shape of the zero mode is multiplied with an arbitrary
coefficient, we normalize it to $\sum_k \binnedZeroMode_k^2 = 1$.}
\begin{equation}
  \label{eq:shift_prodAmp}
  \prodAmp^\text{fit}_k
  = \prodAmp^\text{phys}_k + \zeroModeCoefficient\, \binnedZeroMode_k
  \quad\text{with}\quad
  \binnedZeroMode_k
  \equiv \tilde{\Delta}(m_k).
\end{equation}

Since the ambiguity represented by~\zeroModeCoefficient in
\cref{eq:shift_prodAmp} leaves the intensity and therefore the
likelihood function unchanged, it can only be resolved by imposing
prior knowledge about the dynamic isobar amplitudes.\footnote{This is
conceptually similar to gauge fixing in gauge symmetries.}  For the
$1^{-+}$ wave, we can safely assume the $\{\prodAmp^\text{phys}_k\}$
to contain the \Prho resonance.  We resolve the zero-mode ambiguity by
determining that value of~\zeroModeCoefficient that minimizes the
deviation of~$\{\prodAmp^\text{fit}_k\}$ from the \Prho Breit-Wigner
amplitude.  The deviation is measured by the residuals
\begin{equation}
  \label{eq:residuals}
  \delta_k(\zeroModeCoefficient)
  \equiv \prodAmp_k^\text{fit} - \zeroModeCoefficient\, \binnedZeroMode_k - \hat{\prodAmp}_a(m_k; \mThreePi, \tpr),
\end{equation}
where $\hat{\prodAmp}_a(m_k; \mThreePi, \tpr)$ is the Breit-Wigner
amplitude for the \Prho resonance as given by
\cref{eq:bw_model_isobar}.  To determine~\zeroModeCoefficient and
thereby resolve the zero-mode ambiguity, we minimize the Mahalanobis
distance\footnote{Since the~$\delta_k$ are complex-valued, while the
covariance matrix describes real-valued quantities, the sum over the
\mTwoPi~interval indices~$k$ and~$l$ runs implicitly also over the
real and imaginary parts of the corresponding amplitudes (see also
Sec.~IV~B in \refCite{Akhunzyanov:2018lqa}).}
\begin{equation}
  \label{eq:zero_mode_chi2}
  \chi^2(\zeroModeCoefficient)
  = \sum_{k, l} \delta_k(\zeroModeCoefficient)\, \coma_{k l}^{-1}\, \delta_l(\zeroModeCoefficient),
\end{equation}
where \coma is the covariance matrix of the
~$\{\prodAmp_k^\text{fit}\}$ (see \cref{sec:coma_manipulations}).
Since we cannot exclude contributions from excited \Prho*~states at
higher~\mTwoPi, we limit the sum in \cref{eq:zero_mode_chi2} to those
\mTwoPi~intervals~$k$ and~$l$ where $\mTwoPi <
\SI{1.12}{\GeVcc}$.\footnote{The covariance matrix is first cut to
this \mTwoPi~range and then inverted.}

Imposing the \Prho Breit-Wigner amplitude in \cref{eq:residuals} by
minimizing the $\chi^2$~function in \cref{eq:zero_mode_chi2}, we
determine only one complex-valued degree of freedom, \ie
\zeroModeCoefficient, from the data.  The real-valued zero-mode shape
$\tilde{\Delta}(m_{i j})$ that enters \cref{eq:zero_mode_chi2} is
fixed and given by \cref{eq:zero_mode_shape}.  It is important to note
that this procedure does not lead to a circular argument, \ie we do
not get out what we put in.  This is because our procedure that
resolves the zero-mode ambiguity cannot artificially generate a fake
\Prho resonance signal in the \twoPi isobar amplitude, \ie a circular
structure in the \Argand, if the \Prho is not contained in the data.
This has been verified through Monte Carlo
studies~\cite{Krinner:2017dba,Krinner:2018bwg}.  In other words, we do
not demand that the dynamic isobar amplitude is described entirely by
the \Prho as in the conventional PWA method.  Instead, we only
impose the \Prho to be a part of the amplitude.  We hence fix a single
complex value, \ie\ \zeroModeCoefficient, using minimal assumptions on
the shape of the dynamic isobar amplitude.  This still leaves $2 (n -
1)$ complex-valued degrees of freedom, where $n$~is the number of
\mTwoPi~intervals, that are determined from the data.  The
freed-isobar PWA thus yields much more information on the dynamic
isobar amplitude in the \wave{1}{-+}{1}{+}{\pipiPF}{P} wave than the
conventional PWA method, which only returns a single complex-valued
parameter, \ie the transition amplitude, for this wave.

In \cref{fig:fixingSpinExotic}, we show the result of the freed-isobar
PWA (green data points) and the corresponding zero-mode corrected
amplitude (blue data points) for an exemplary $(\mThreePi, \tpr)$
cell.  The green points are shifted away from their physical value by
the zero mode with an accidental value for the
coefficient~\zeroModeCoefficient, which is different in every
$(\mThreePi, \tpr)$ cell.  The gray histogram
represents the \Prho Breit-Wigner
amplitude~$\hat{\prodAmp}_a(m_k; \mThreePi, \tpr)$ used in
\cref{eq:residuals,eq:zero_mode_chi2} to resolve the zero-mode
ambiguity.  Since the zero-mode shape given in
\cref{eq:zero_mode_shape} is real valued, all green points in the
Argand diagram are shifted in parallel directions to the blue points.
It should be noted that only data for $\mTwoPi < \SI{1.12}{\GeVcc}$
were used to resolve the ambiguity.  The resulting zero-mode corrected
amplitude agrees well with the \Prho Breit-Wigner also in the mass
region $\mTwoPi > \SI{1.12}{\GeVcc}$.  This method to resolve the
ambiguity is used to produce \cref{fig:twoDplots,fig:slices}.  Note
that the zero mode does not influence the \mThreePi~intensity
distributions obtained by summing the contributions of the
freed-isobar transition amplitudes $\{\prodAmp_{a, k}\}$ from all
\mTwoPi~intervals coherently (see \cref{fig:coherent}).

\begin{figure}[tbp]
  \centering%
  \subfloat[][]{%
    \includegraphics[width=\twoPlotWidth]{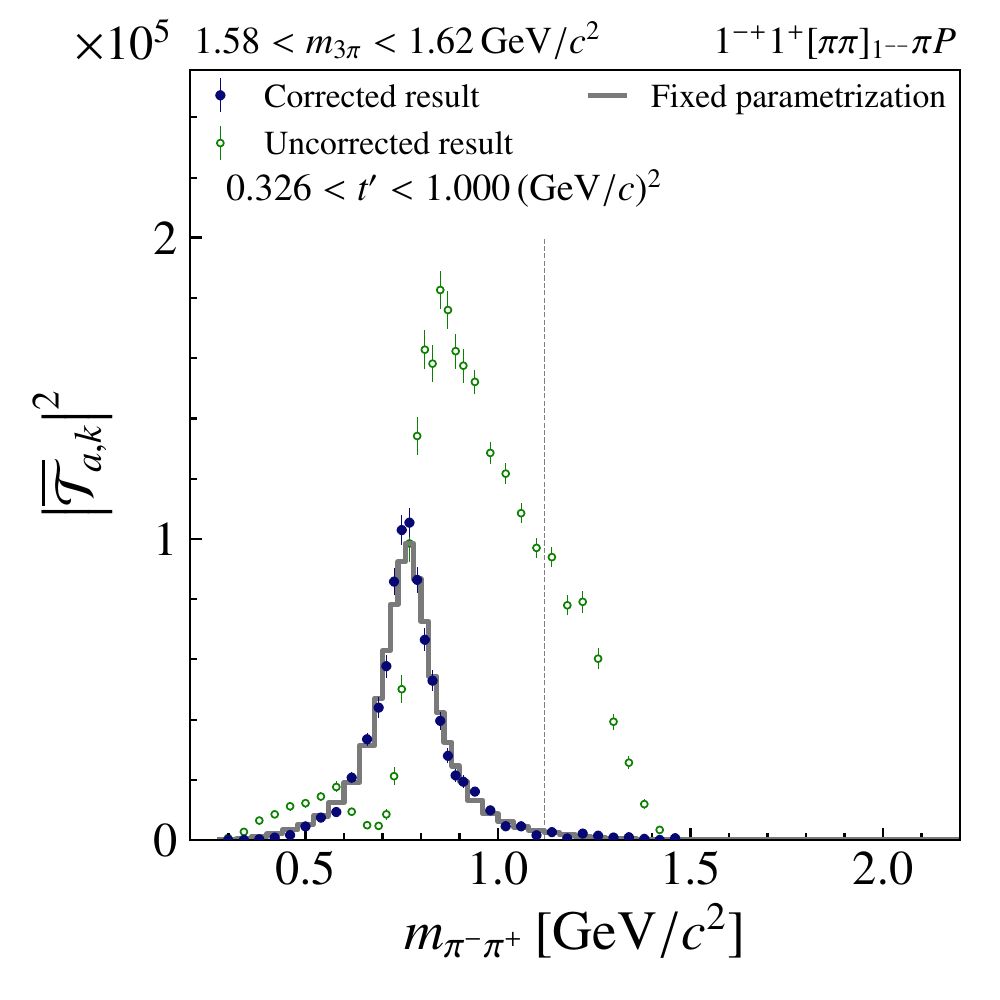}%
    \label{fig:fixingSpinExotic_int}%
  }%
  \newLineOrHspace{\twoPlotSpacing}%
  \subfloat[][]{%
    \includegraphics[width=\twoPlotWidth]{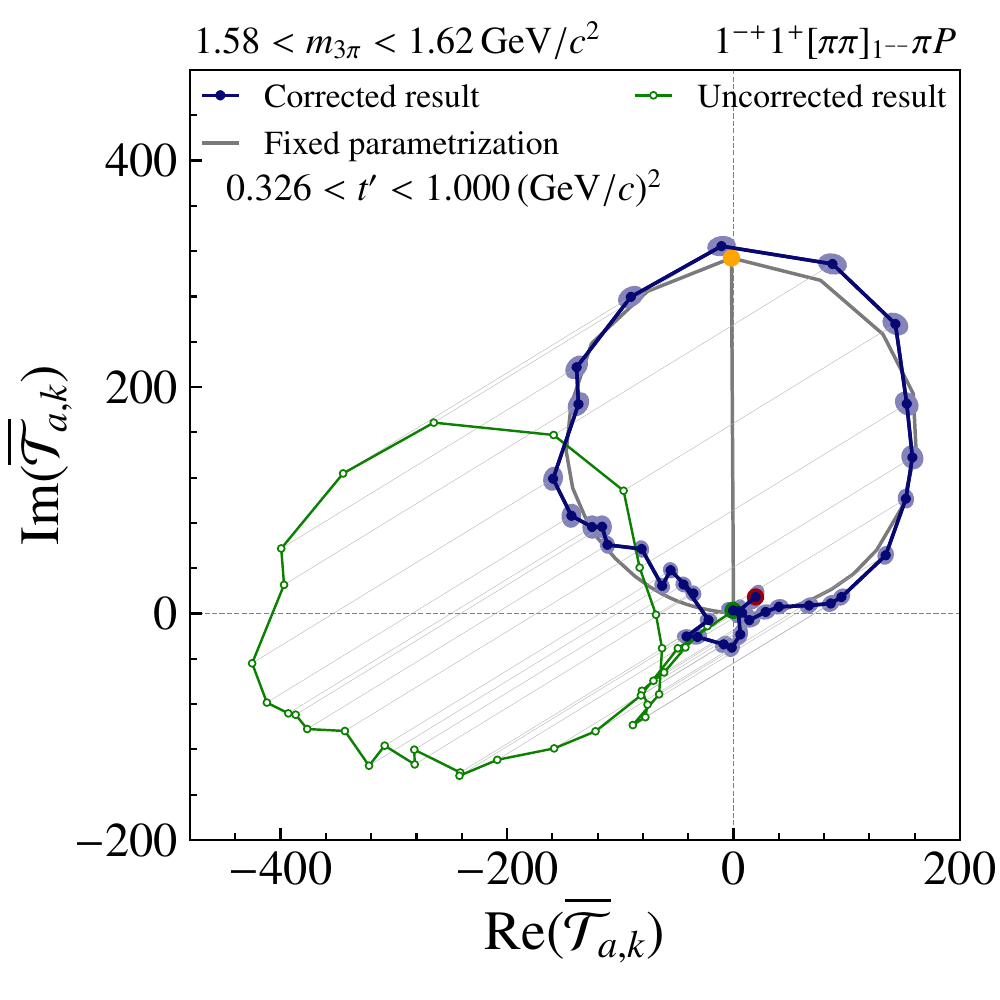}%
    \label{fig:fixingSpinExotic_argand}%
  }%
  \caption{Similar to \cref{fig:slices} but showing the effect of
    resolving the zero-mode ambiguity in the
    \wave{1}{-+}{1}{+}{\pipiPF}{P} wave using, as an example, the
    highest \tpr~bin and an \mThreePi~bin in the \PpiOne[1600]
    resonance region.  \subfloatLabel{fig:fixingSpinExotic_int}~shows
    the intensity distribution and
    \subfloatLabel{fig:fixingSpinExotic_argand}~the corresponding
    \Argand as a function of~\mTwoPi.  The green data points
    correspond to the set~$\{\prodAmp^\text{fit}_k\}$ of binned
    transition amplitudes [defined in
    \cref{eq:dyn_amp_freed,eq:prod_amp_freed}] as extracted in the
    freed-isobar PWA\@.  The blue data points represent the transition
    amplitudes $\{\prodAmp^\text{fit}_k - \zeroModeCoefficient\,
    \binnedZeroMode_k\}$ after resolving the zero-mode ambiguity by
    minimizing \cref{eq:zero_mode_chi2}.  The upper limit
    $\mTwoPi=\SI{1.12}{\GeVcc}$ of the range used in this minimization
    is indicated by the vertical line in
    \subfloatLabel{fig:fixingSpinExotic_int}.  The gray histogram
    represents the \Prho Breit-Wigner amplitude used to resolve the
    zero-mode ambiguity.  In the \Argand, corresponding green and blue
    data points are connected by gray lines representing the
    $\{\zeroModeCoefficient\, \binnedZeroMode_k\}$ values.}%
  \label{fig:fixingSpinExotic}%
\end{figure}

We have studied several constraints to resolve the zero-mode
ambiguity.  In a first study, we extend the fit range to the full
kinematically allowed \mTwoPi~range.  In a second study, we allow the
\Prho resonance parameters to float in the fit that resolves the
ambiguity.  In this approach, we either use a single set of \Prho
resonance parameters while simultaneously fitting all $(\mThreePi,
\tpr)$ cells or we allow for different \Prho resonance parameters in
each $(\mThreePi, \tpr)$ cell.  The resulting \Prho resonance
parameters obtained using the above approaches are discussed in
\cref{sec::fipwa_exotic}.  All these approaches yield similar results
for the zero-mode corrected amplitudes (see also discussion in
\cref{sec:freed_isobar}).

In a third study, we try to resolve the zero mode by minimizing the
variation of the zero-mode corrected amplitudes between neighboring
\mThreePi~bins.  This means we do not make any assumption on the
\mTwoPi~dependence of the amplitudes and just require a smooth
behavior as a function of~\mThreePi.  The resulting zero-mode
corrected amplitudes still exhibit a clear \Prho resonance signal but
the method induces a considerable bias towards small intensities of
the dynamic isobar amplitudes across all \mThreePi~and \mTwoPi~bins.
Because of this observed bias, we do do not use this approach.

\section{Preparation of the covariance matrix}%
\label{sec:coma_manipulations}

In the freed-isobar PWA, the zero-mode
coefficient~\zeroModeCoefficient in \cref{eq:shift_prodAmp} is a
nuisance parameter that contains no physical information and is not
constrained by the data (see \cref{app:zeroMode}).  But
since~\zeroModeCoefficient mixes with all other fit parameters, it
influences the uncertainties of these parameters.  We thus want to
remove the corresponding uncertainties from the covariance
matrix~\coma of the extracted transition
amplitudes~$\{\prodAmp_k^\text{fit}\}$ that we obtain from the
minimizing algorithm.  Since the zero-mode $\tilde{\Delta}(m_{i j})$
in \cref{eq:zero_mode_shape} is real valued, the ambiguity does not
mix real and imaginary parts of the transition amplitudes.  We
therefore remove the uncertainty that corresponds to the zero-mode
individually from the covariance matrices~$\coma_{\!\Re}$
and~$\coma_{\!\Im}$ of real and imaginary parts of the transition
amplitudes.  Those entries of the covariance that mix real and
imaginary parts of the~$\{\prodAmp_k^\text{fit}\}$ are unaffected by
the zero mode.  To this end, we define a projection
operator~$\mathbf{P}$ that acts on the covariance matrix according to
\begin{equation}
  \label{eq:intermediary_coma}
  \coma_{\!\Re,\, \Im} \to \mathbf{P} \cdot \coma_{\!\Re,\, \Im} \cdot \mathbf{P}.
\end{equation}
This projection operator is an $n \times n$ real-valued matrix, where
$n$~is the number of \mTwoPi~intervals.  It is defined as
\begin{equation}
  \label{eq:projection_op}
  \mathbf{P}_{k l}
  \equiv \delta_{k l} - \binnedZeroMode_k\, \binnedZeroMode_l,
\end{equation}
so that
\begin{equation}
  \label{eq:projection_op_prop}
  \mathbf{P} \cdot \vec\binnedZeroMode
  = \vec{0},
  \quad
  \mathbf{P} \cdot \vec\binnedZeroMode_\perp
  = \vec\binnedZeroMode_\perp,
  \quad\text{and}\quad
  \mathbf{P} \cdot \mathbf{P}
  = \mathbf{P}.
\end{equation}
Here, $\vec\binnedZeroMode_\perp$ is an arbitrary direction in
the space of the~$\{\prodAmp_k^\text{fit}\}$ that is orthogonal to
the zero mode~$\vec\binnedZeroMode$.\footnote{$\vec\binnedZeroMode$~is
the vector that has the~$\{\binnedZeroMode_k\}$ as components.}

By construction, the covariance matrices~$\coma_{\!\Re,\, \Im}$ in
\cref{eq:intermediary_coma}, from which we removed the uncertainties
corresponding to the zero mode, have eigenvectors in the direction of
the zero mode with an eigenvalue of zero.  However, using these
matrices in \cref{eq:zero_mode_chi2} would render the
$\chi^2$~function completely independent of the zero-mode
coefficient~\zeroModeCoefficient so that we would not be able to
determine~\zeroModeCoefficient by minimizing this function.  We hence
reinsert the zero mode as an eigenvector weighted with an arbitrary
positive coefficient~\comaCoefficient, \ie we perform the
transformation
\begin{equation}
  \coma_{\!\Re,\, \Im}
  \to \mathbf{P} \cdot \coma_{\!\Re,\, \Im} \cdot \mathbf{P}
  + \comaCoefficient\, \vec\binnedZeroMode \otimes \vec\binnedZeroMode.
\end{equation}
Doing so, we ensure that the zero mode~$\vec\binnedZeroMode$ is an
exact eigenvector of the covariance matrix~\coma and therefore
the determination of~\zeroModeCoefficient is independent from the
determination of all other fit parameters in \cref{eq:zero_mode_chi2}.
For this reason, the solutions for the zero-mode
coefficient~\zeroModeCoefficient and the other fit parameters in
\cref{eq:residuals} are also independent of the particular choice for
the value of~\comaCoefficient.  We verified numerically that this
holds over 17~orders of magnitude for the value of~\comaCoefficient.
\section*{Acknowledgements}%
\label{sec:acknowledgements}

We gratefully acknowledge the support of the CERN management and staff
as well as the skills and efforts of the technicians of the
collaborating institutions.  We also gratefully acknowledge the
computing resources provided by the Computational Center for Particle
and Astrophysics (C2PAP).  This work was made possible by the
financial support of our funding agencies:
BMBF - Bundesministerium f\"ur Bildung und Forschung (Germany);
DFG - German Research Foundation, Transregional Research Center TR110 (Germany);
FP7, HadronPhysics3, Grant 283286 (European Union);
MEYS, Grant LM20150581 (Czech Republic);
B. Sen fund (India);
CERN-RFBR Grant 12-02-91500;
FCT, Grants CERN/FIS-PAR/0007/2017 and  CERN/FIS-PAR/0022/2019 (Portugal);
MEXT and JSPS, Grants 18002006, 20540299, 18540281 and 26247032, the Daiko and Yamada Foundations (Japan);
the Ministry of Science and Technology (Taiwan);
the Israel Academy of Sciences and Humanities (Israel);
Tomsk Polytechnic University Competitiveness Enhancement Program (Russia);
the National Science Foundation, Grant no. PHY-1506416 (USA);
NCN, Grant 2017/26/M/ST2/00498 (Poland);
C. C., M. H., A. Kerbizi and T. T. were supported by the European Union's Horizon 2020 research and innovation programme under grant agreement STRONG-2020-No 824093;
W. D. and M. Faessler were supported by the DFG cluster of excellence 'Origin and Structure of the Universe' (Germany);
M. Gorzellik was supported by the DFG Research Training Group Programmes 1102 and 2044 (Germany);
P.-J. L. was supported by ANR, France with P2IO LabEx (ANR-10-LBX-0038) in the framework "Investissements d'Avenir" (ANR-11-IDEX-003-01).

\bibliographystyle{utphys_bgrube}
\providecommand{\href}[2]{#2}\begingroup\raggedright\endgroup
\clearpage
\section*{\protect\huge\protect\textsc{Supplemental Material}}
\addcontentsline{toc}{section}{\protect\textsc{Supplemental Material}}
\clearpage
\clearpage{}%
In this Supplemental Material\ifMultiColumnLayout{ to
\refCite{paper4}}{}, we provide additional information.

In \cref{suppl:zeroMode}, we give details on the calculation of the
zero mode in the spin-exotic \wave{1}{-+}{1}{+}{\pipiPF}{P} wave that
is discussed in \ifMultiColumnLayout{Appendix~B of
\refCite{paper4}}{\cref{app:zeroMode}}.  In
\cref{suppl:freed_isobar_comp}, we show the comparison between
freed-isobar and conventional PWA similar to
\ifMultiColumnLayout{Figs.~8 and~9 in
\refCite{paper4}}{\cref{fig:coherent,fig:modeledComparison}} for the
\tpr~bins that are not shown in the paper.  Finally, in
\cref{suppl:freed_isobar_dyn_amp}, we show the \pipiPF dynamic isobar
amplitude in the \wave{1}{-+}{1}{+}{\pipiPF}{P} wave as a function
of~\mTwoPi similar to \ifMultiColumnLayout{Fig.~6 in
\refCite{paper4}}{\cref{fig:slices}} for all $(\mThreePi, \tpr)$ cells
the freed-isobar PWA was performed in.

\section{Ambiguity in the $\JPC = 1^{-+}$ amplitude in the freed-isobar partial-wave analysis}%
\label{suppl:zeroMode}

In order to derive \ifMultiColumnLayout{Eqs.~(B7) and~(B8) in
Appendix~B of \refCite{paper4}}{\cref{eq:HF_angles_x,eq:HF_angles_y}},
we consider the decay of a particle~$X$ into three particles~1, 2,
and~3.  We want to express momentum vectors defined in the helicity
rest frame~(HF) of the (13)~two-particle subsystem using variables
defined in the Gottfried-Jackson rest frame~(GJ) of the
(123)~system~$X$.

In the Gottfried-Jackson frame,
\begin{equation}
  \label{eq:mom_sum_GJ}
  \vec{p}_1 + \vec{p}_2 + \vec{p}_3
  = \vec{0},
\end{equation}
where $\vec{p}_i$ is the three-momentum of particle~$i$.  Since we
want to transform from the Gottfried-Jackson into the (13)~helicity
frame, we consider the three-momentum vector
\begin{equation}
  \vec{p}_{13}
  = \vec{p}_1 + \vec{p}_3
\end{equation}
of the (13)~subsystem in the Gottfried-Jackson frame.  The $z$~axis of
the (13)~helicity frame is given by the direction of~$\vec{p}_{13}$,
\ie
\begin{equation}
  \label{eq:ezHF}
  \ezHF
  = \frac{\vec{p}_{13}}{p_{13}},
\end{equation}
where $p_{13} \equiv \Abs{\vec{p}_{13}}$.  The $y$~axis is given by
\begin{equation}
  \label{eq:eyHF}
  \eyHF
  = \frac{\ezGJ \times \ezHF}{\Abs{\ezGJ \times \ezHF}}
\end{equation}
and the $x$~axis completes the right-handed orthonormal basis, \ie
\begin{equation}
  \label{eq:exHF}
  \exHF
  = \eyHF \times \ezHF.
\end{equation}
While the normalization of the $x$~and $z$~unit vectors of the
helicity frame is trivial, we find the normalization of the $y$~unit
vector to be
\begin{equation}
  \label{eq:norm_eyHF}
  \Abs{\ezGJ \times \ezHF}
  = \sin\thetaGJ.
\end{equation}
Here, \thetaGJ is the polar angle of~$\vec{p}_{13}$ in the
Gottfried-Jackson frame; the corresponding azimuthal angle is \phiGJ.

Using \cref{eq:eyHF,eq:norm_eyHF}, we can rewrite \cref{eq:exHF}:
\begin{align}
  \exHF
  &= \frac{1}{\sin\thetaGJ}\, \rBrk{\ezGJ \times \ezHF} \times \ezHF \nonumber \\
  &= \frac{1}{\sin\thetaGJ}\, \sBrk{ \rBrk{\ezGJ \cdot \ezHF}\, \ezHF - \rBrk{\ezHF \cdot \ezHF}\, \ezGJ} \nonumber \\
  \label{eq:exHF_2}
  &= \frac{\cos\thetaGJ}{\sin\thetaGJ}\, \ezHF - \frac{1}{\sin\thetaGJ}\, \ezGJ.
\end{align}
Based on the expressions for the basis vectors of the helicity frame
in the Gottfried-Jackson frame in
\cref{eq:exHF_2,eq:eyHF,eq:norm_eyHF,eq:ezHF}, we can write down the
corresponding rotation matrix~$\mathbf{R}$.  The rows of this matrix
are given by the helicity-frame basis vectors, \ie
\begin{align}
  \mathbf{R}
  &= \rBrk{\exHF, \eyHF, \ezHF}^\intercal \nonumber \\
  \label{eq:orth_trafo}
  &= \rBrk{\frac{\cos\thetaGJ}{\sin\thetaGJ}\, \frac{\vec{p}_{13}}{\Abs{\vec{p}_{13}}} - \frac{1}{\sin\thetaGJ}\ezGJ,
    \frac{\ezGJ \times \vec{p}_{13}}{\sin\thetaGJ\, \Abs{\vec{p}_{13}}},
    \frac{\vec{p}_{13}}{\Abs{\vec{p}_{13}}}}^\intercal.
\end{align}
Applying this matrix to a momentum vector~$\vec{p}$ defined in the
Gottfried-Jackson frame yields a vector
\begin{equation}
  \label{eq:rot_HF}
  \vec{p}\,'
  = \mathbf{R} \cdot \vec{p}
\end{equation}
that is defined in a rotated reference frame that differs from the
(13)~helicity frame merely by a Lorentz boost along
$\vec{p}_{13}$~direction into the rest frame of the (13)~subsystem.
Since the $\vec{p}_{13}$~direction corresponds to the $z$~direction in
the rotated frame, the boost does not affect the $x$~and
$y$~components of~$\vec{p}\,'$ and hence the corresponding components
of the vector~$\vec{p}\,^\HF$ defined in the helicity frame are given
by
\begin{equation}
  \label{eq:v_boost_trans}
  p_x^\HF
  = p_x'
  \quad\text{and}\quad
  p_y^\HF
  = p_y'.
\end{equation}
However, the boost mixes the $z$~component~$p_z'$ and the energy~$E$
that corresponds to~$\vec{p}$ and~$\vec{p}\,'$.  Therefore,
\begin{equation}
  \label{eq:v_boost_z}
  p_z^\HF
  = \rBrk{\frac{E_{13}}{m_{13}}\, p_z' - \frac{p_{13}}{m_{13}}\, E},
\end{equation}
where $\gamma = E_{13} / m_{13}$ and $\beta \gamma = p_{13} / m_{13}$
are the usual relativistic factors of the Lorentz boost.  Here,
$E_{13}$ and~$p_{13}$ are energy and momentum of the (13)~subsystem in
the Gottfried-Jackson frame and~$m_{13}$ is the corresponding
invariant mass.  Using
\cref{eq:orth_trafo,eq:rot_HF,eq:v_boost_trans,eq:v_boost_z}, we can
write all components of $\vec{p}\,^\HF$:
\begin{align}
  \label{eq:vxHF}
  p_x^\HF
  &= \frac{\cos\thetaGJ}{\sin\thetaGJ}\, \frac{\vec{p}_{13} \cdot \vec{p}}{p_{13}} - \frac{1}{\sin\thetaGJ}\, p_z \\
  \label{eq:vyHF}
  p_y^\HF
  &= \frac{\rBrk{\ezGJ \times \vec{p}_{13}} \cdot \vec{p}}{\sin\thetaGJ\, p_{13}} \\
  \label{eq:vzHF}
  p_z^\HF
  &= \frac{E_{13}}{m_{13}}\, \frac{\vec{p}_{13} \cdot \vec{p}}{p_{13}} - \frac{p_{13}}{m_{13}}\, E.
\end{align}

In order to derive \ifMultiColumnLayout{Eqs.~(B7) and~(B8) in
Appendix~B of
\refCite{paper4}}{\cref{eq:HF_angles_x,eq:HF_angles_y}}, we
specifically need the expressions for~$p_{1, x}^\HF$ and~$p_{1,
y}^\HF$.  Due to \cref{eq:mom_sum_GJ}
\begin{equation}
  \vec{p}_1
  = -\vec{p}_2 - \vec{p}_3
  = -\vec{p}_{23}
\end{equation}
and in particular
\begin{equation}
  p_{1, z}
  = - p_{23, z}
  = -p_{23}\, \cosThetaGJbose,
\end{equation}
where~$p_{23}$ and~\thetaGJbose are the magnitude and the polar angle
of~$\vec{p}_{23}$ in the Gottfried-Jackson frame; the corresponding
azimuthal angle is~\phiGJbose.  Using the above relations and
\cref{eq:vxHF}, we find
\begin{equation}
  \label{eq:p1xHF}
  p_{1, x}^\HF
  = \frac{\cosThetaGJbose}{\sin\thetaGJ}\, p_{23} - \frac{\cosThetaGJ}{\sin\thetaGJ}\, \frac{\vec{p}_{13} \cdot \vec{p}_{23}}{p_{13}}.
\end{equation}
We make use of the properties of the combination of dot and the cross
products when calculating~$p_{1, y}^\HF$ via \cref{eq:vyHF}:
\begin{align}
  \rBrk{\ezGJ \times \vec{p}_{13}} \cdot \vec{p}_1
  &= -\rBrk{\ezGJ \times \vec{p}_{13}} \cdot \vec{p}_{23} \\
  &= -\rBrk{\vec{p}_{13} \times \vec{p}_{23}} \cdot \ezGJ \\
  &= \rBrk{\vec{p}_{23} \times \vec{p}_{13}} \cdot \ezGJ.
\end{align}
We thus need the $z$~component of $\vec{p}_{23} \times \vec{p}_{13}$,
which is given by
\begin{multlineOrEq}
  \rBrk{\vec{p}_{23} \times \vec{p}_{13}}_z
  = p_{13}\, p_{23}\, \sin\thetaGJ\, \sin\thetaGJbose\, \big( \cos\phiGJ\, \sin\phiGJbose \newLineOrNot
    - \sin\phiGJ\, \cos\phiGJbose \big).
\end{multlineOrEq}
This simplifies to
\begin{equation}
  \rBrk{\vec{p}_{23} \times \vec{p}_{13}}_z
  = p_{13}\, p_{23}\, \sin\thetaGJ\, \sin\thetaGJbose\, \sin\rBrk{\phiGJ - \phiGJbose},
\end{equation}
leading to
\begin{equation}
  \label{eq:p1yHF}
  p_{1, y}^\HF
  = p_{23}\, \sin\thetaGJbose\, \sin\rBrk{\phiGJ - \phiGJbose}.
\end{equation}
Since the momenta~$p_{13}$ and~$p_{23}$ are the breakup
momenta~$Q_{13}$ and~$Q_{23}$ as defined in
\ifMultiColumnLayout{Eq.~(B9) in Appendix~B of
\refCite{paper4}}{\cref{eq:breakupThree}}, \cref{eq:p1xHF,eq:p1yHF}
are identical to \ifMultiColumnLayout{Eqs.~(B7) and~(B8) in Appendix~B
of \refCite{paper4}}{\cref{eq:HF_angles_x,eq:HF_angles_y}}.

\clearpage
\ifMultiColumnLayout{\onecolumngrid}{}

\section{Comparison of the freed-isobar with the conventional partial-wave analysis}%
\label{suppl:freed_isobar_comp}

In \cref{fig:coherent2}, we show the comparison between the coherently
summed result of the freed-isobar PWA for the second and third
\tpr~bin (similar to \ifMultiColumnLayout{Fig.~8 in
\refCite{paper4}}{\cref{fig:coherent}}).

\begin{figure}[!h]
\centering%
  \subfloat[][]{%
    \includegraphics[width=\twoPlotWidth]{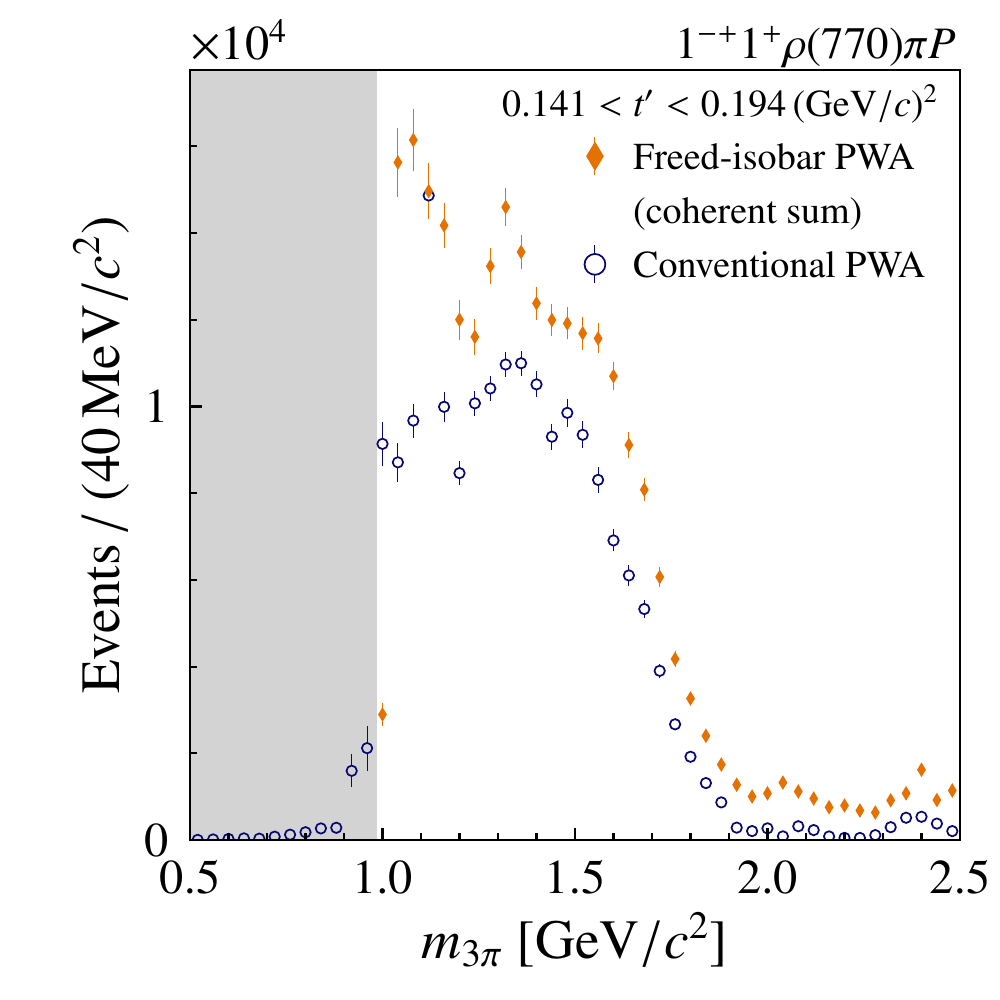}%
  }%
  \hspace*{\twoPlotSpacing}%
  \subfloat[][]{%
    \includegraphics[width=\twoPlotWidth]{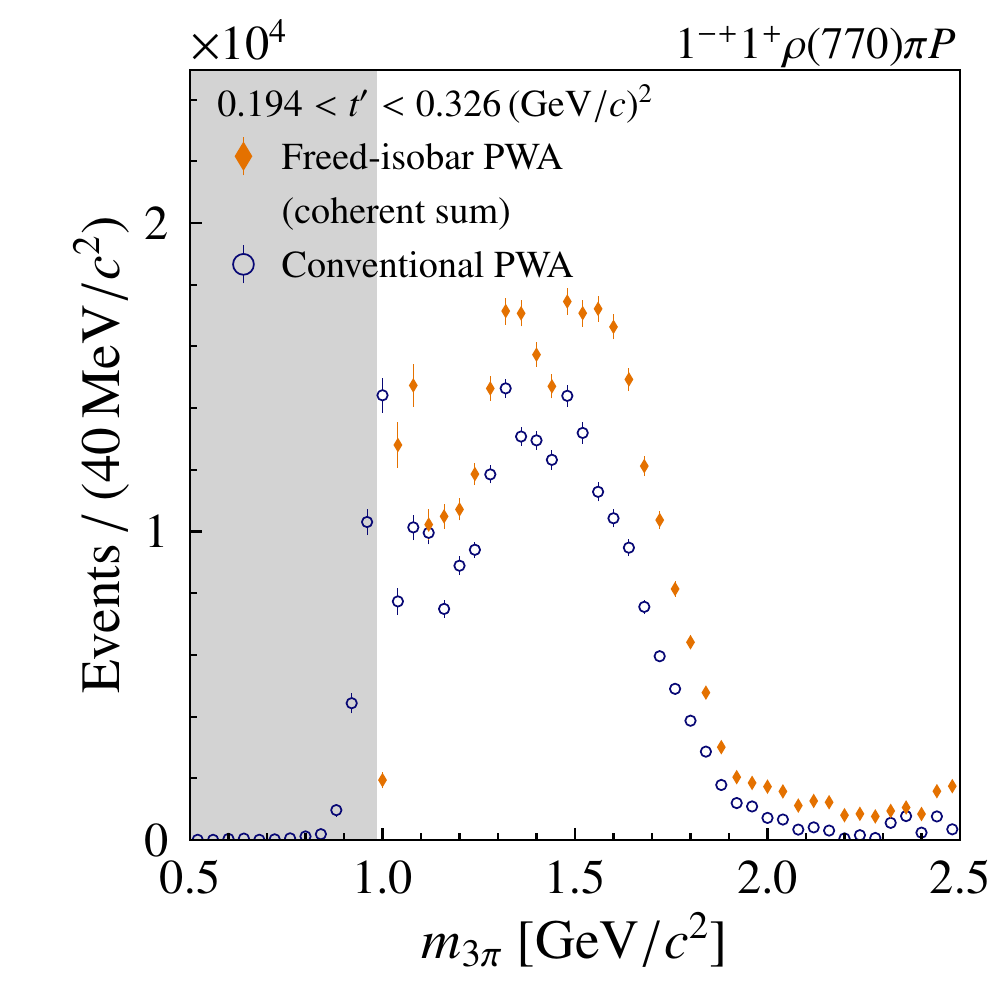}%
  } %
  \caption{Comparison of the \mThreePi~intensity distributions of the
    \wave{1}{-+}{1}{+}{\pipiPF}{P} wave from the freed-isobar PWA
    (orange data points) and of the \wave{1}{-+}{1}{+}{\Prho}{P} wave
    from the conventional PWA (blue data points), similar to
    \ifMultiColumnLayout{Fig.~8 in
    \refCite{paper4}}{\cref{fig:coherent}}.}%
  \label{fig:coherent2}
\end{figure}

In \cref{fig:modelComparison2}, we show the results of the
Breit-Wigner resonance-model fit to the results of the freed-isobar
PWA for the spin-exotic wave similar to \ifMultiColumnLayout{Fig.~9 in
\refCite{paper4}}{\cref{fig:modeledComparison}} for the three lowest
\tpr~bins.

\begin{figure}[!h]
  \centering%
  \subfloat[][]{%
    \includegraphics[width=\twoPlotWidth]{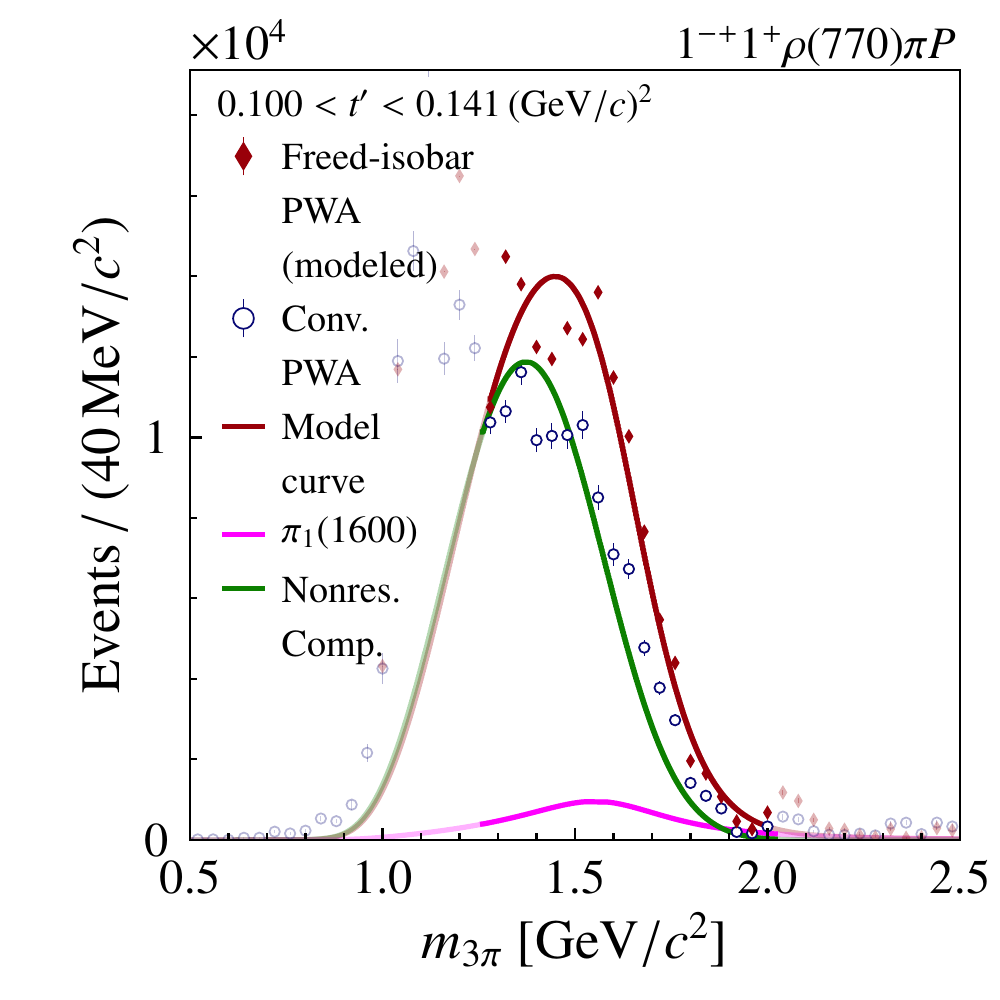}%
  }%
  \hspace*{\twoPlotSpacing}%
  \subfloat[][]{%
    \includegraphics[width=\twoPlotWidth]{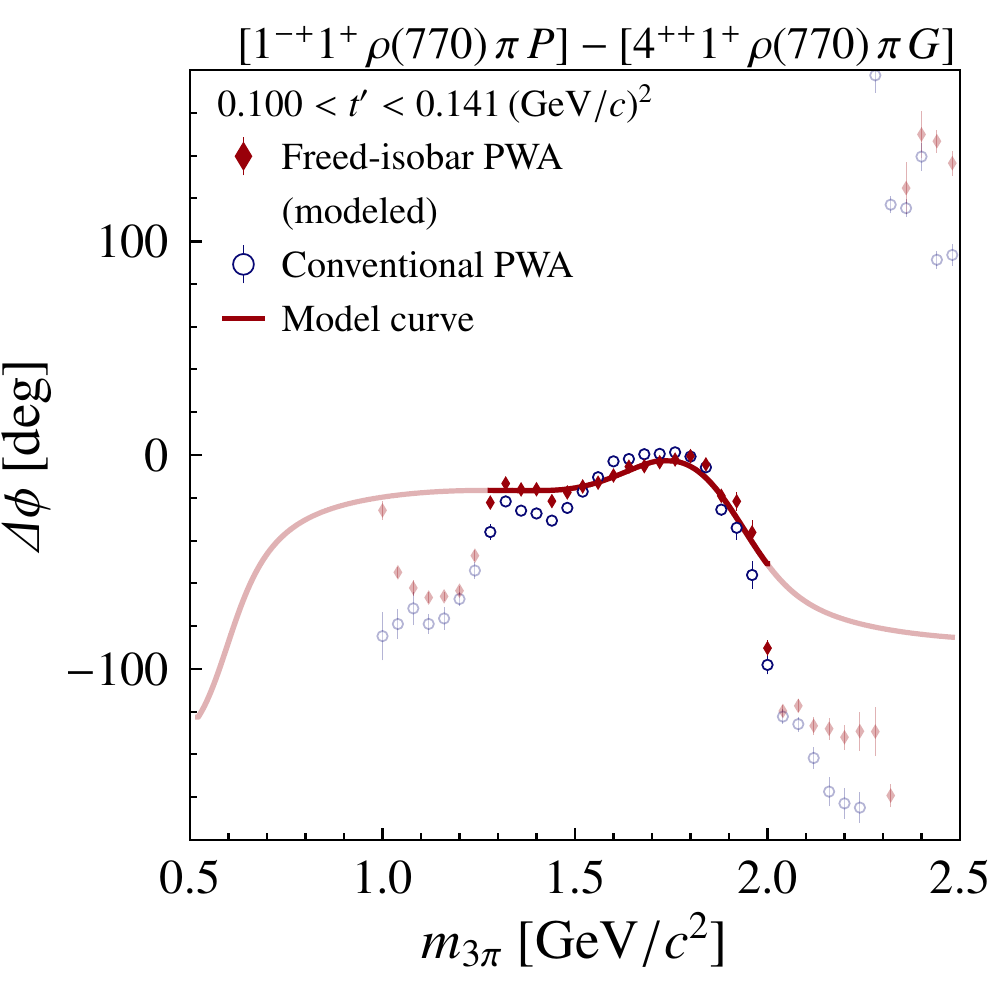}%
  }%
  \\%
  \subfloat[][]{%
    \includegraphics[width=\twoPlotWidth]{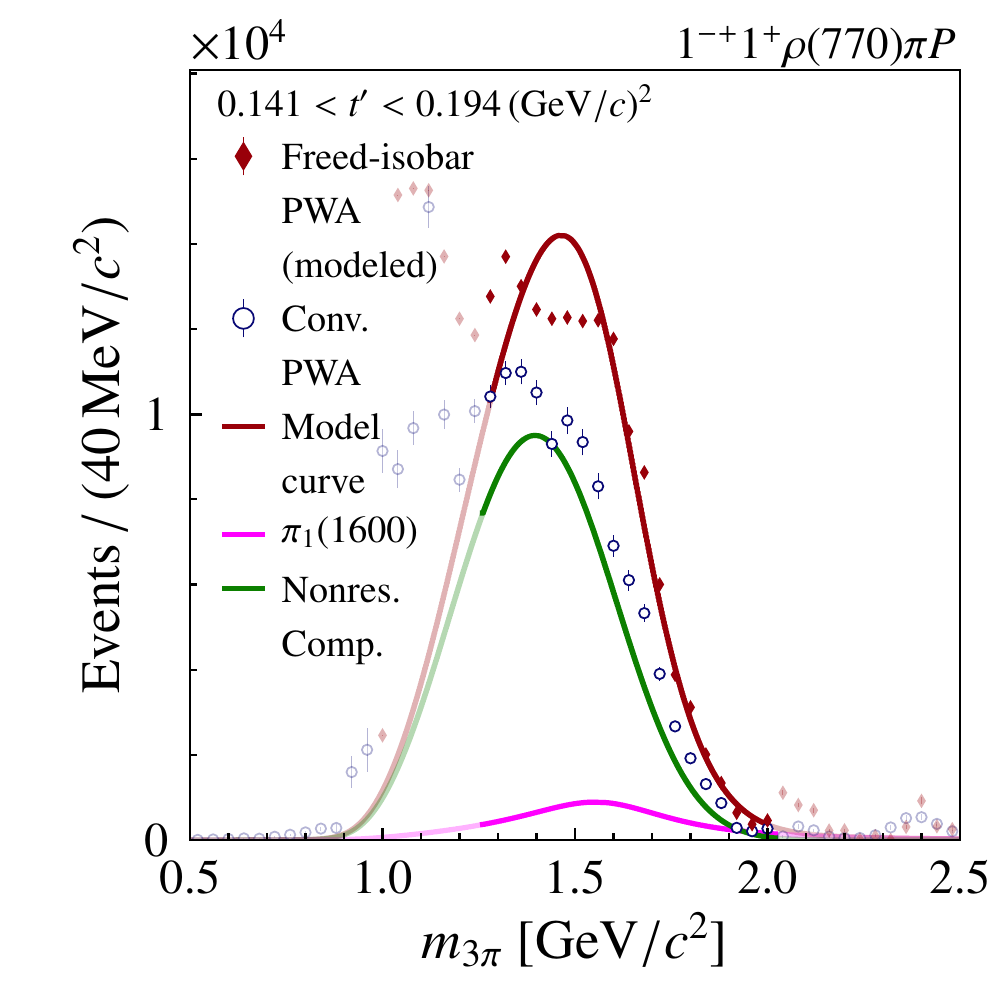}%
  }%
  \hspace*{\twoPlotSpacing}%
  \subfloat[][]{%
    \includegraphics[width=\twoPlotWidth]{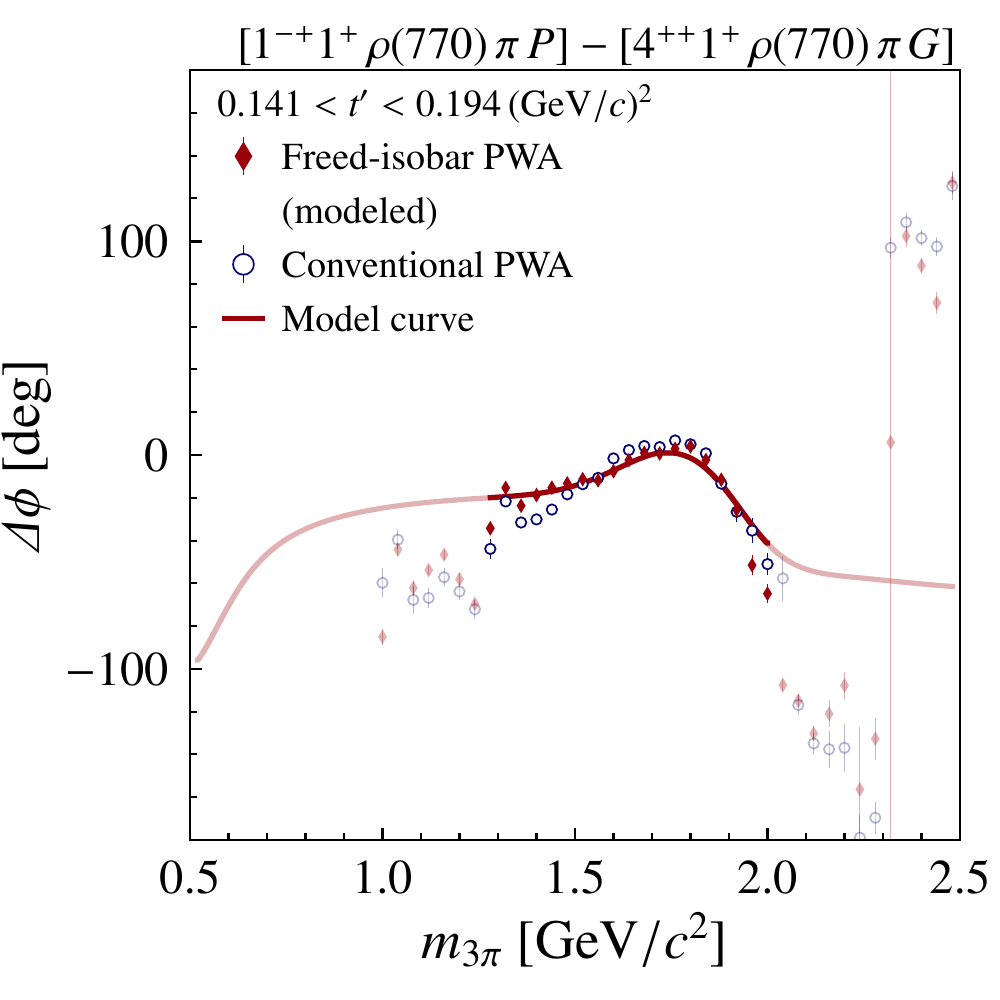}%
  }%
  \\%
  \subfloat[][]{%
    \includegraphics[width=\twoPlotWidth]{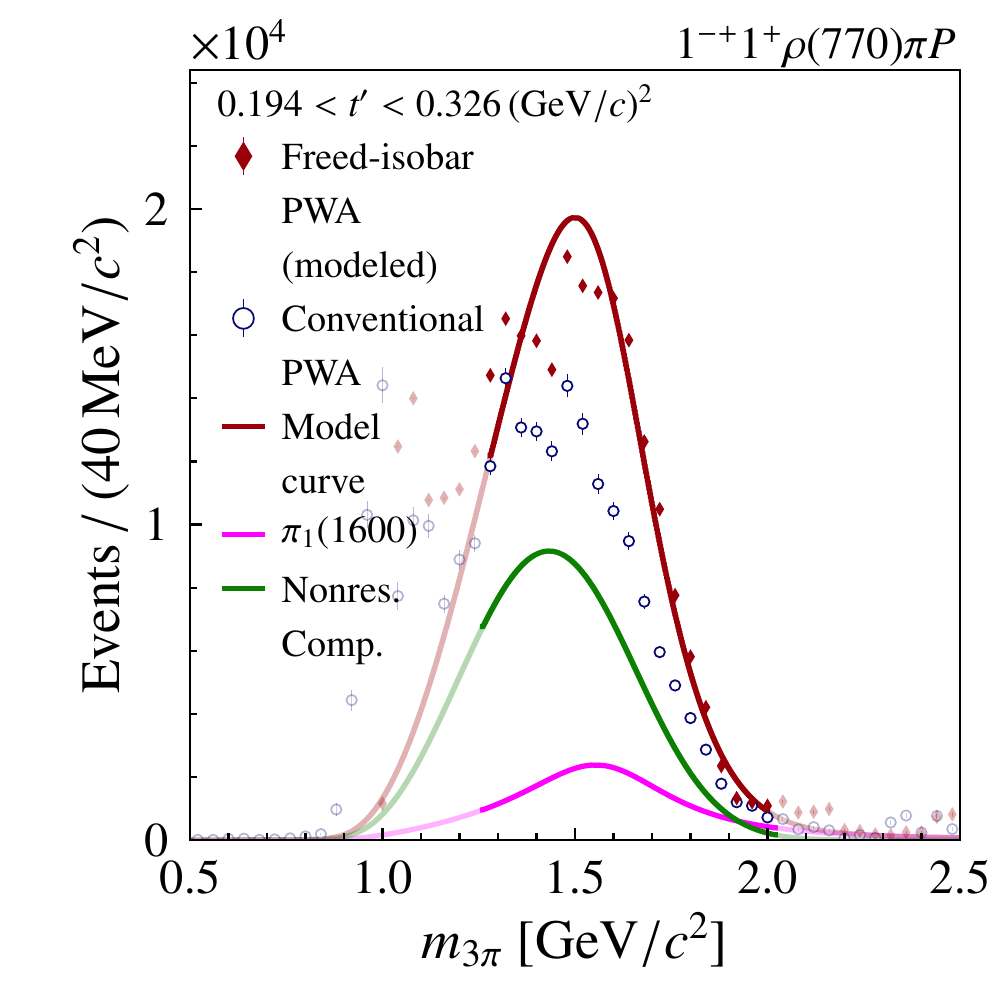}%
  }%
  \hspace*{\twoPlotSpacing}%
  \subfloat[][]{%
    \includegraphics[width=\twoPlotWidth]{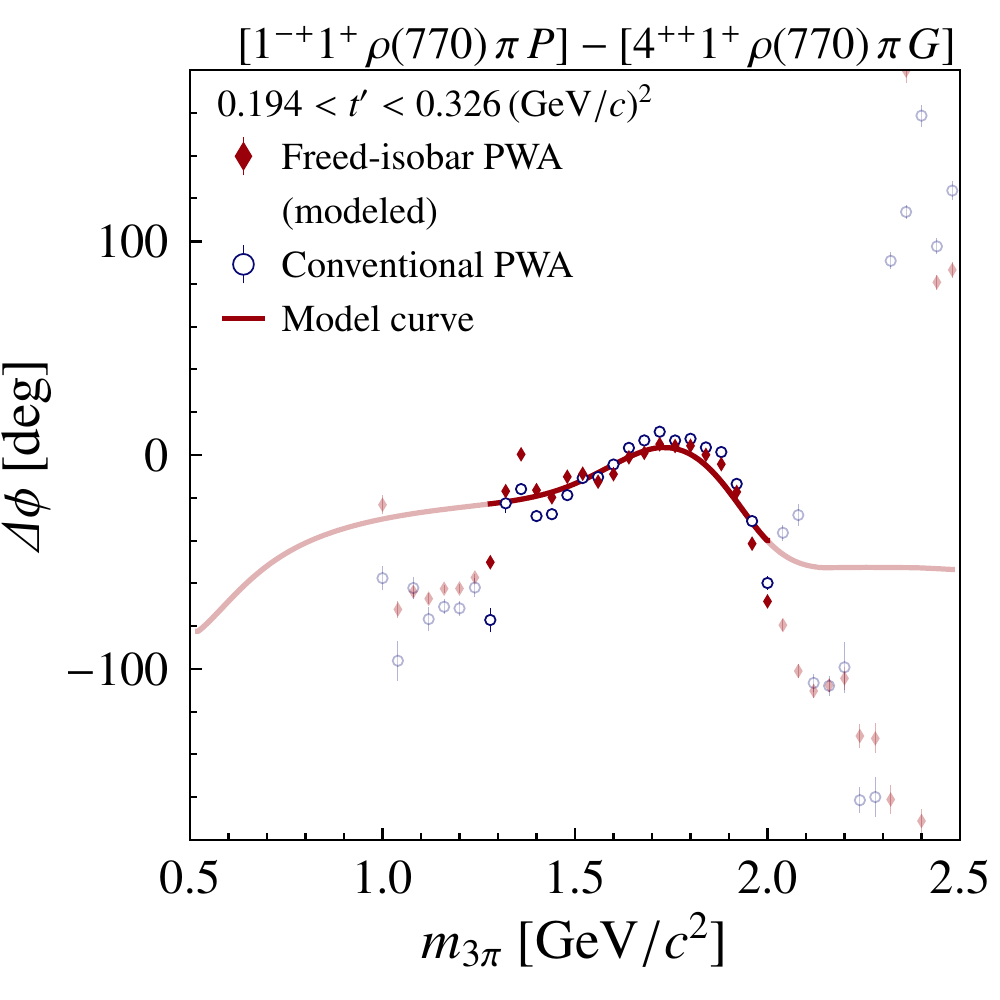}%
  }%
  \caption{Intensity distribution and phase of the spin-exotic wave
    \wrt the \wave{4}{++}{1}{+}{\Prho}{G} wave in the three lowest
    \tpr~bins, similar to \ifMultiColumnLayout{Fig.~9 in
    \refCite{paper4}}{\cref{fig:modeledComparison}}.}%
  \label{fig:modelComparison2}
\end{figure}

\clearpage
\section{Dynamic isobar amplitudes from freed-isobar partial-wave analysis}%
\label{suppl:freed_isobar_dyn_amp}
\newcommand*{\submissionFormat}{}  %

In addition to the three $(\mThreePi, \tpr)$ cells shown in
\ifMultiColumnLayout{Fig.~6 in \refCite{paper4}}{\cref{fig:slices}},
we show in \Crefrange{fig:slices_first_0}{fig:slices_11_3} the \pipiPF
dynamic isobar amplitude in the \wave{1}{-+}{1}{+}{\pipiPF}{P} wave as
a function of~\mTwoPi for all 152~$(\mThreePi, \tpr)$ cells with
$\mThreePi > \SI{0.98}{\GeVcc}$.

\newcommand*{\figFileNameOffset}{17}  %
\newcommand*{\figFileNameStride}{13}  %
\foreach \tbin\tname in {%
  0/\SIvalRange{0.100}{\tpr}{0.141}{\GeVcsq},%
  1/\SIvalRange{0.141}{\tpr}{0.194}{\GeVcsq},%
  2/\SIvalRange{0.194}{\tpr}{0.326}{\GeVcsq},%
  3/\SIvalRange{0.326}{\tpr}{1.000}{\GeVcsq}%
}%
{%
\subsection{\pipiPF dynamic isobar amplitude for \tname}

\begin{figure}[h]
  \centering%
  \subfloat[][]{%
    \newcommand*{\figureFilePath}{\ifSubmissionFormat{%
      fig\the\numexpr\tbin*\figFileNameStride+\figFileNameOffset\relax a}{%
      2pi_fits/supplemental/1mp1p1mmP_int_m12_t\tbin}}%
    \includegraphics[width=\twoPlotWidth]{\figureFilePath}%
  }%
  \hspace*{\twoPlotSpacing}%
  \subfloat[][]{%
    \newcommand*{\figureFilePath}{\ifSubmissionFormat{%
      fig\the\numexpr\tbin*\figFileNameStride+\figFileNameOffset\relax b}{%
      2pi_fits/supplemental/1mp1p1mmP_arg_m12_t\tbin}}%
    \includegraphics[width=\twoPlotWidth]{\figureFilePath}%
  }%
  \\%
  \subfloat[][]{%
    \newcommand*{\figureFilePath}{\ifSubmissionFormat{%
      fig\the\numexpr\tbin*\figFileNameStride+\figFileNameOffset\relax c}{%
      2pi_fits/supplemental/1mp1p1mmP_int_m13_t\tbin}}%
    \includegraphics[width=\twoPlotWidth]{\figureFilePath}%
  }%
  \hspace*{\twoPlotSpacing}%
  \subfloat[][]{%
    \newcommand*{\figureFilePath}{\ifSubmissionFormat{%
      fig\the\numexpr\tbin*\figFileNameStride+\figFileNameOffset\relax d}{%
      2pi_fits/supplemental/1mp1p1mmP_arg_m13_t\tbin}}%
    \includegraphics[width=\twoPlotWidth]{\figureFilePath}%
  }%
  \caption{Intensity distributions and \Argands similar to
    \ifMultiColumnLayout{Fig.~6 in
    \refCite{paper4}}{\cref{fig:slices}} for
    \SIvalRange{0.98}{\mThreePi}{1.06}{\GeVcc} and \tname.}%
  \label{fig:slices_first_\tbin}
\end{figure}

\clearpage
\foreach \mbin\mlow\mup in {%
  0/1.06/1.18,%
  1/1.18/1.30,%
  2/1.30/1.42,%
  3/1.42/1.54,%
  4/1.54/1.66,%
  5/1.66/1.78,%
  6/1.78/1.90,%
  7/1.90/2.02,%
  8/2.02/2.14,%
  9/2.14/2.26,%
  10/2.26/2.38,%
  11/2.38/2.50%
}%
{%
\begin{figure}[t]
  \centering%
  \subfloat[][]{%
    \newcommand*{\figureFilePath}{\ifSubmissionFormat{%
      fig\the\numexpr\tbin*\figFileNameStride+\mbin+1+\figFileNameOffset\relax a}{%
      2pi_fits/supplemental/1mp1p1mmP_int_m\the\numexpr\mbin*3+14_t\tbin}}%
    \includegraphics[width=\twoPlotWidth]{\figureFilePath}%
  }%
  \hspace*{\twoPlotSpacing}%
  \subfloat[][]{%
    \newcommand*{\figureFilePath}{\ifSubmissionFormat{%
      fig\the\numexpr\tbin*\figFileNameStride+\mbin+1+\figFileNameOffset\relax b}{%
      2pi_fits/supplemental/1mp1p1mmP_arg_m\the\numexpr\mbin*3+14_t\tbin}}%
    \includegraphics[width=\twoPlotWidth]{\figureFilePath}%
  }%
  \\%
  \subfloat[][]{%
    \newcommand*{\figureFilePath}{\ifSubmissionFormat{%
      fig\the\numexpr\tbin*\figFileNameStride+\mbin+1+\figFileNameOffset\relax c}{%
      2pi_fits/supplemental/1mp1p1mmP_int_m\the\numexpr\mbin*3+15_t\tbin}}%
    \includegraphics[width=\twoPlotWidth]{\figureFilePath}%
  }%
  \hspace*{\twoPlotSpacing}%
  \subfloat[][]{%
    \newcommand*{\figureFilePath}{\ifSubmissionFormat{%
      fig\the\numexpr\tbin*\figFileNameStride+\mbin+1+\figFileNameOffset\relax d}{%
      2pi_fits/supplemental/1mp1p1mmP_arg_m\the\numexpr\mbin*3+15_t\tbin}}%
    \includegraphics[width=\twoPlotWidth]{\figureFilePath}%
  }%
  \\%
  \subfloat[][]{%
    \newcommand*{\figureFilePath}{\ifSubmissionFormat{%
      fig\the\numexpr\tbin*\figFileNameStride+\mbin+1+\figFileNameOffset\relax e}{%
      2pi_fits/supplemental/1mp1p1mmP_int_m\the\numexpr\mbin*3+16_t\tbin}}%
    \includegraphics[width=\twoPlotWidth]{\figureFilePath}%
  }%
  \hspace*{\twoPlotSpacing}%
  \subfloat[][]{%
    \newcommand*{\figureFilePath}{\ifSubmissionFormat{%
      fig\the\numexpr\tbin*\figFileNameStride+\mbin+1+\figFileNameOffset\relax f}{%
      2pi_fits/supplemental/1mp1p1mmP_arg_m\the\numexpr\mbin*3+16_t\tbin}}%
    \includegraphics[width=\twoPlotWidth]{\figureFilePath}%
  }%
  \caption{Intensity distribution and \Argands similar to
    \ifMultiColumnLayout{Fig.~6 in
    \refCite{paper4}}{\cref{fig:slices}} for
    \SIvalRange{\mlow}{\mThreePi}{\mup}{\GeVcc} and \tname.}%
  \label{fig:slices_\mbin_\tbin}
\end{figure}
\clearpage
}}

\clearpage{}%

\end{document}